\documentclass[11pt]{article}  

\usepackage[top=1in,bottom=1in,left=1in,right=1in]{geometry}

\usepackage{graphicx}

\usepackage{amssymb,amsfonts,amsmath}
\usepackage{comment}
\usepackage{cite}
\usepackage{subfigure}           
\usepackage{caption}
\usepackage{xr}                
\usepackage[T1]{fontenc}
\usepackage[utf8]{inputenc}

\usepackage{mathtools}
\newcommand{\defeq}{\vcentcolon=}

\newcommand{\sref}[1]{Section~\ref{#1}}
\newcommand{\tref}[1]{Table~\ref{#1}}
\newcommand{\fref}[1]{Figure~\ref{#1}}

\def\bfmath#1{\mathchoice
        {\mbox{\boldmath$#1$}}%
        {\mbox{\boldmath$#1$}}%
        {\mbox{\boldmath$\scriptstyle#1$}}%
        {\mbox{\boldmath$\scriptscriptstyle#1$}}}%

\def\fall#1#2{(#1)_{#2\downarrow}}
\def\rise#1#2{(#1)_{#2\uparrow}}

\def\bbE{\mathbb{E}}
\def\bbP{\mathbb{P}}

\def\eps{\varepsilon}

\def\modelOne{\textsc{Model~1}}
\def\modelTwo{\textsc{Model~2}}
\def\modelGravel{\textsc{Model~3}}
\def\modelTennessen{\textsc{Model~4}}

\begin{document}
\captionsetup{aboveskip=-0.2\normalbaselineskip}

\begin{center}  
{\LARGE Distortion of genealogical properties when the sample is very large}

\renewcommand*{\thefootnote}{\fnsymbol{footnote}}

\vspace{8mm}
Anand Bhaskar$^a$,\hspace{5mm} Andrew G. Clark$^b$,\hspace{5mm} Yun S. Song$^{a,c,}$\footnote{To whom correspondence may be addressed: yss@cs.berkeley.edu}
\vspace{4mm}

$^a${Computer Science Division, University of California, Berkeley, CA 94720, USA}\\
$^b${Department of Molecular Biology and Genetics, Cornell University, Ithaca, NY 14853, USA}\\
$^c${Department of Statistics, University of California, Berkeley, CA 94720, USA}
\end{center}

\vspace{3mm}
\begin{abstract} 
	Study sample sizes in human genetics are growing rapidly, and in due course it will become routine to analyze samples with hundreds of thousands if not millions of individuals.  In addition to posing computational challenges, such large sample sizes call for carefully re-examining the theoretical foundation underlying commonly-used analytical tools.  Here, we study the accuracy of the coalescent, a central model for studying the ancestry of a sample of individuals.  The coalescent arises as a limit of a large class of random mating models and it is an accurate approximation to the original model provided that the population size is sufficiently larger than the sample size.  We develop a method for performing exact computation in the discrete-time Wright-Fisher (DTWF) model and compare several key genealogical quantities of interest with the coalescent predictions.  For realistic demographic scenarios, we find that there are a significant number of multiple- and simultaneous-merger events under the DTWF model, which are absent in the coalescent by construction.  Furthermore, for large sample sizes, there are noticeable differences in the expected number of rare variants between the coalescent and the DTWF model. To balance the tradeoff between accuracy and computational efficiency, we propose a hybrid algorithm that utilizes the DTWF model for the recent past and the coalescent for the more distant past.  Our results demonstrate that the hybrid method with only a handful of generations of the DTWF model leads to a frequency spectrum that is quite close to the prediction of the full DTWF model.
\end{abstract}

\section{Introduction}
Human genetics has entered a new era where the study sample sizes regularly exceed ten thousand, a number commonly cited as the effective population size of humans \cite{takahata:1993,erlich:1996,harding:1997,harpending:1998}.  A consistent finding arising from recent large-sample studies \cite{coventry:2010, nelson:2012, tennessen:2012, fu:2012} is that human genomes harbor a substantial excess of rare variants compared to that predicted using previously applied demographic models.  
For example, Nelson \emph{et al.} \cite{nelson:2012} found that over 70\% of single nucleotide variants are singletons and doubletons, which corresponds to a minor allele frequency on the order of 0.01\% for their study sample.  There are several factors that may contribute to the discrepancy between observations in the data and theoretical predictions, including the following possible explanations:
\begin{enumerate}
	\item Previously applied demographic models are wrong.
      In particular, the observed polymorphism patterns are indicative of a recent rapid growth of the effective population size, much more rapid than in previously applied demographic models.  This conclusion would be consistent with historical records of census population size \cite{keinan:2012}.    
	\item Population substructure \cite{ptak:2002, stadler:2009} and natural selection have distorted the observed polymorphism patterns while previous demographic inference studies have failed to adequately account for these factors.
	\item Theoretical predictions for a given demographic model are inaccurate when the sample size is very large.             
   Coalescent theory, which arises as a limit of a large class of discrete-time random mating models, provides an accurate approximation to the original discrete-time model only if the effective population size is sufficiently larger than the sample size. Violation of this assumption may distort genealogical properties in a way that may inflate rare variants relative to the predictions of coalescent theory.          
\end{enumerate}                                     

The goal of this paper is to investigate the last possibility in detail, by examining the deviation between the coalescent and a well-known discrete-time random model, namely the Wright-Fisher (WF) model. 

Kingman's coalescent \cite{kingman:1982:SPA, kingman:1982:EPS, kingman:1982:JAP}, henceforth simply referred to as the coalescent, is a central model in modern population genetics for studying the ancestry of a sample of individuals taken from a large randomly mating population. The coalescent is a continuous-time Markov process that can be constructed as a scaling limit of a discrete-time Wright-Fisher (DTWF) model, by taking the population size to infinity while rescaling the unit of time by the population size. The dynamics of a DTWF model can be complicated, in which multiple sets of lineages can find common ancestors in a single generation. In contrast, at most two lineages can find a common ancestor at any given time under the coalescent, and hence it is a mathematically and algorithmically more tractable model. The coalescent is an excellent approximation to the original discrete-time model if, for all times, the population size is sufficiently large relative to the number of ancestral lineages of a sample, in which case multiple and simultaneous mergers of lineages in a single generation are unlikely.

In this paper, we investigate whether the coalescent continues to be a good approximation to the DTWF model in the case where the sample size increases to the point where the coalescent assumptions may be violated.  We compare the two models under certain demographic scenarios previously considered in the literature, including the case of recent rapid population growth for humans \cite{gravel:2011, tennessen:2012}.  We examine several key genealogical statistics of interest such as the number of multiple and simultaneous mergers in the DTWF model, the number of lineages as a function of time (NLFT), and the sample frequency spectrum. A key feature of our work is that all our results, under both the coalescent and the  DTWF model, are based on \emph{exact} deterministic computations rather than Monte-Carlo simulations.

To perform exact computation in the DTWF model, we exploit the Markov property of the model and devise dynamic programming algorithms to compute various genealogical quantities of interest exactly.        
These algorithms are computationally expensive, so we also consider a hybrid method that uses the DTWF model for the recent past and the coalescent for the more distant past.  We demonstrate that this hybrid approach produces substantially more accurate predictions than does the coalescent, while being more efficient than performing computation in the full DTWF model.

\section{Results}\label{sec:results}

\subsection{Demographic models}\label{sec:models}

\begin{figure}[t]
	\begin{center}
	\subfigure[]{\label{fig:pop_growth_model1}
    	\includegraphics[width=0.45\textwidth]{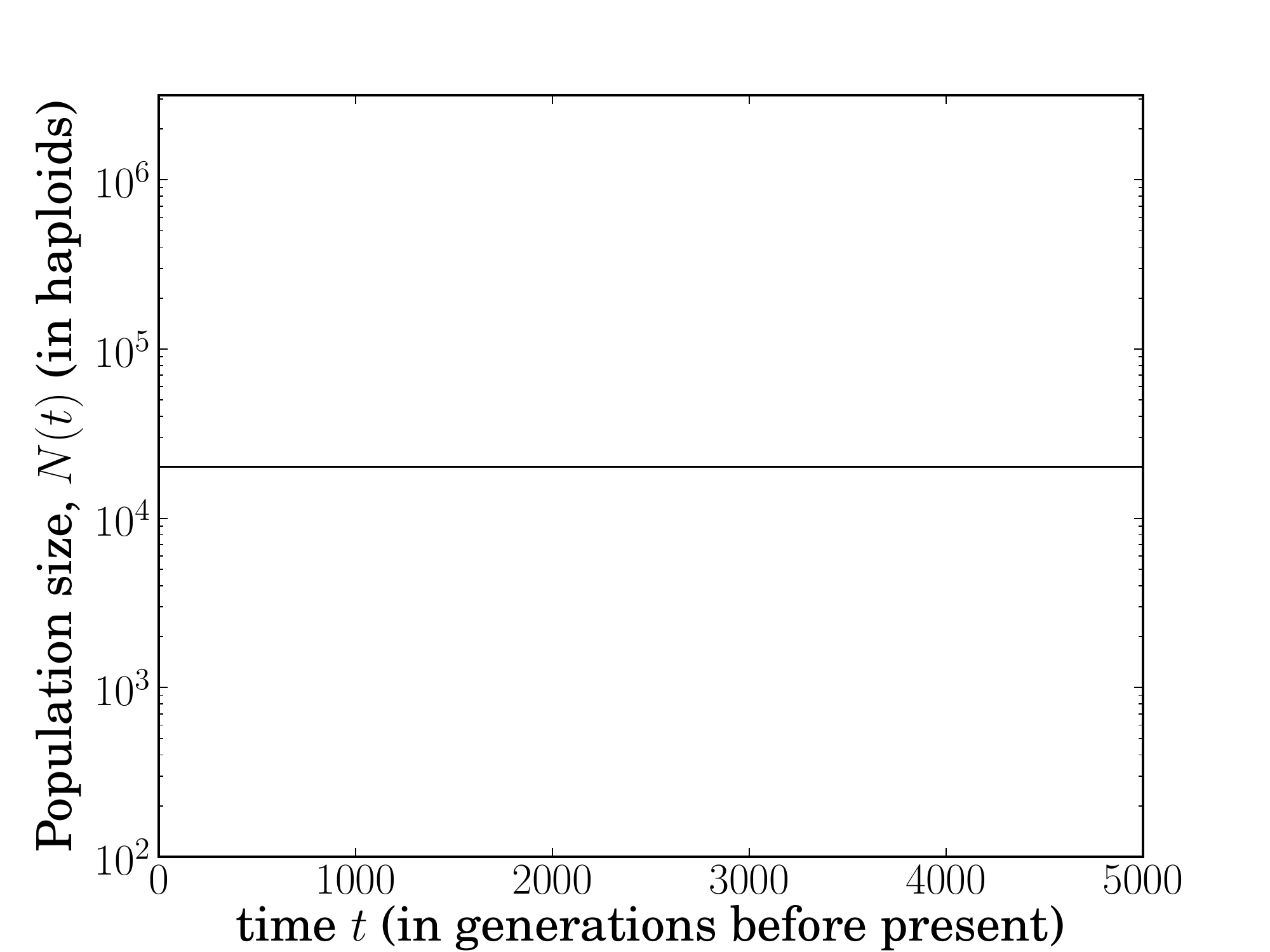}
    }
	\subfigure[]{\label{fig:pop_growth_model2}
    	\includegraphics[width=0.45\textwidth]{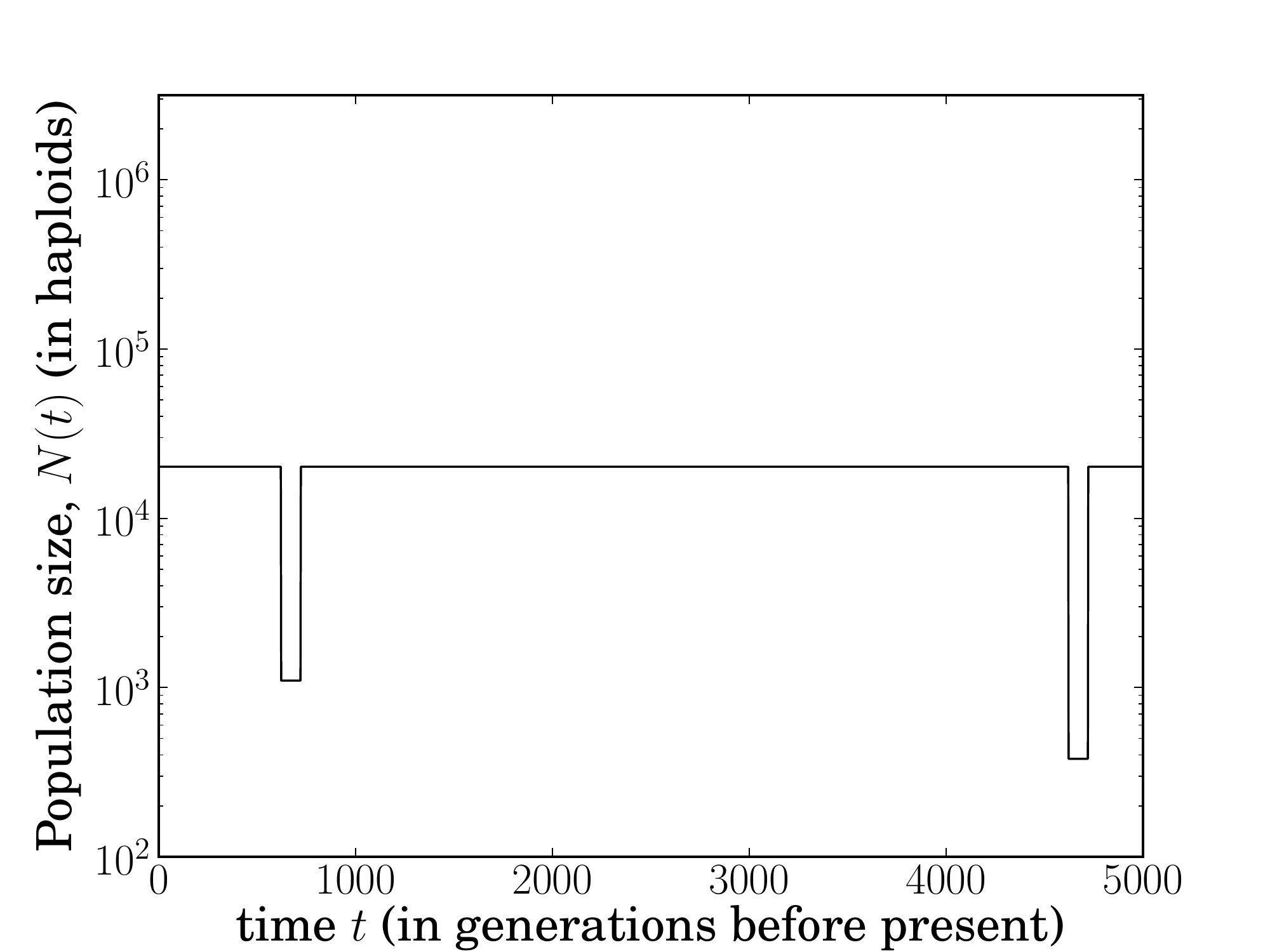}
    }
	\subfigure[]{\label{fig:pop_growth_model3}
	    	\includegraphics[width=0.45\textwidth]{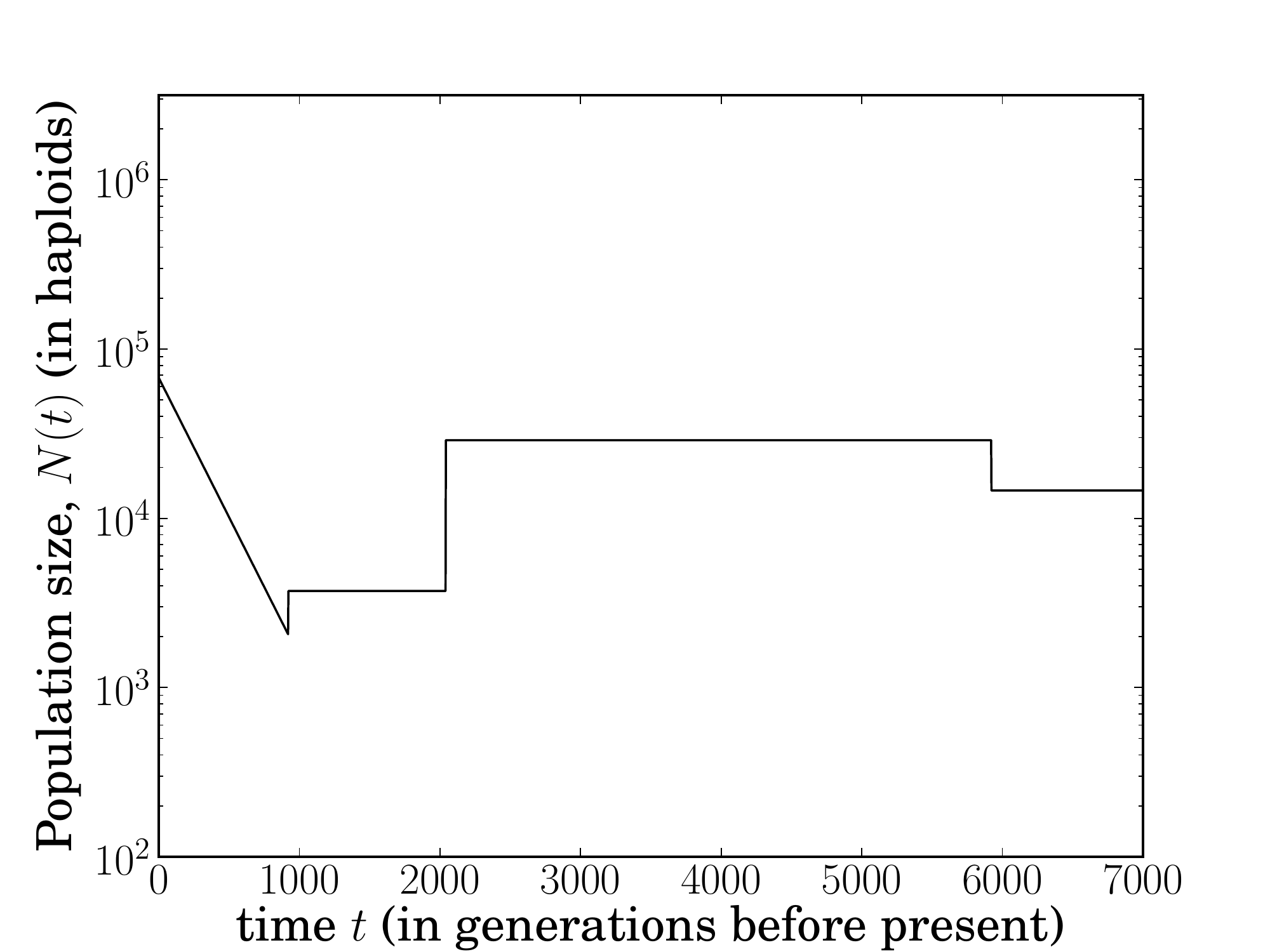}
    }
	\subfigure[]{\label{fig:pop_growth_model4}
	    	\includegraphics[width=0.45\textwidth]{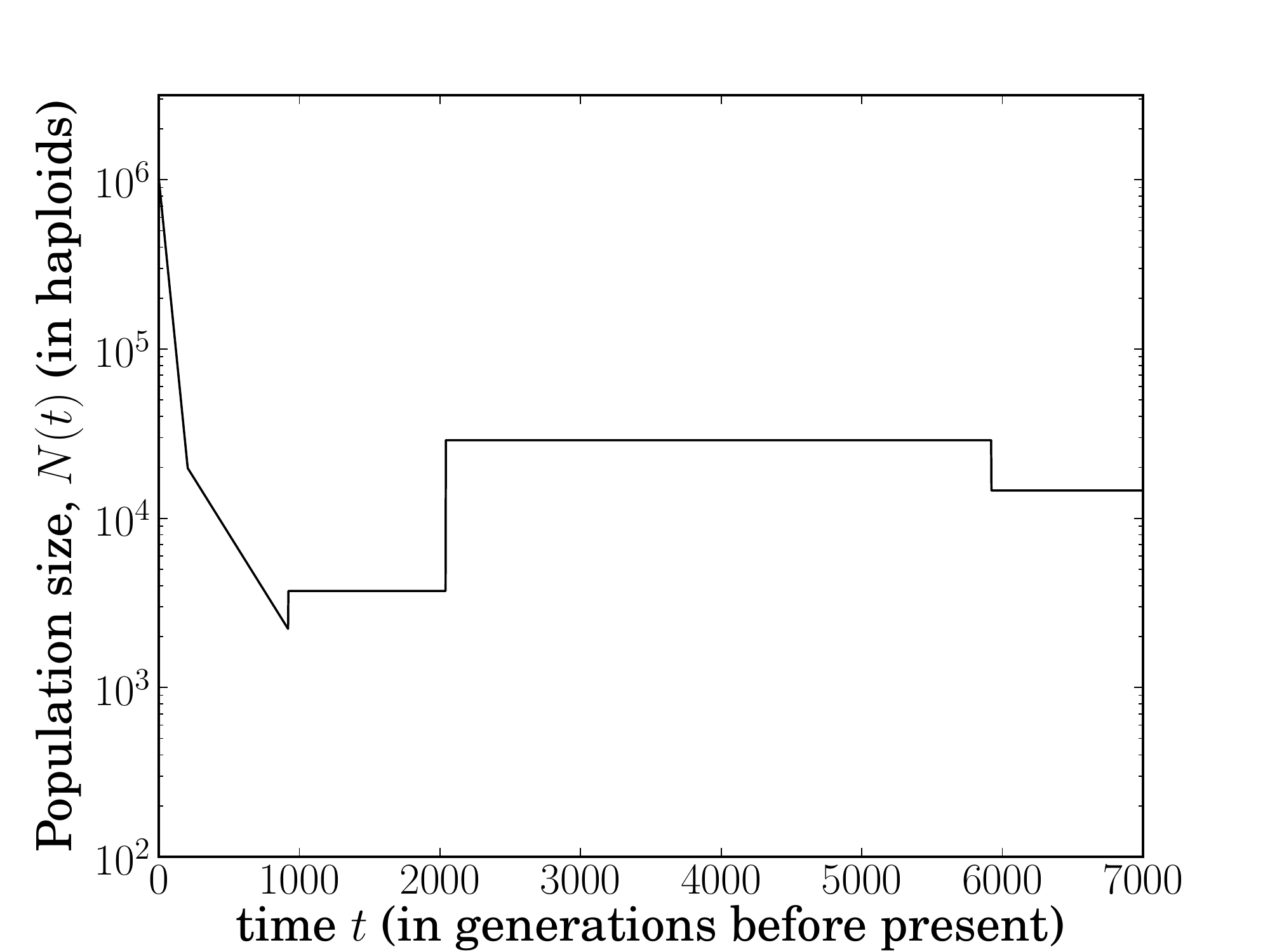}
    }
	\end{center}
    \caption{{\bf Demographic models examined in this work.} Each graph shows the effective population size $N_t$ (in haploids) as a function of time (number of generations ago) in a Wright-Fisher model of random mating. The present time corresponds to $t=0$. The demographic models are labeled as \subref{fig:pop_growth_model1} \modelOne, \subref{fig:pop_growth_model2} \modelTwo, \subref{fig:pop_growth_model3} \modelGravel, and \subref{fig:pop_growth_model4} \modelTennessen.}
    \label{fig:pop_growth_models}
\end{figure}

In addition to the case of a constant population size, we consider three models of variable population size. The details of the demographic models we consider are provided below and illustrated in \fref{fig:pop_growth_models}:
\begin{itemize}
	\item \modelOne: A constant population size of 10,085 diploid individuals.
	\item \modelTwo: Proposed in \cite{keinan:2007}, this model has two population bottlenecks, the most recent of which lasted for 100 generations starting from 620 generations in the past, and a more ancient bottleneck lasting 100 generations, starting from 4,620 generations in the past. Further back in time, the population size is fixed at 10,085 diploids.
	\item \modelGravel: This demographic model was inferred by Gravel \emph{et al.} \cite{gravel:2011} for the CEU subpopulation from the 1000 Genomes \cite{1000G:2010} exon pilot data.  In this model, a population expansion in the last 920 generations occurs at a rate of 0.38\% per generation.
	\item \modelTennessen: This demographic model was inferred by Tennessen \emph{et al.} \cite[Figure 2B]{tennessen:2012} for the CEU subpopulation from exome-sequencing of 2440 individuals. The ancient demography is similar to that in \modelGravel. However, following the most recent bottleneck, there are two epochs of exponential expansion in the most recent 920 generations -- a slower expansion phase for 716 generations at 0.307\% per generation, followed by a rapid expansion rate of 1.95\% per generation for 204 generations.
\end{itemize}

Using the above four demographic models, we examine deviations in the following quantities between the coalescent and the DTWF model:
\begin{itemize}
    \item[(a)] Multiple and simultaneous mergers in the DTWF model.
    \item[(b)] Number of lineages as a function of time (NLFT).
    \item[(c)] Expected sample frequency spectrum.
\end{itemize}

\subsection{Multiple and simultaneous mergers in the DTWF model}

For a given demographic model in the DTWF framework, it becomes more likely that multiple lineages may be lost in a single generation as the sample size $n$ increases. The first-order approximations used in the derivation of the coalescent from the DTWF model assume that the sample size $n$ is on the order of $\sqrt{N}$, with $N$ being the population size. For example, consider a sample of size $n = 250$ with an effective population size of $N = 20{,}000$ haploids. \fref{fig:prob_drops_n250} shows the probability distribution of the number of parents of the sample in the previous generation. There is a high probability that the sample will have less than $n - 1$ parents in the previous generation, an event which is ignored in the asymptotic calculation used in the coalescent derivation from the DTWF model.
\fref{fig:fraction_lineages_lost_to_non_2drops} shows the expected fraction of lineages (relative to $n-1$) that are lost due to either multiple or simultaneous mergers, from the present up to time $t$ in the past.  
\tref{tab:fraction_lineages_lost_to_non_2drops} shows the numerical values of the expected fraction as $t\to\infty$.
The sharp jump in the plot for \modelTwo~(Figure S\ref{fig:fraction_lineages_lost_to_non_2drops_model2}) corresponds to the beginning (backwards in time) of population bottlenecks when the population size declines substantially, thus instantaneously increasing the rate at which lineages find common ancestors and are lost. 
For small sample sizes relative to the population size, it is unlikely for more than one lineage to be lost in a single generation, as can be seen in the plots for $n = 20$ and $n=200$.
In contrast, for large sample sizes ($n = 2 \times 10^4$), almost all the lineages are lost in generations when more than one lineage is lost.  

When multiple lineages are lost in a single generation of the DTWF model, there are several ways this could happen. For example, suppose 2 out of $m$ lineages are lost in one generation. This could be the result of 3 lineages finding the same parent in the previous generation, or two pairs of lineages each finding a common parent, with the two parents being different.  In general,  there are $S(m, j)$ different ways that $m$ labeled lineages can have $j$ distinct parents in the previous generation, where $S(m, j)$ is the Stirling number of the second kind, counting the number of ways of partitioning a set of $m$ labeled objects into $j$ non-empty subsets. 
A particular pattern of mergers of $m$ lineages which leads to $j$ distinct parents, where $\lceil \frac{m}{2} \rceil \leq j \leq m$, is illustrated in \fref{fig:pairwise_simultaneous_merger}.  Here,  $m-j$ pairs of lineages each find a common parent distinct from all other parents, and the remaining $2j-m$ lineages do not merge with any other lineages.  There are $j$ ancestral lineages left after this type of merger. We call this an \emph{$(m-j)$-pairwise-simultaneous merger}. 
For $k\geq 2$, we use the term \emph{$k$-merger} to denote an event where exactly $k$ lineages find the same common parent in the previous generation.    It is possible to have several multiple merger events in a single generation.  
For example, a $j$-pairwise-simultaneous merger is equivalent to there being exactly $j$ 2-merger events and no other merger events in a single generation.

In the coalescent, since at most 2 lineages find a common ancestor in any given time, the only kind of possible merger is a single 2-merger (or equivalently, a 1-pairwise-simultaneous merger). On the other hand, in a DTWF model with $m$ lineages at a given time, there are $\frac{1}{2} {m \choose 2} {m - 2 \choose 2}$ possible 2-pairwise-simultaneous mergers, and ${m \choose 3}$ possible 3-mergers, yielding the following expression for the total number of different ways for $m$ lineages to find $m-2$ distinct parents in the previous generation: 
\begin{equation} 
	S(m, m-2) = {m \choose 3} + \frac{1}{2} {m \choose 2} {m - 2 \choose 2}.
	 \label{eq:S(m,m-2)} 
\end{equation} 
Since the second term is $O(m^4)$ while the first term is $O(m^3)$, for large $m$ we expect 2-pairwise-simultaneous mergers to be the dominant reason for losing 2 lineages in a single generation.

\fref{fig:fraction_lineages_lost_to_simultaneous_mergers_non_2drops} illustrates the ratio of the sum of the expected number of lineages lost due to $k$-pairwise-simultaneous mergers, for $k \geq 2$, to the results shown in \fref{fig:fraction_lineages_lost_to_non_2drops}, the expected number of lineages lost due to multiple or simultaneous mergers, from the present up to time $t$ in the past.  As \fref{fig:fraction_lineages_lost_to_simultaneous_mergers_non_2drops} shows, a substantial fraction of the lineages that are lost in generations with multiple lost lineages (i.e. in generations with mergers forbidden in the coalescent) are due to pairwise simultaneous mergers.
 Incidentally, that the curves for $n=20$ starts out near $0.93$ can be attributed to the fact that the ratio of the second term in the right hand side of \eqref{eq:S(m,m-2)} to $S(m,m-2)$ is $\frac{51}{55}$ for $m=20$.

The expected fraction (relative to $n-1$) of lineages lost due to $k$-mergers is shown in \tref{tab:k-mergers}. A substantial number of lineages are lost to 3-mergers in \modelOne, \modelTwo~and \modelGravel~for $n = 2 \times 10^4$ because the sample size is of the same order as the population size at time $0$.  Even in \modelTennessen,  about 1.9\% of lineages participate in 3-mergers. \fref{fig:fraction_3mergers_function_of_time} shows the fraction of 3-mergers up to time $t$ relative to the total expected number of 3-mergers as $t \to \infty$. As expected, in \modelOne, \modelTwo, and \modelGravel, due to the large sample size relative to the population size at time $0$, a substantial portion of the 3-mergers take place very early when the number of surviving lineages drops quickly. It is rather surprising that in \modelTennessen, where there is a rapid exponential population growth, a large fraction of the 3-mergers in fact take place during this period of growth. In particular, more than 25\% of the expected 3-mergers for $n = 2 \times 10^4$ occur in the most recent 32 generations when the effective population size is at least $5.5 \times 10^5$.

\begin{table}[t]
\caption{\textbf{Expected percentage of lineages (relative to $n-1$, where $n$ is the sample size) lost due to $k$-mergers in \textsc{Models 1--4}.} 
In \modelOne~and \modelTwo~ for $n = 2 \times 10^4$, a substantial number of lineages are involved in 3-mergers, and more than 6\% of the lineages are involved in 4-mergers, because the sample size is of the same order as the current population size. Even in \modelGravel~and \modelTennessen~for $n = 2 \times 10^4$, around 9\% and 2\% of the lineages participate in 3-mergers, respectively.}
\label{tab:k-mergers}
\begin{center}        
{\scriptsize
\begin{tabular}{cccccccccccccccccc}\hline
        &   \multicolumn{2}{c}{\modelOne} &   \multicolumn{2}{c}{\modelTwo}	&   \multicolumn{2}{c}{\modelGravel} 	&   \multicolumn{2}{c}{\modelTennessen}	\\
    $k$ &   $n = 2 \times 10^3$  &   $n = 2 \times 10^4$  &   $n = 2 \times 10^3$  &   $n = 2 \times 10^4$  &   $n = 2 \times 10^3$  &   $n = 2 \times 10^4$     &	$n = 2 \times 10^3$	&	$n = 2 \times 10^4$	\\
    \hline
    2   &   96.71\% 	&   68.94\% 	&   96.66\%     	&   68.93\% 	&   98.77\%     &   89.99\%	    &	98.96\%		&	98.11\%		\\
    3	&   3.21\%  	&   22.93\% 	&   3.26\%      	&   22.93\% 	&   1.22\%      &   9.25\%      &	1.03\%		&	1.87\%   	\\
    4   &   0.08\%  	&   6.36\%  	&   0.08\%      	&   6.36\%  	&   0.01\%      &   0.72\%      &	0.01\%		&	0.02\%    	\\
	\hline
\end{tabular}
}
\end{center}
\end{table}

Based on the results described above, one would expect that the number of ancestral lineages remaining at a given time decreases more rapidly under the DTWF model than under the coalescent, and we investigate this quantity next.

\subsection{Number of lineages as a function of time}

Here, we compare the expected number of lineages as a function of time (NLFT) in the coalescent and in the DTWF model.  In what follows, we let $A^C_n(t)$ and $A^D_n(t)$ denote the random variables for the number of lineages at generation $t$ in the coalescent and the DTWF model, respectively, starting with a sample of size $n$ at time 0.
Under the coalescent, the expectation and standard deviation of the NLFT, $\bbE[A^C_n(t)]$ and $\sigma(A^C_n(t))$, can be computed exactly in a numerically-stable fashion for an arbitrary variable population size model as described in \sref{sec:methods}. An algorithm to compute $\bbE[A^D_n(t)]$ and $\sigma(A^D_n(t))$ under the DTWF model is also described there.

For the four demographic models considered, \fref{fig:wf_fraction_lineages_surviving} shows the expectation and standard deviation of the NLFT under the DTWF model, 
while Figures~\ref{fig:fraction_lineages_surviving_WF_vs_coalescent} and \ref{fig:stddev_lineages_surviving_WF_vs_coalescent} show the relative differences in the expectation and standard deviation, respectively, of the NLFT in the coalescent with respect to the NLFT in the DTWF model.   
For large sample sizes under \modelOne, \modelTwo, and \modelGravel, it can be seen that the lineages are lost at a faster rate in the DTWF model than in the coalescent. This pattern is consistent with the fact that these demographic models exhibit a substantial number of 3-mergers in the DTWF model for large sample sizes (see \tref{tab:k-mergers}), though the deviation in the expected NLFT is still substantially less than the expected number of 3-mergers. The deviation disappears after about 1000 generations when enough time has passed for the number of ancestral lineages to become sufficiently small that the coalescent approximation holds.

For \modelTennessen, the expected NLFT in the coalescent provides a fairly good approximation to that in the DTWF model for all times and for all sample sizes considered. This is because the population size remains much larger than the number of ancestral lineages at all times.

\subsection{Expected sample frequency spectrum}

Given a sample of $n$ haploid (or $n/2$ diploid) individuals, a common summary of the sample used in various population genetic analyses is the sample frequency spectrum, $\bfmath{\hat{\tau}}_n = (\hat{\tau}_{n,1},\ldots,\hat{\tau}_{n,n-1})$. Under the infinite-sites model of mutation, the $k$th entry $\hat{\tau}_{n,k}$ corresponds to the number of polymorphic sites in the sample that have $k$ derived alleles and $n - k$ ancestral alleles, where $1 \leq k \leq n-1$. For a sample of $n$ haploids randomly drawn from the population, we denote the expected value of $\hat{\tau}_{n,k}$ in the coalescent and the DTWF models by $\tau^{C}_{n,k}$ and $\tau^{D}_{n,k}$, respectively.  In the case of a constant population size, $\tau^{C}_{n,k}$ under the infinite-sites model of mutation is given exactly by the expression
\begin{align}
    \tau^{C}_{n,k} = \frac{\theta}{k},
\end{align}                    
where $\theta$ denotes a population-scaled mutation rate. (Mutations arise according to a Poisson process with intensity $\theta/2$ in each lineage, independently of all other lineages.)  For variable population size models, the results of Polanski and Kimmel \cite{polanski:2003-1} can be used to compute the expected sample frequency spectrum numerically-stably under the coalescent.  In \sref{sec:methods}, we develop an algorithm to compute the expected sample frequency spectrum under the DTWF model, denoted by $\bfmath{\tau}^D_n = (\tau^D_{n,1},\ldots,\tau^D_{n,n-1})$.

\begin{figure}[t]    
	\begin{center}
	\subfigure[]{\label{fig:wf_vs_coal_first2Entries_unnormalized_model1}
		\includegraphics[width=0.45\textwidth]{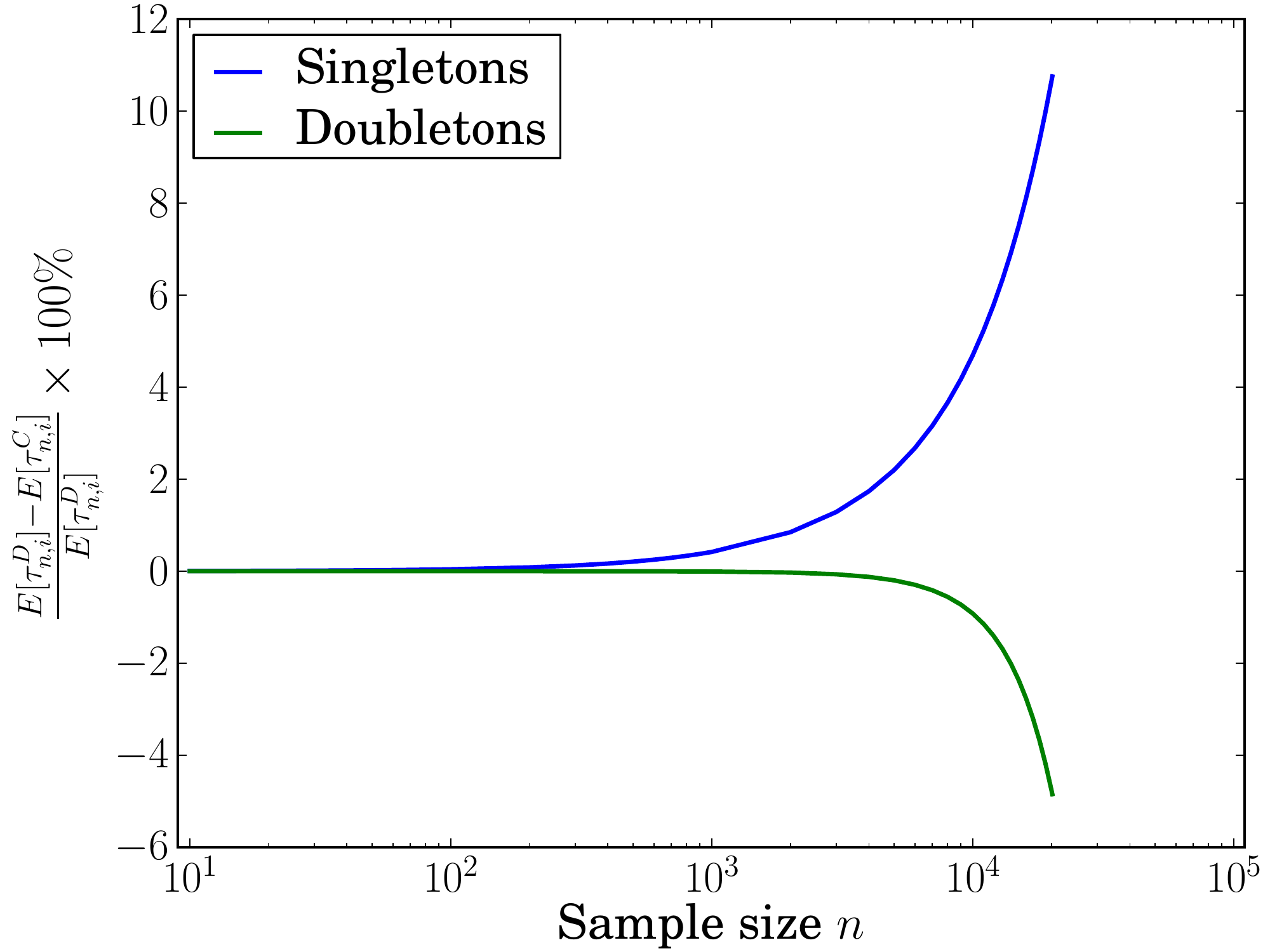}
	}
	\subfigure[]{\label{fig:wf_vs_coal_first2Entries_unnormalized_model2}
		\includegraphics[width=0.45\textwidth]{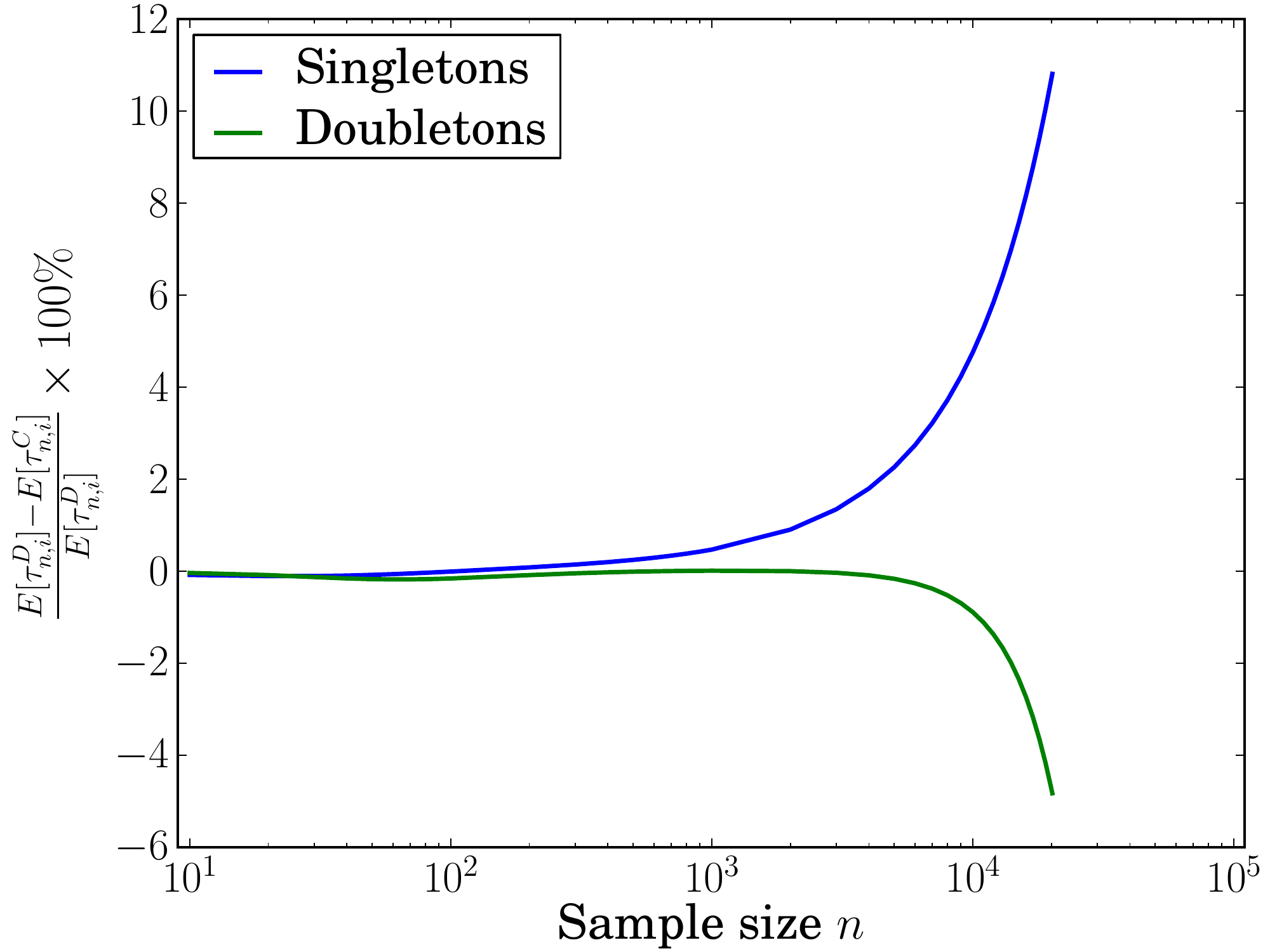}
	}
 	\subfigure[]{\label{fig:wf_vs_coal_first2Entries_unnormalized_model3} 
	    \includegraphics[width=0.45\textwidth]{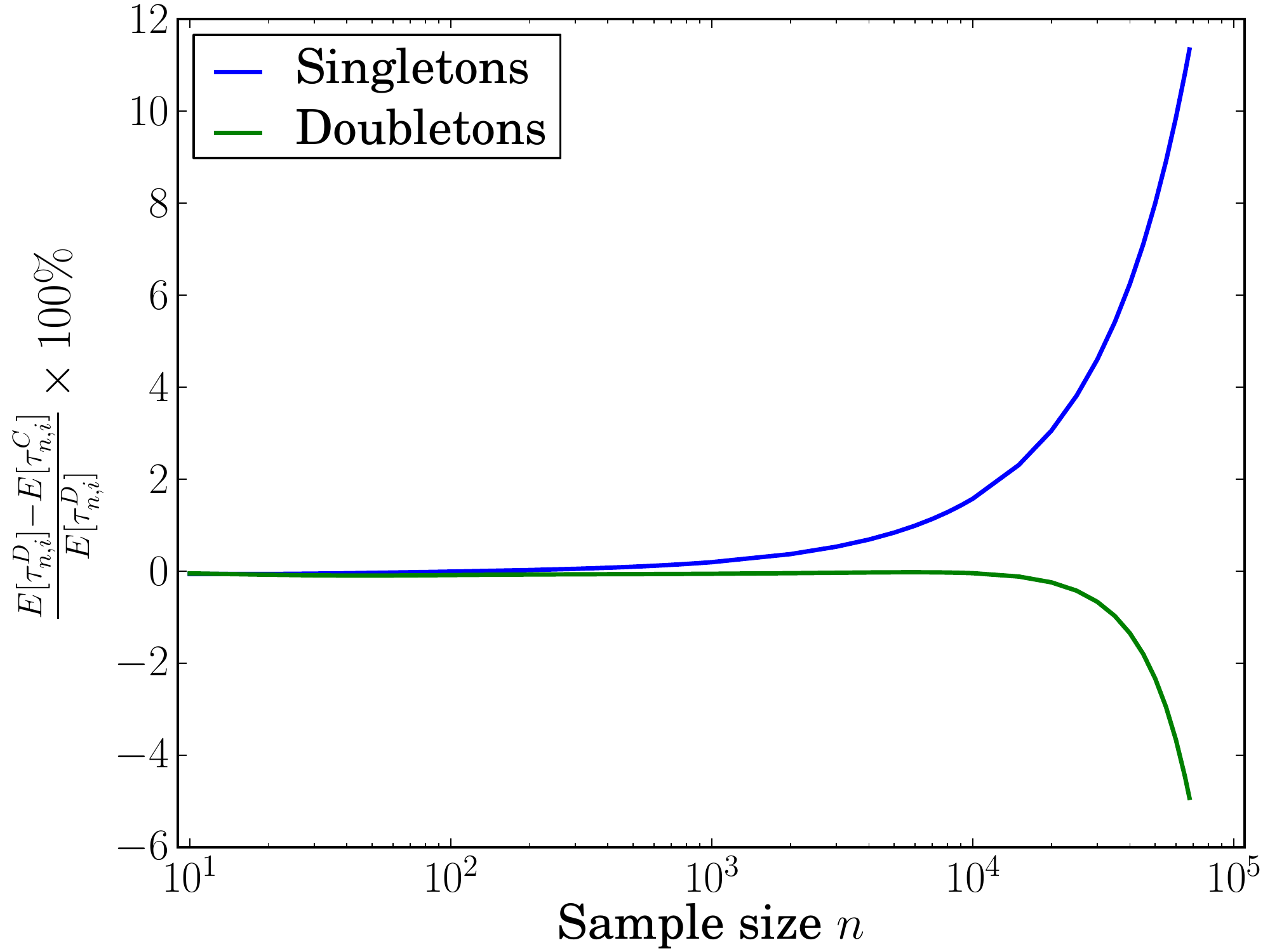}	    
	}
	\subfigure[]{\label{fig:wf_vs_coal_first2Entries_unnormalized_model4} 
		\includegraphics[width=0.45\textwidth]{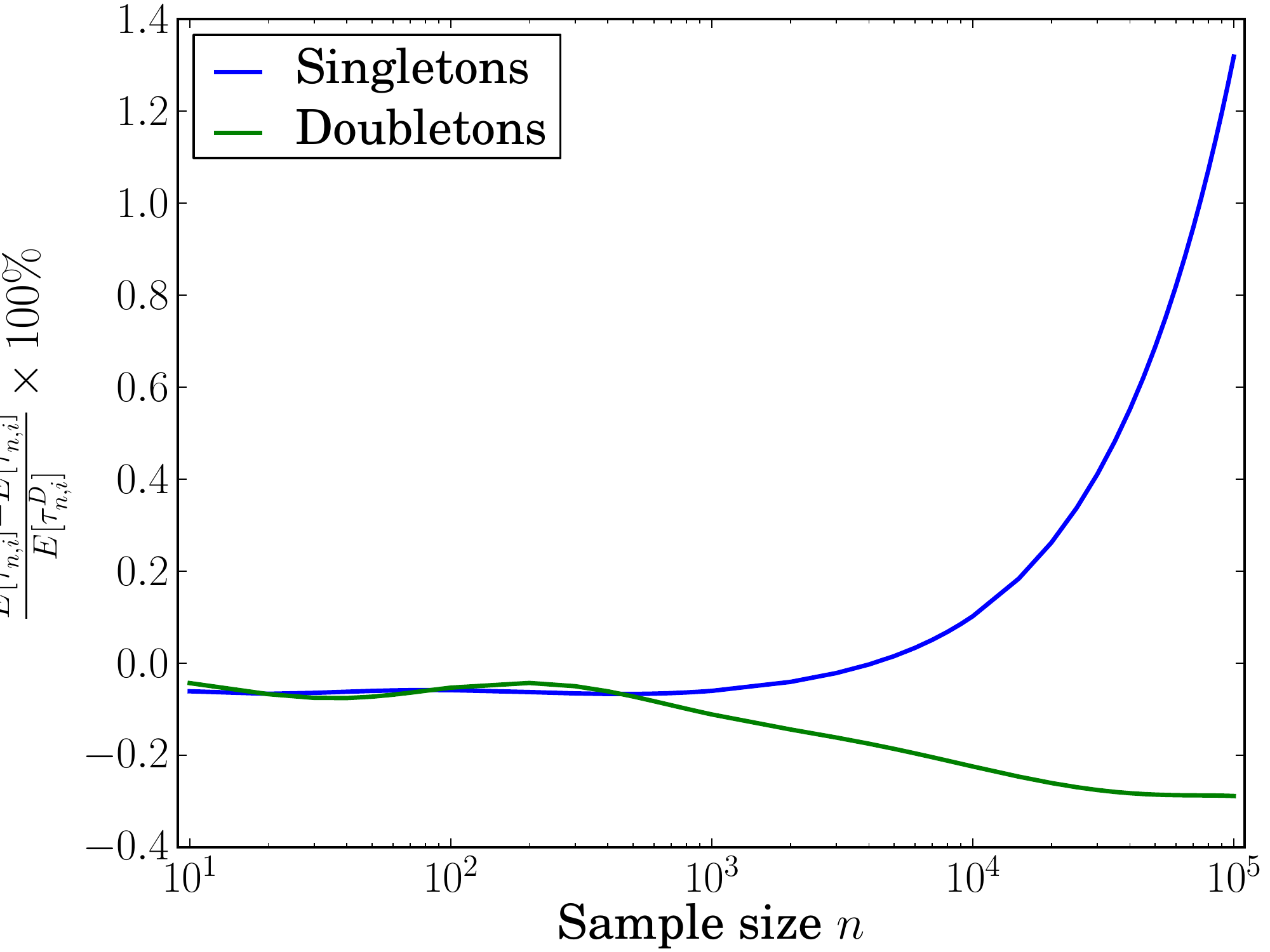} 
	}
	\end{center}
	\caption{{\bf The percentage relative error in the number of singletons and doubletons between the coalescent and DTWF models, as a function of the sample size $n$.}
   When the sample size is comparable to the current population size,  the number of singletons predicted by the DTWF model is \emph{larger} than the coalescent prediction by as much as 11\%, while the number of doubletons predicted by the DTWF model is \emph{smaller} than the coalescent prediction by about 4.8\%.  In \modelTennessen, we could not consider a sample size comparable to the population size ($10^6$) because of computational burden, but we expect a similar extent of deviation as in \textsc{Models 1--3} as $n$ increases.
\subref{fig:wf_vs_coal_first2Entries_unnormalized_model1} \modelOne.
\subref{fig:wf_vs_coal_first2Entries_unnormalized_model2} \modelTwo.
\subref{fig:wf_vs_coal_first2Entries_unnormalized_model3} \modelGravel.
\subref{fig:wf_vs_coal_first2Entries_unnormalized_model4} \modelTennessen.
	}
	\label{fig:freq_spectrum_WF_vs_coalescent_singletons_doubletons}
\end{figure}

\fref{fig:freq_spectrum_WF_vs_coalescent_singletons_doubletons} illustrates the relative difference between the coalescent and the DTWF model in the number of singletons ($\tau_{n,1}$) and doubletons ($\tau_{n,2}$) as a function of the sample size ($n$). 
As the figure shows, the number of singletons predicted by the DTWF model is \emph{larger} than the coalescent prediction by as much as 11\% when the sample size is comparable to the current population size.  For \modelTennessen~ (\fref{fig:wf_vs_coal_first2Entries_unnormalized_model4}), we could not consider sample sizes $> 10^5$ because of computational burden, but the results for \textsc{Models 1--3} suggest that we should expect to observe $\geq 10\%$ deviation when the sample size $n$ is increased to $10^6$, the current population size in \modelTennessen.
The deviation in the number of doubletons is also significant when the sample size is comparable to the current population size; the DTWF prediction for doubletons is \emph{smaller} than the coalescent prediction by about $4.8\%$.

The findings described above are especially important given that rare variants comprise a large fraction of segregating sites when the sample size is large.
In \fref{fig:cdf_freq_spectrum_WF}, we plot the cumulative distribution of the frequency spectrum in the DTWF model for \textsc{Models} 1--4. The number of singletons in \textsc{Models 3} and \textsc{4} is higher than in \textsc{Models 1} and \textsc{2} due to exponential population growth. The rapid population expansion in \modelTennessen~results in about 51\% of the segregating sites being singletons and over 80\% of the segregating sites having less than 5 copies of the derived allele in a sample of size $n = 2 \times 10^4$.
\fref{fig:coal_freq_spectrum_top_entries} shows the expected proportion of rare variants (derived allele frequency $\leq 0.01\%$) as a function of the sample size $n$ for \textsc{Models 3} and \textsc{4} under the coalescent. It can be seen that as $n$ approaches the current population size, the proportion of rare variants increases substantially. \fref{fig:coal_freq_spectrum_singletons} shows the expected proportion of segregating sites that are singletons as a function of $n$ for \textsc{Models 1--4} under the coalescent.  For small sample sizes (say, $n< 100$), the proportion of singletons in \textsc{Model 3} and \textsc{4} (which incorporate rapid recent population expansion) is not much larger than that in \textsc{Models 1}~and \textsc{2}. However, the difference increases considerably as the sample size goes beyond a few hundred individuals, illustrating the need for large sample sizes to infer recent population expansion from frequency spectrum data.

\subsection{A hybrid method for computing the frequency spectrum}

As detailed in \sref{sec:methods}, computation in the DTWF model is substantially more involved than in the coalescent.   In particular,  while computation in the coalescent depends only on the number of piecewise-exponential epochs and not the duration of each epoch,  the runtime of our dynamic programming algorithm for the DTWF model actually depends on the number of generations over which the algorithm is run.                                                             
Since noticeable deviation between the DTWF model and the coalescent arises when the number of ancestral lineages is not negligible compared to the population size, a reasonable tradeoff between accuracy and runtime would be to use the DTWF model for the recent past and the coalescent for the more distant past (when the number of ancestral lineages has decreased sufficiently).

\begin{figure}[t]
		\begin{center} 
		\subfigure[]{\label{fig:wf_first2Entries_unnormalized_modelGravel_t1}
			\includegraphics[width=0.45\textwidth]{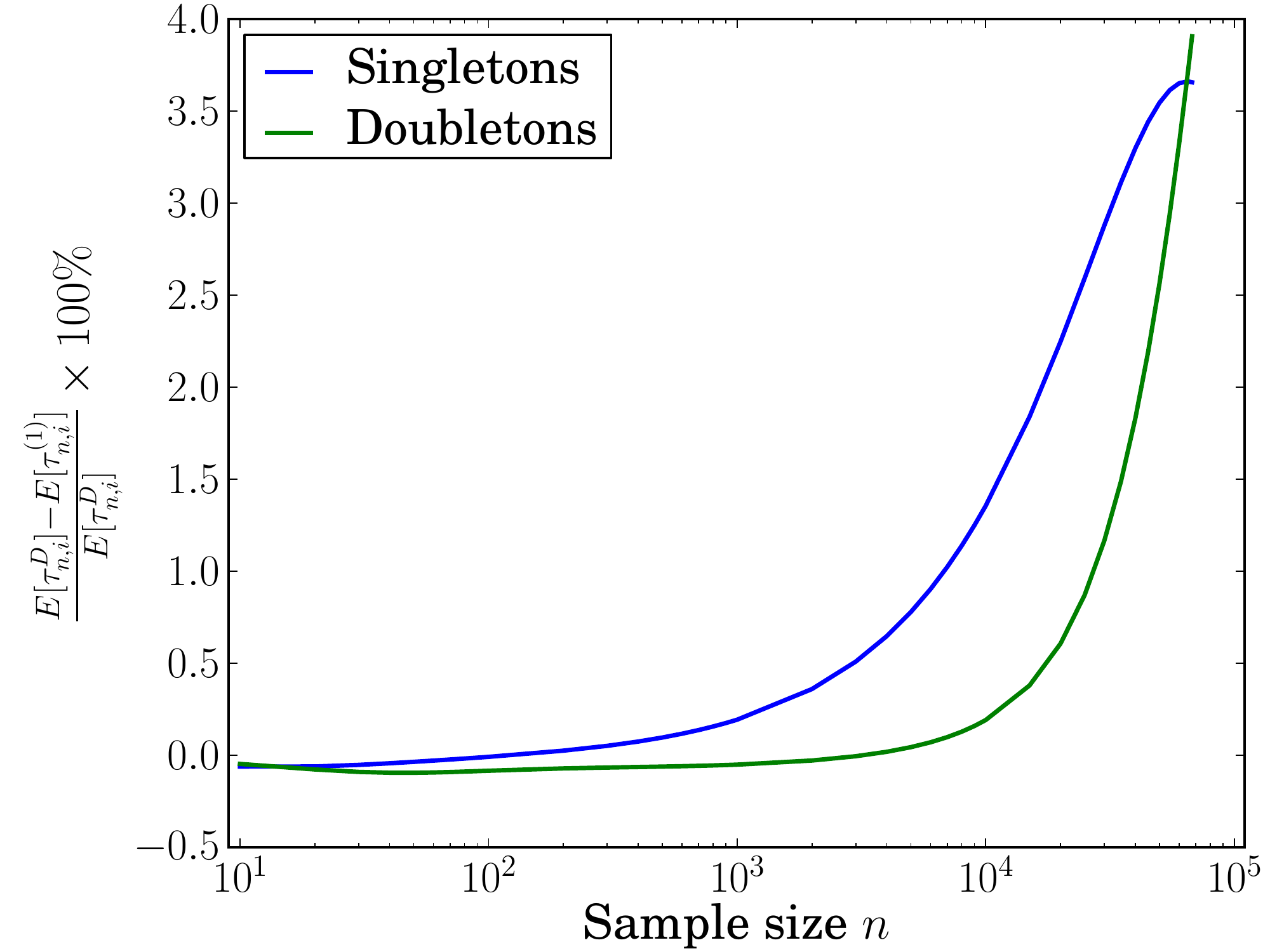}
		}
		\subfigure[]{\label{fig:wf_first2Entries_unnormalized_modelGravel_t5} 
			\includegraphics[width=0.45\textwidth]{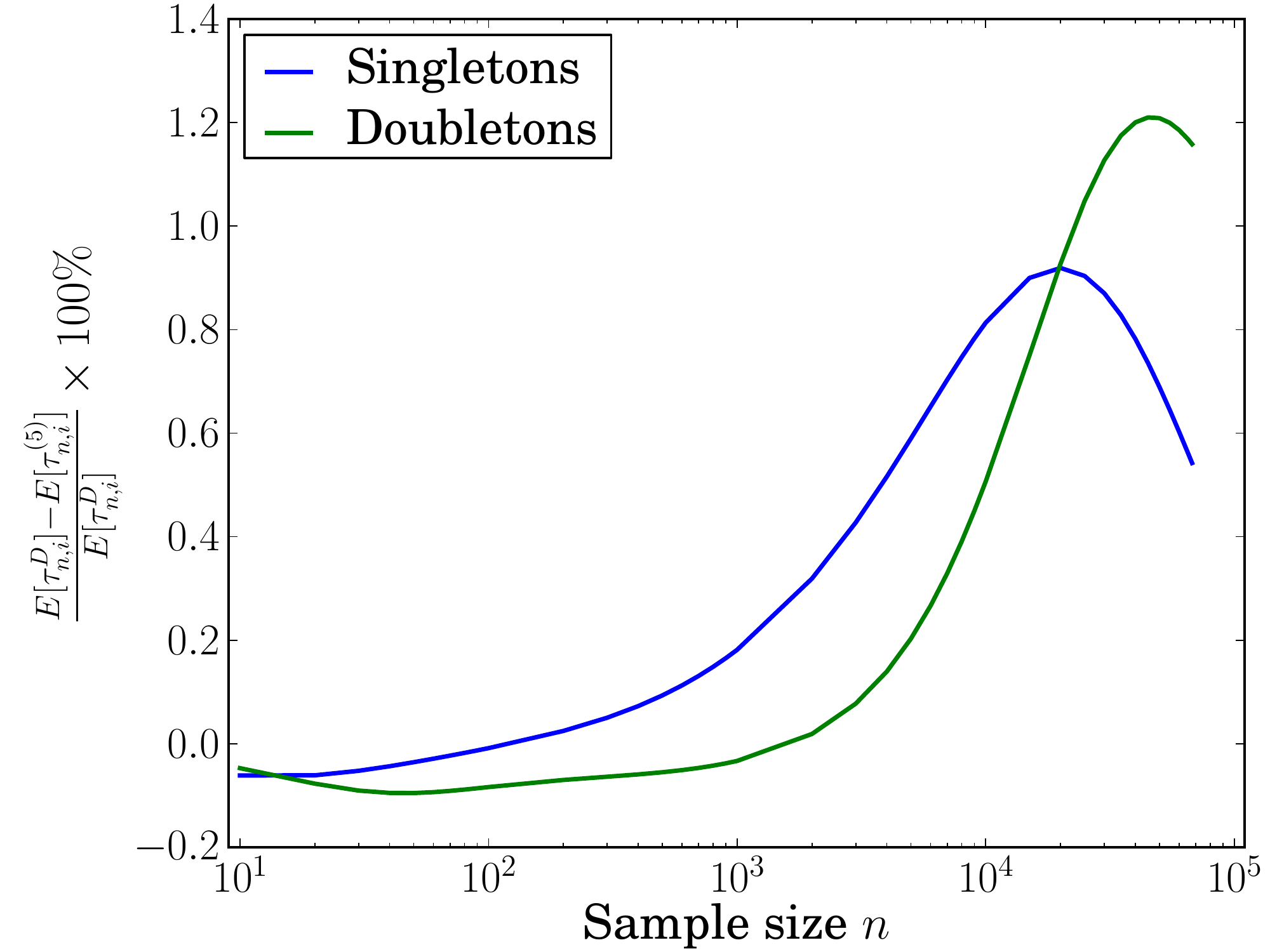} 
		}
		\subfigure[]{\label{fig:wf_first2Entries_unnormalized_modelGravel_t10} 
			\includegraphics[width=0.45\textwidth]{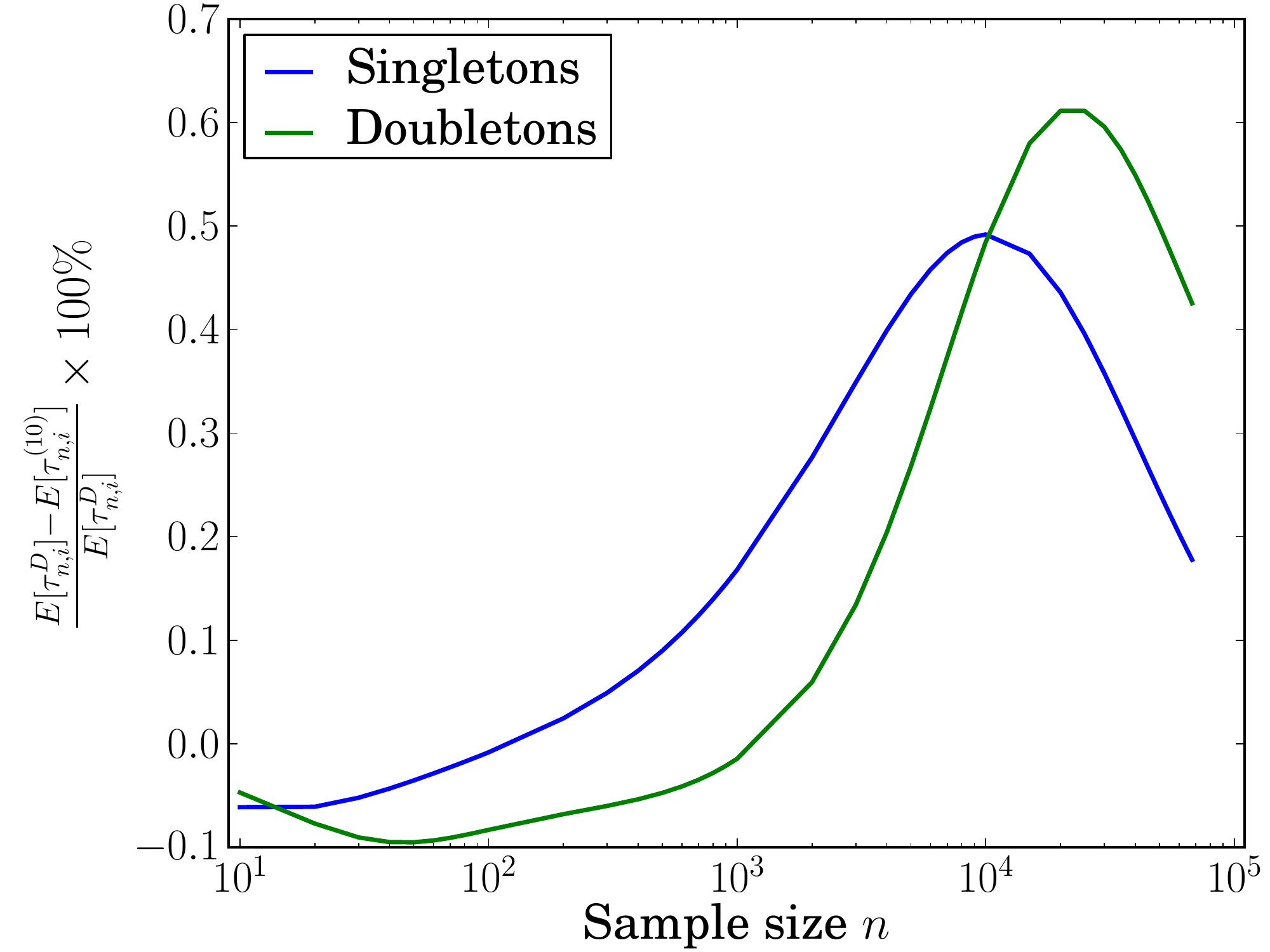} 
		}
		\subfigure[]{\label{fig:wf_first2Entries_unnormalized_modelGravel_t25} 
			\includegraphics[width=0.45\textwidth]{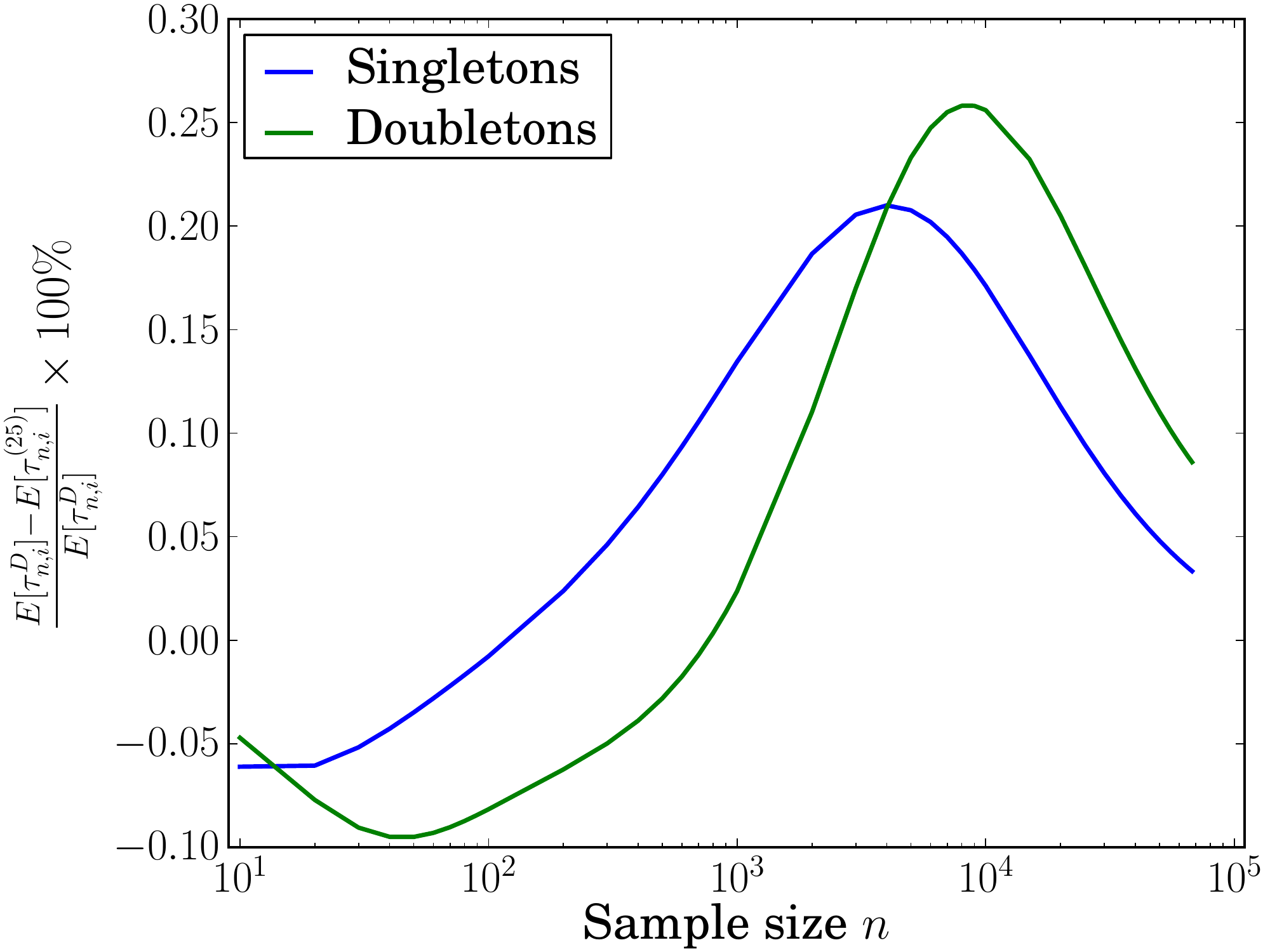} 
		}     
		\end{center}
		\caption{{\bf The percentage relative error in the number of singletons and doubletons between the computations under the full DTWF model and a hybrid algorithm which uses the coalescent model in generations $t > t_s$ for } \textsc{\modelGravel}.  The case of $t_s=0$ corresponds to using the coalescent model only, illustrated in \fref{fig:wf_vs_coal_first2Entries_unnormalized_model3}.  We use $\bfmath{\tau}^{(t_s)}$ to denote the frequency spectrum using the hybrid algorithm with a switching time of $t_s$.  As $t_s$ increases, we see that the deviation between the computation under the DTWF model and the hybrid algorithm decreases monotonically, suggesting that one could use such an algorithm to efficiently approximate the frequency spectrum under the DTWF model.
		The values of $t_s$ for the plots are
		\subref{fig:wf_first2Entries_unnormalized_modelGravel_t1} $t_s = 1$, 
		\subref{fig:wf_first2Entries_unnormalized_modelGravel_t5} $t_s = 5$,
		\subref{fig:wf_first2Entries_unnormalized_modelGravel_t10} $t_s = 10$,
		\subref{fig:wf_first2Entries_unnormalized_modelGravel_t25} $t_s = 25$.
		}
		\label{fig:freq_spectrum_WF_hybrid_singletons_doubletons}
\end{figure}

To explore this idea, we implemented a hybrid method for computing the frequency spectrum which, for a specified switching generation $t_s$, uses the full DTWF model for generations $0 < t \leq t_s$, followed by the coalescent for generations $t > t_s$.  In particular, for $t_s = 0$, this algorithm computes the frequency spectrum under the coalescent, while for $t_s = \infty$, it computes the frequency spectrum under the full DTWF model.  As \fref{fig:freq_spectrum_WF_hybrid_singletons_doubletons} illustrates for \modelGravel, the difference in the frequency spectrum between the full DTWF model and the hybrid algorithm decreases rapidly as $t_s$ increases.  With $t_s=5$ generations, the largest deviation in the number of singletons is less than $1\%$, which is a substantial reduction from $11\%$ for $t_s=0$ (\fref{fig:wf_vs_coal_first2Entries_unnormalized_model3}).

\section{Methods}                                                                                                                              
\label{sec:methods}
The computation of the various genealogical quantities in the DTWF model, such as the number of simultaneous- and multiple-mergers, the NLFT, and the expected frequency spectrum, rely on the Markov property of the DTWF model. By considering the types and counts of the mergers occurring in the previous generation, one can write down one-step recurrence equations relating these genealogical quantities over time and solve these recurrences by dynamic programming.  The details are provided below.
	%
	%
\subsection{Notation}
Let $N_t$ be the number of (haploid) individuals at generation $t$ in a DTWF model, where $t = 0$ corresponds to the present and $t$ is increasing going back in the past. Let $p^{(t)}_{n,m}$ denote the probability that a sample of $n$ labeled individuals at generation $t$ has $m$ distinct ancestors at generation $t+1$. Due to the random mating in a DTWF model, $p^{(t)}_{n,m}$ is given by the expression,
\begin{align*}
	p^{(t)}_{n,m} = S(n, m) {\fall{N_{t+1}}{m} \over N_{t+1}^n},
\end{align*}
where $S(n, m)$ is a Stirling number of the second kind which counts the number of ways of partitioning a set of $n$ labeled objects into $m$ non-empty unlabeled subsets, and $\fall{x}{k}$ denotes the falling factorial of $x$,
\begin{align}
\fall{x}{k} = 
\begin{cases} 
x (x - 1) \cdots (x - k + 1),	& \mbox{if } k \geq 1,	\\
1,	& \mbox{if } k = 0.
\end{cases}
\end{align}
The probabilities $p^{(t)}_{n,m}$ can be computed using dynamic programming without having to explicitly compute the Stirling numbers $S(n, m)$ using the following recursion for $1 \leq m \leq n$,
\begin{align}
p^{(t)}_{n,m} = \frac{N_{t+1} - m + 1}{N_{t+1}} p^{(t)}_{n-1,m-1} + \frac{m}{N_{t+1}} p^{(t)}_{n-1,m},	\label{eq:rec_p_n_m}
\end{align}
with the convention that $p^{(t)}_{0,0} \defeq 1$, and $p^{(t)}_{n,m} = 0$ for all other values of $m$ and $n$.
\eqref{eq:rec_p_n_m} can be derived by noting that $S(n,m) = S(n-1,m-1) + m S(n-1,m)$ for $1 \leq m \leq n$.

For an algorithmic reason that will become clear presently, we assume that there is a critical generation $t_c$ such that $N_t = N$ (some constant) for all $t > t_c$.  This assumption is not so restrictive since for sufficiently large $t$, there will be only 1 lineage left with high probability, and the genealogical properties we study will not be affected.  For $t > t_c$, we drop the dependence on $t$ in the probabilities $p^{(t)}_{n,m}$, and simply write them as $p_{n,m}$.

\subsection{Expected number of lineages as a function of time  under the DTWF model}\label{sec:wf_nlft}
By conditioning on the number of ancestral lineages in the previous generation, it is easy to write a recurrence relation for the distribution of the number of lineages at generation $t$ given that there are $n$ lineages at time 0,
\begin{align}
\bbP[A^D_n(t) = m] = 
\begin{cases} 
\sum_{k = m}^{n}p^{(t - 1)}_{k,m} \bbP[A^D_n(t - 1) = k],    &   \mbox{if } t > 0,   \\
\delta_{n,m},   &    \mbox{if } t = 0,
\end{cases} \label{eq:nlft_df}
\end{align}
where $\delta_{n,m} = 1$ if $n = m$, and $\delta_{n,m} = 0$ otherwise.
One can then compute the expectation $\bbE[A^D_n(t)]$ using \eqref{eq:nlft_df}.

\subsection{Expected number of multiple- and simultaneous-mergers in the DTWF model}
Let $M^{(t)}_{n,k}$ be the random variable denoting the number of $k$-mergers that occur in a genealogical tree starting with a sample of size $n$ observed at generation $t$. The expected number of $k$-mergers in a sample of size $n$ observed at present, $\bbE[M^{(0)}_{n,k}]$, can be computed by conditioning on the mergers that occur between generation $t$ and $t+1$. We then have the following recurrence relations for $\bbE[M^{(t)}_{n,k}]$,
\begin{align}
    \bbE[M^{(t)}_{n,k}] = 
    \begin{cases}
        {n \choose k}\sum_{m=k+1}^n p^{(t)}_{k,1} p^{(t)}_{n-k,m-k} {N_{t+1} - m + k \over N_{t+1}} + \sum_{m=1}^{n} p^{(t)}_{n,m} \bbE[M^{(t+1)}_{m,k}],  &   \text{if } k < n,   \\
        p^{(t)}_{k, 1} + p^{(t)}_{k,k} \bbE[M^{(t+1)}_{k,k}], &   \text{if } k = n.        
    \end{cases}\label{eq:exp_k_mergers_varpop}
\end{align}
If the population size is constant, or for $t > t_c$ when the population size remains fixed in the past, we can drop the dependence on $t$ in the notation $\bbE[M^{(t)}_{n,k}]$, and derive the following recurrence relations and boundary conditions for $\bbE[M_{n,k}]$:
\begin{align}
    \bbE[M_{n,k}] = 
    \begin{cases}
        {n \choose k}\sum_{m=k+1}^n {p_{k,1} p_{n-k,m-k} \over 1 - p_{n,n}} {N - m + k \over N} + \sum_{m=1}^{n-1} {p_{n,m} \over 1 - p_{n,n}} \bbE[M_{m,k}],  &   \text{if } k < n,   \\
        {p_{k,1} \over 1 - p_{k,k}},  &   \text{if } k = n.        
    \end{cases}\label{eq:exp_k_mergers}
\end{align}
One can write similar recurrence relations for the expected number of $k$-simultaneous mergers by conditioning on the mergers that occur during each generation of reproduction.

\subsection{Sample frequency spectrum under the DTWF model}\label{sec:methods_freq_spectrum}
The expected frequency spectrum of a randomly drawn sample of $n$ individuals is denoted $\bfmath{\tau}_n = (\tau_{n,1}, \ldots, \tau_{n,n-1})$, where $\tau_{n,k}$ corresponds to the number of polymorphic sites in the sample having $k$ derived alleles and $n-k$ ancestral alleles for $1 \leq k < n$ under an infinite-sites model of mutation.
For a given sample of individuals observed at present, the ancestral process in the DTWF model generates a genealogical tree, where the root of the tree is the most recent common ancestor (MRCA) of the sample, with the individuals in the sample forming the leaves of the tree. 
For the DTWF model, we can use dynamic programming to efficiently calculate $\tau_{n,k}$ as follows.
Let $\gamma^{(t)}_{a, b}$ be a random variable denoting the total branch length (in number of generations) of a subtree that subtends a particular set of $a$ labeled individuals in a larger set of $a+b$ individuals observed at time $t$. Then by the exchangeability of the individuals in the sample, the definition of $\gamma$, and linearity of expectation, we have,
\begin{align}
    \tau_{n,k} = \frac{\theta}{2} {n \choose k} \bbE[\gamma^{(0)}_{k, n - k}],    \label{eq:kth_entry_freq_spectrum}
\end{align}
since there are ${n \choose k}$ subsamples of $k$ individuals out of the $n$ individuals in the original sample.
By conditioning on the mergers between lineages that take place between generations $t$ and $t+1$, we get the following recurrence relations for $\bbE[\gamma^{(t)}_{a,b}]$,
\begin{align}
\bbE[\gamma^{(t)}_{a,b}] &= 
\begin{cases}
\sum_{j=1}^{a} \sum_{k=1}^{b} p^{(t)}_{a,j} p^{(t)}_{b,k} {\fall{N_{t+1}-j}{k} \over \fall{N_{t+1}}{k}} \bbE[\gamma^{(t+1)}_{j,k}]   & \text{if } a > 1,	\\
1 + \sum_{m=1}^{b} {N_{t+1} - m \over N_{t+1}} p^{(t)}_{b,m} \bbE[\gamma^{(t+1)}_{1,m}]	& \text{if } a = 1 \text{ and } b > 1,	\\
1 + p^{(t)}_{2,2} \bbE[\gamma^{(t+1)}_{1,1}] 	&	\text{if } a = b = 1.
\end{cases}\label{eq:dp_freq_spectrum_varpop}
\end{align}
If the population size is constant, or for $t > t_c$ when the population size remains fixed in the past, we can drop the dependence on $t$ in the notation $\gamma^{(t)}_{a,b}$, and by conditioning on the previous genealogical event, we can derive the following recurrence relations and boundary conditions for $\bbE[\gamma_{a,b}]$,
\begin{align}
\bbE[\gamma_{a,b}] &= 
\begin{cases}
\sum_{j=1}^{a} \sum_{k=1}^{b} (1 - \delta_{j,a} \delta_{k,b}) {p_{a,j} p_{b,k} \over 1 - p_{a+b, a+b}} {\fall{N-j}{k} \over \fall{N}{k}} \bbE[\gamma_{j,k}]   & \text{if } a > 1,	\\
{1 \over 1 - p_{b+1,b+1}} + \sum_{m=1}^{b-1} {N - m \over N} {p_{b,m} \over 1 - p_{b+1,b+1}} \bbE[\gamma_{1,m}]	& \text{if } a = 1 \text{ and } b > 1,	\\
N	&	\text{if } a = b = 1.
\end{cases}\label{eq:dp_freq_spectrum}
\end{align}

From recurrence relations \eqref{eq:dp_freq_spectrum_varpop} and \eqref{eq:dp_freq_spectrum}, the expected frequency spectrum for a sample of size $n$ can be computed in $O(n^4)$ and $O(n^4 t_c)$ time for the constant and variable population cases respectively. However, if one truncates the summation range for the indices $j$ and $k$ in \eqref{eq:dp_freq_spectrum_varpop} and \eqref{eq:dp_freq_spectrum} to only those $j, k$ values where $p^{(t)}_{a,j}$ and $p^{(t)}_{a,j}$ (respectively, $p_{a,j}$ and $p_{b,k}$) are greater than some small tolerance parameter $\eps > 0$, the time complexity of the above dynamic programs can be improved to $\widetilde{O}(n^2)$ and $\widetilde{O}(n^2 t_c)$, where the $\widetilde{O}$ notation signifies the dependence of the quantities on the truncation parameter $\eps$.

We used \eqref{eq:kth_entry_freq_spectrum} along with a truncation parameter of $\eps = 10^{-120}$ to compute the expected frequency spectra values presented in the Results. Upon decreasing this threshold further, we did not observe any change to the frequency spectra, suggesting that the computed answers are accurate. 

\subsection{Expected number of lineages as a function of time under the coalescent}\label{sec:coal_nlft}
Suppose we have a panmictic population with size $N(t)$ at time $t$, evolving according to Kingman's coalescent.
If we sample $n$ lineages at time 0 and let $A^C_n(t)$ denote the number of ancestral lineages of this sample surviving at time $t$, then we have the following expression for the probability distribution function of $A^C_n(t)$ \cite{takahata:1985,tavare:1984},
\begin{align}
	\bbP[A^C_n(t) = m] = \sum_{i = m}^{n} e^{-{i \choose 2} \Omega(t)} (-1)^{i - m} {(2i - 1) \rise{m}{(i-1)} \fall{n}{i} \over m! (i-m)! \rise{n}{i}},   \label{eq:pr_a_n_t}	
\end{align}
where
\begin{align}
	\Omega(t) = \int_0^t {N(0) \over N(\tau)} d\tau. \label{eq:omega_t}
\end{align}
The summation in \eqref{eq:pr_a_n_t} has terms with alternating signs, and this leads to a loss of numerical precision due to catastrophic cancellation for even moderate sample sizes \cite{maruvka:2011}. Hence, computing the expectation of the NLFT, $\bbE[A_n(t)]$, by naively using \eqref{eq:pr_a_n_t} will not produce reliable answers. However, using a formula of Tavar\'e \cite[equation 5.11]{tavare:1984}, one gets the following closed-form expressions for the expectation and variance of $A_n(t)$ that are numerically stable to evaluate,
\begin{align}
    \bbE[A_n^C(t)] &=  \sum_{i = 1}^{n} e^{-{i \choose 2} \Omega(t)} (2i - 1) {\fall{n}{i} \over \rise{n}{i}},  \label{eq:exp_An} \\
\text{Var}\left(A_n^C(t)\right) &= \sum_{i = 1}^{n} e^{-{i \choose 2} \Omega(t)} (2i - 1) (i^2 - i + 1) {\fall{n}{i} \over \rise{n}{i}} - \left[ \sum_{i = 1}^{n} e^{-{i \choose 2} \Omega(t)} (2i - 1) {\fall{n}{i} \over \rise{n}{i}} \right]^2.    \label{eq:var_An}
\end{align}
Each term in the summations in \eqref{eq:exp_An} and \eqref{eq:var_An} is positive, and hence poses no numerical problems for evaluation. Furthermore, the terms in the sum decay rapidly for large $i$ due to the exponential functions involved.

\section{Discussion}      

Several analyses of genomic sequence variation in large samples of humans \cite{keinan:2012, nelson:2012, tennessen:2012, fu:2012} have found a substantial excess of rare variation compared to those predicted using previously applied demographic models.  The inference in these studies is that these results are consistent with a rapid growth of the effective population size in the recent past (much more rapid than in previously applied demographic models), a conclusion consistent with historical records of census population size \cite{keinan:2012}.  These studies also employed sample sizes that would appear to be large enough to violate assumptions of the coalescent, potentially distorting genealogical properties in a way that may inflate rare variation relative to the predictions of coalescent theory.   In this paper, we have investigated this issue by developing a method for performing exact computation in the discrete-time Wright-Fisher model of random mating.  We have studied the deviation between the coalescent and the Wright-Fisher model for several key genealogical quantities that are used for population genomic inference. 

For realistic demographic scenarios, our results show that there are a significant number of multiple- and simultaneous-merger events under the Wright-Fisher model that are ignored by construction of the coalescent.   
Furthermore, there are noticeable differences in the expected number of rare variants between the coalescent and the DTWF model, and we expect the difference to become amplified as the sample size increases.   
A number of demographic inference methods are based on fitting the expected frequency spectra under the coalescent \cite{coventry:2010, nelson:2012, tennessen:2012} or the diffusion process \cite{gravel:2011, gutenkunst:2009, lukic:2011, lukic:2012} to observed data.  For instance, the exponential growth parameters in \textsc{Models 3} \cite{gravel:2011} and \textsc{4} \cite{tennessen:2012} were inferred using a likelihood method based on the diffusion process approximation to the DTWF model, by fitting the predicted frequency spectrum to polymorphism patterns observed in a sample size of 876 individuals and 2,440 individuals, respectively.   Since the diffusion process approximation to the DTWF model is equivalent to the coalescent approximation, the differences in the frequency spectrum (see \fref{fig:freq_spectrum_WF_vs_coalescent_singletons_doubletons}) between the coalescent and the DTWF model indicate that we might infer different demographies if the analysis were done using the DTWF model.

To balance the tradeoff between accuracy and computational efficiency, we have proposed a hybrid algorithm that uses the DTWF model for the recent past and the coalescent for the more distant past.  This hybrid algorithm provides a way to obtain more accurate predictions of the frequency spectrum than in the coalescent, while being computationally more efficient than considering the full DTWF model.  We leave the exploration of this method for demographic inference as future research.

Wakeley and Takahashi \cite{wakeley:2003} have provided asymptotically accurate expressions (as the effective population size $N \to \infty$) for the number of singletons and the number of segregating sites under a variant of the DTWF model which allows for a larger number of offspring than the effective population size, assuming that the effective population size stays constant over time.
Fu \cite{fu:2006} has also examined the accuracy of the standard coalescent model and proposed an alternative continuous-time ``exact'' coalescent model applicable in the regime when $N(N-1) \cdots(N-n+1) \times N^{-n} \gg 0$, where $N$ denotes the effective population size and $n$ the sample size. That work was restricted to the case of a constant population size, while in this paper we have considered realistic demographic scenarios inferred from recent large-scale population genomic studies. Moreover, for some of the demographic scenarios and sample sizes considered here, the assumption in Fu's work \cite{fu:2006} that $N(N-1) \cdots(N-n+1) \times N^{-n} \gg 0$ is violated.
Wakeley \emph{et al.} \cite{wakeley:2012} have shown that it is difficult to reject the coalescent even for data generated using \emph{fixed} pedigrees with random genetic assortment. Our work is complementary to that study and compares the coalescent to the DTWF random mating model.

In this paper, we have focused on the DTWF model for simplicity.  However, it is known that under some weak conditions on the limiting probabilities of a 2-merger and a 3-merger, a large family of exchangeable random mating models converge to the same coalescent limit as the unit of time is rescaled appropriately and the population size gets large  \cite{mohle:2001, mohle:2003}.  The rate of convergence to the coalescent differs between different random mating models \cite{bhaskar:2009}, and hence the accuracy of coalescent predictions for large sample sizes depends on the random mating model being considered.   The deviation from the coalescent could be amplified for other  random-mating models.  It would be interesting to consider the accuracy of the coalescent for other random mating models of relevance to human genetics; e.g.,  models in which generations overlap and the distribution of offspring number more closely reflects the observed pattern for human populations (for example, the Swedish family data of Low and Clarke \cite{low:1991} or the Saguenay-Lac-Saint-Jean population considered by Moreau \emph{et al.} \cite{moreau:2011}).
Furthermore, it would be interesting to compare discrete-time random models and the coalescent with respect to haplotype sharing (identity-by-descent and identity-by-state), linkage disequilibrium, and natural selection when the sample size is very large. For example, Davies \emph{et al.} \cite{davies:2007} employed simulations to demonstrate that for a constant population size model, recombination and gene conversion can increase the number of ancestral lineages of a sample of chromosomes to the extent that multiple and simultaneous mergers in the DTWF model can lead to substantial differences from the coalescent model in the rates of coalescence and in the number of sequences carrying ancestral material. It would be interesting to perform such comparisons for more realistic demographic models for humans.

We will soon enter an era where it will become routine to analyze samples with hundreds of thousands if not millions of individuals. For these large sample sizes, the standard coalescent will no longer serve as an adequate model for evolution. The DTWF model is mathematically cumbersome to work with, which was one of the original motivations for adopting the coalescent for modern population genetics analyses. However, for these very large sample sizes, we will need to develop new mathematically and computationally tractable stochastic processes that better approximate realistic models of human population evolution, and under which we can efficiently compute genealogical quantities like we have been able to under the coalescent.

\section*{Acknowledgments}
We would like to thank Jack Kamm for useful discussions. This research is supported in part by NIH grants R01-HG003229 and R01-GM094402, and a Packard Fellowship for Science and Engineering.

\bibliographystyle{abbrv}   
\bibliography{refs}  
\newpage                         
\makeatletter     
\renewcommand{\thefigure}{S\@arabic\c@figure}
\renewcommand{\thetable}{S\@arabic\c@table}      
\renewcommand{\theequation}{S\arabic{equation}}
\makeatother 

\setcounter{figure}{0}
\setcounter{table}{0}
\setcounter{equation}{0}                               

%

%
%
%
%
%
%
%

\section*{Supporting Table and Figures}   

\begin{table*}[h]
\caption{\textbf{Expected percentage of lineages (relative to $n-1$, where $n$ is the sample size) lost due to either multiple or simultaneous mergers for \textsc{Models 1--4}.} 
For large sample sizes ($n = 2 \times 10^3$ or $2 \times 10^4$), in all demographic models, most of the lineages are lost in generations where multiple lineages are lost.
}
\label{tab:fraction_lineages_lost_to_non_2drops}
\begin{center}
\begin{tabular}{cc rr rr rr rr} \hline
    $n$         &   \modelOne     &   \modelTwo     &   \modelGravel 	&   \modelTennessen \\
    \hline
    $20$      			&   0.28\%    	&   2.16\%     	&   0.84\%    	&   0.86\%  	\\
    $2 \times 10^2$	    &   24.97\%		&	30.86\%		&	19.44\%		&	24.79\%		\\
    $2 \times 10^3$     &   91.12\%		&	92.18\%		&	90.81\%		&	92.53\%		\\
    $2 \times 10^4$     &   99.11\%		&	99.22\%		&	99.14\%		&	99.31\%		\\
    \hline    
\end{tabular}
\end{center}
\end{table*}

\hspace{2cm}
\begin{figure*}[ht]          
	\begin{center}
		\includegraphics[width=0.75\textwidth, trim=0mm 0mm 0mm 10mm,clip]{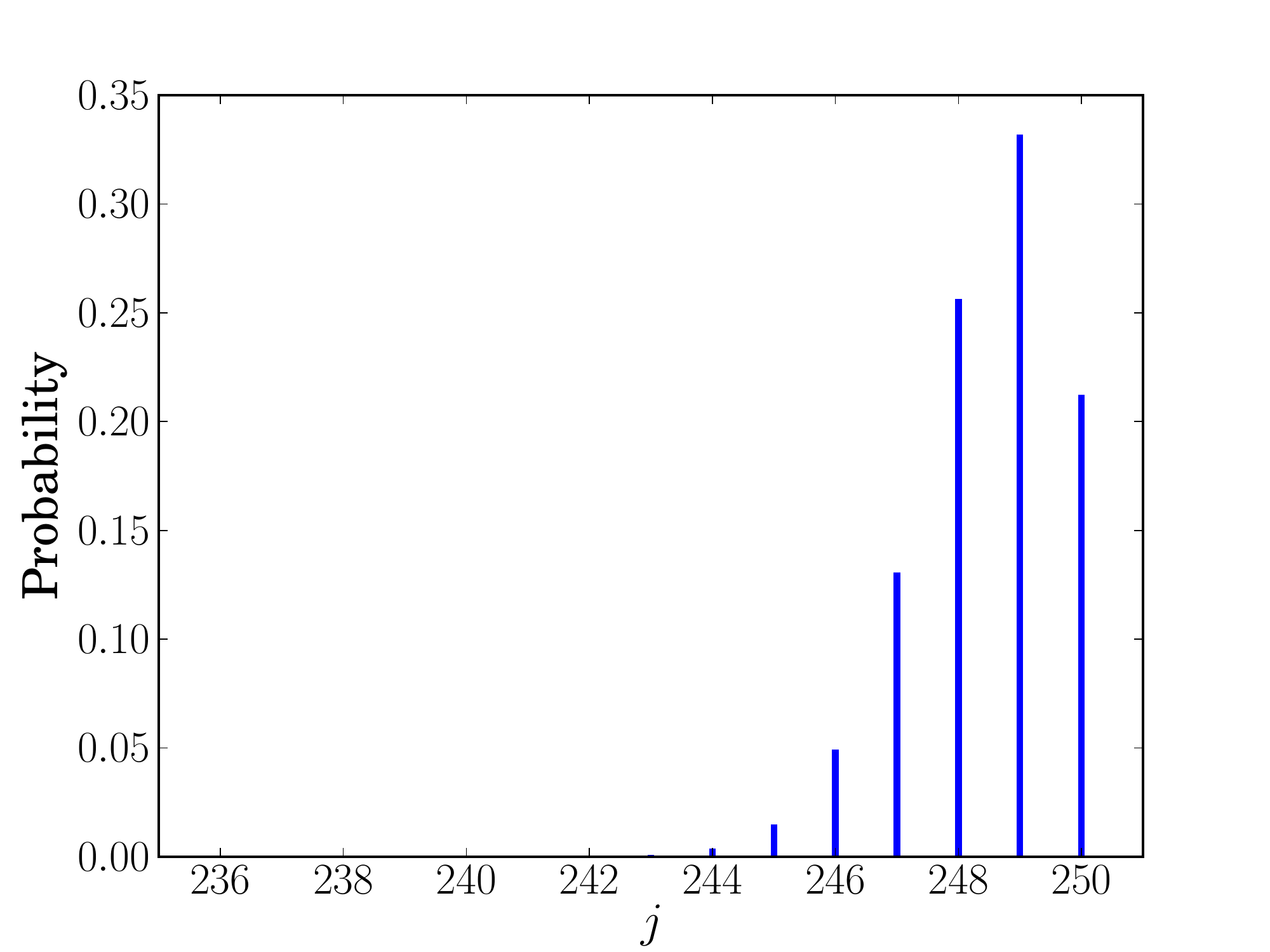}		
	\end{center}
	\caption{{\bf Probability distribution of the number of parents of a sample of size $n = 250$ and population size $N = 20{,}000$ in the Wright-Fisher model.} For each value of $j$ on the $x$-axis, the $y$-axis is the probability that the sample of size $n$ has $j$ parents in the previous generation. There is a substantial probability of the sample having less than $n - 1$ parents in the previous generation, which corresponds to multiple or simultaneous mergers in the ancestral process.}
	\label{fig:prob_drops_n250}
\end{figure*}

\begin{figure*}[ht]
	\begin{center}
 	\subfigure[]{\label{fig:fraction_lineages_lost_to_non_2drops_model1} 
	    \includegraphics[width=0.48\textwidth, trim=10mm 5mm 10mm 5mm,clip]{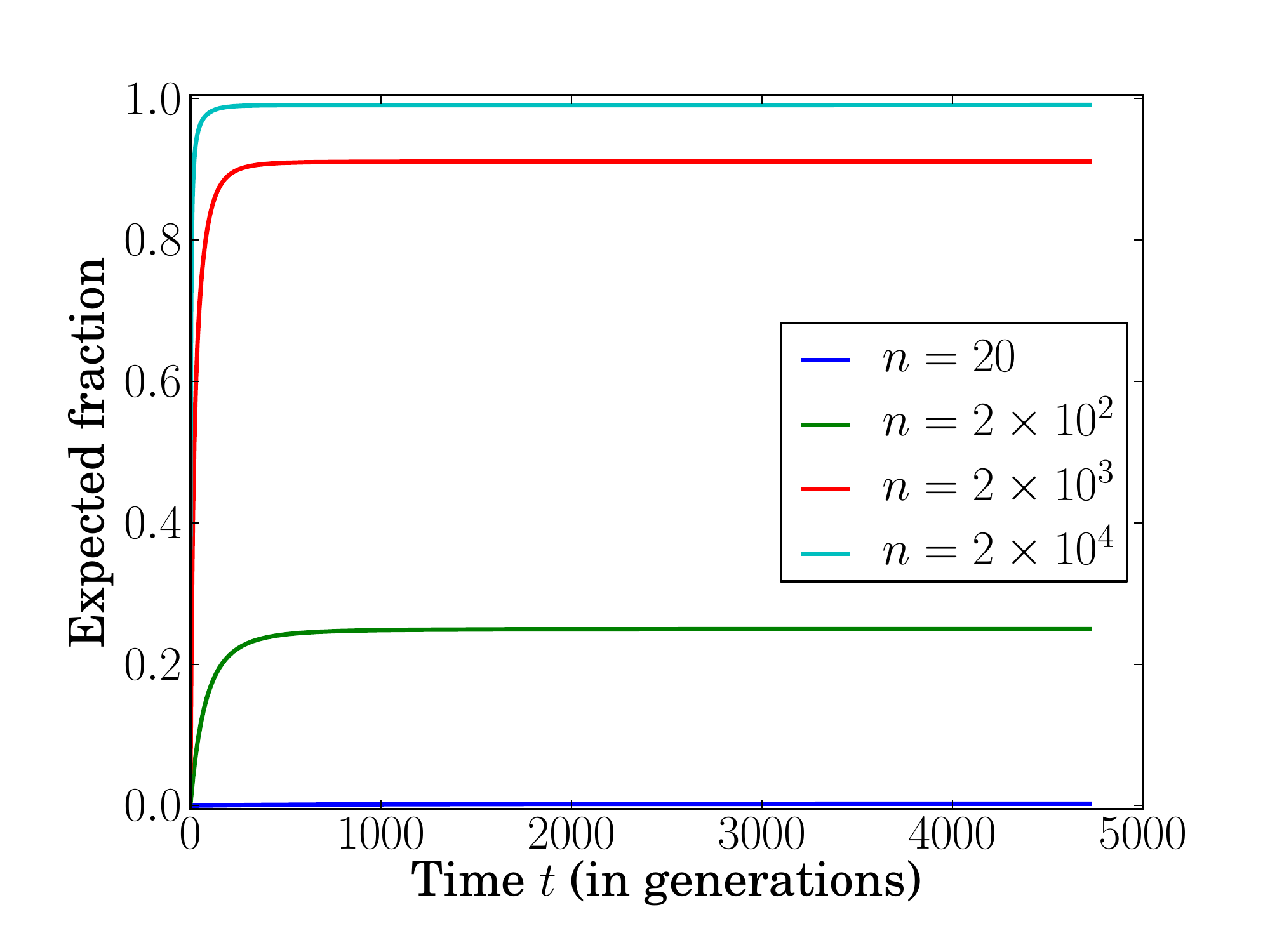}	  
	}
	\subfigure[]{\label{fig:fraction_lineages_lost_to_non_2drops_model2}
		\includegraphics[width=0.48\textwidth, trim=10mm 5mm 10mm 5mm,clip]{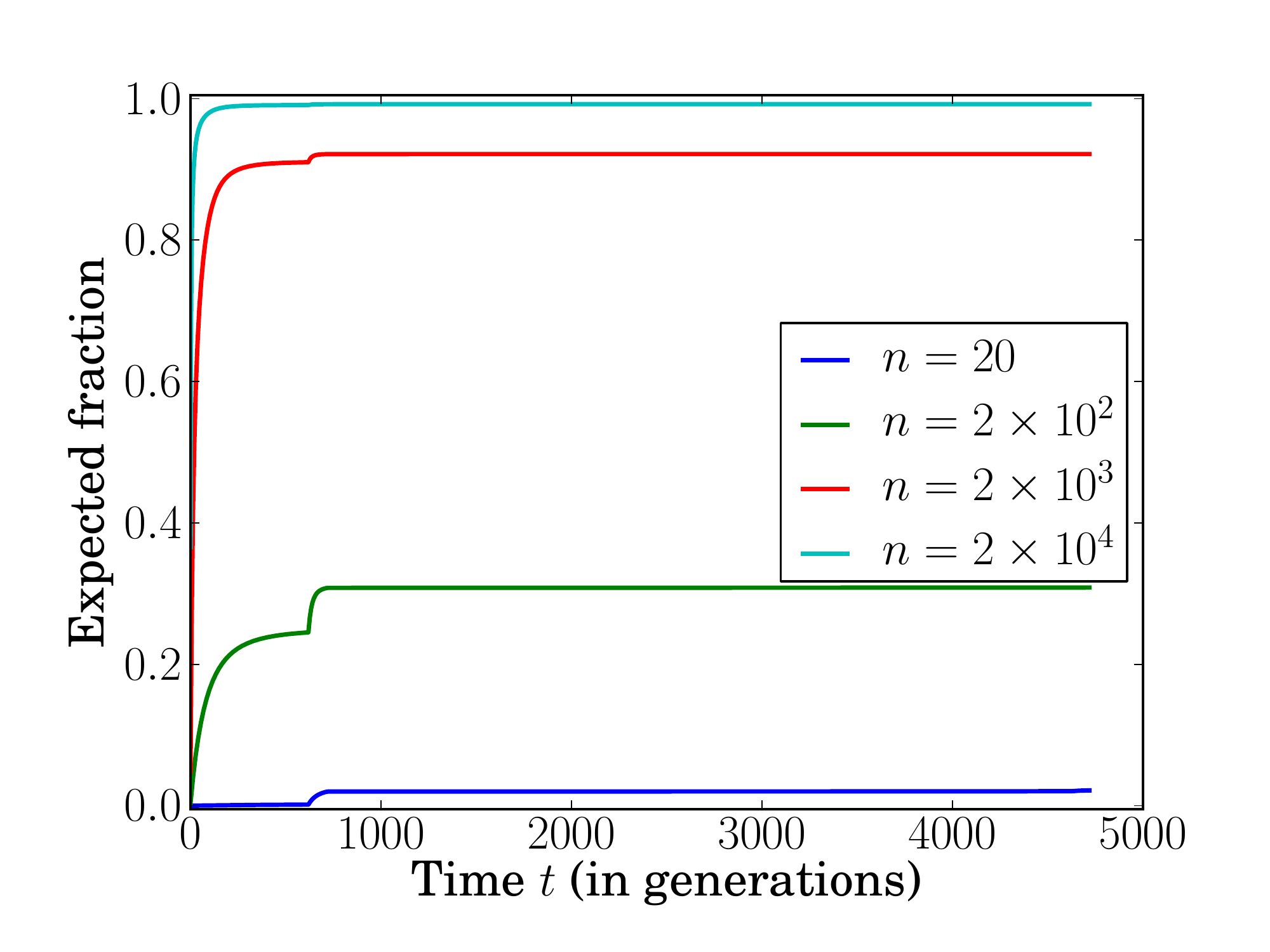}
	}
	\subfigure[]{\label{fig:fraction_lineages_lost_to_non_2drops_model3}
		\includegraphics[width=0.48\textwidth, trim=10mm 5mm 10mm 5mm,clip]{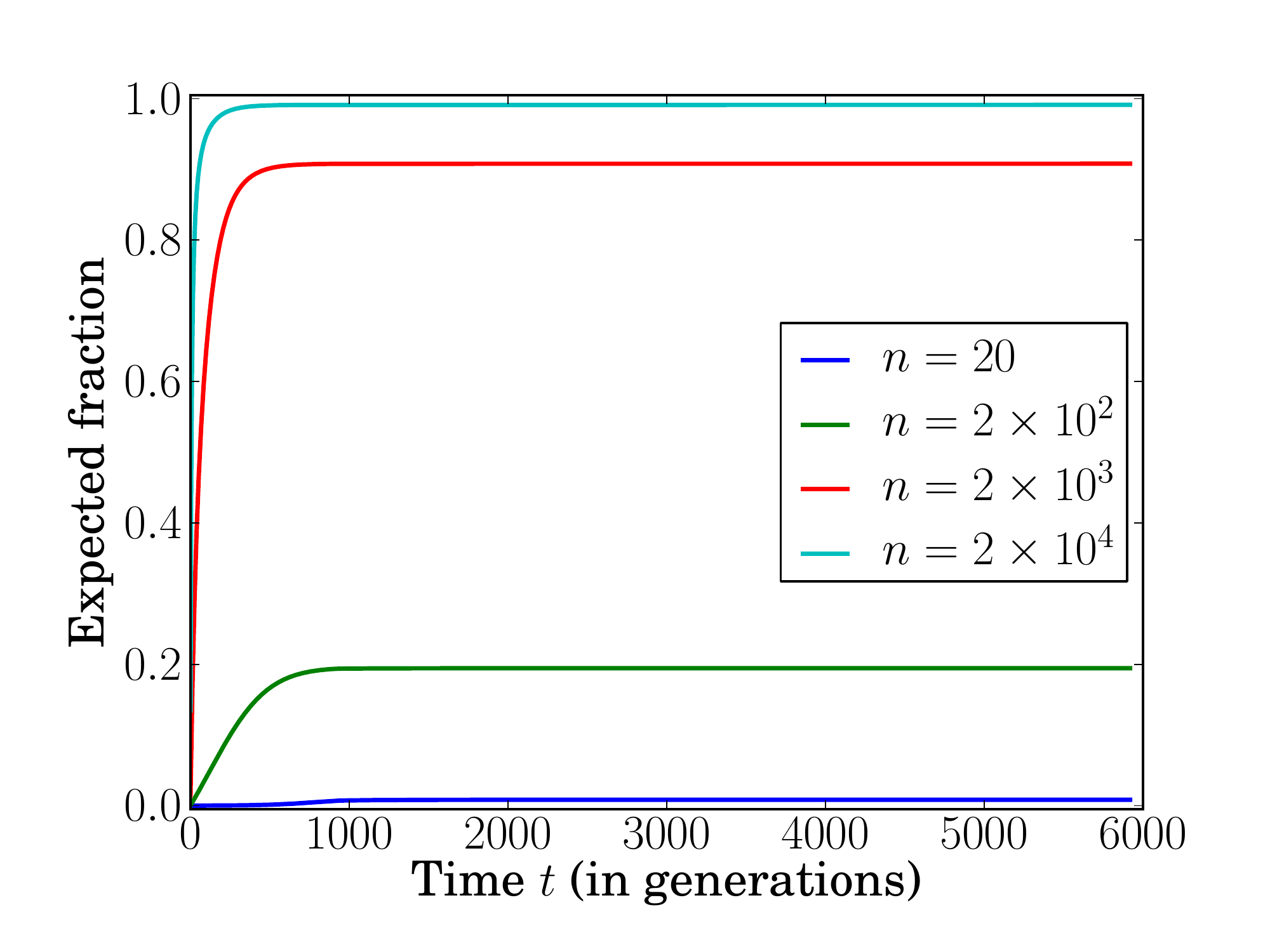}
	}
	\subfigure[]{\label{fig:fraction_lineages_lost_to_non_2drops_model4}
		\includegraphics[width=0.48\textwidth, trim=10mm 5mm 10mm 5mm,clip]{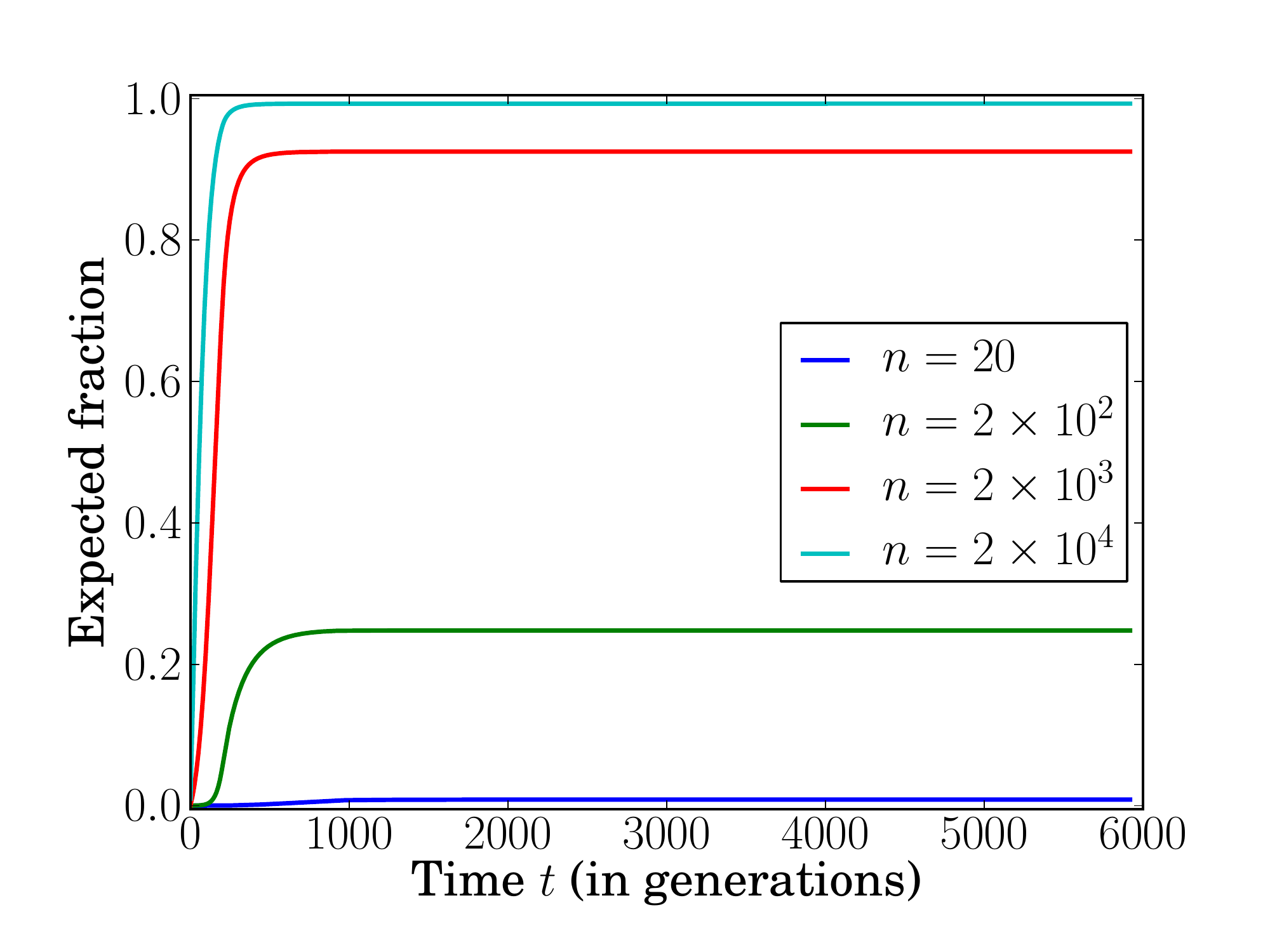}
	}
	\end{center}
    \caption{{\bf Expected proportion of lineages (relative to $n-1$, where $n$ is the sample size at present) that are lost in generations when more than one lineage is lost, from the present
up to time $t$ in the past.} For each generation $t$ on the $x$-axis, the $y$-axis is the expected number of lineages that have been lost due to coalescence events in those generations from $0$ to $t$ where more than one lineage is lost, normalized by the quantity $n-1$, which is the total number of lineages that are lost over all generations. The plots correspond to
\subref{fig:fraction_lineages_lost_to_non_2drops_model1} \modelOne,
\subref{fig:fraction_lineages_lost_to_non_2drops_model2} \modelTwo, 
\subref{fig:fraction_lineages_lost_to_non_2drops_model3} \modelGravel, and
\subref{fig:fraction_lineages_lost_to_non_2drops_model4} \modelTennessen.
The sharp jumps in the plot for \modelTwo (\subref{fig:fraction_lineages_lost_to_non_2drops_model2}) corresponds to the beginning (backwards in time) of population bottlenecks when the population size declines substantially, thus instantaneously increasing the rate at which lineages find common ancestors and are lost. 
For small sample sizes relative to the population size ($n = 20$ and $n=200$), it is unlikely for more than one lineage to be lost in a single generation.
In contrast, for large sample sizes ($n = 2 \times 10^4$), almost all the lineages are lost in generations when more than one lineage is lost.
}
    \label{fig:fraction_lineages_lost_to_non_2drops}
\end{figure*}     

\begin{figure*}[ht]
	\centering
	\includegraphics[]{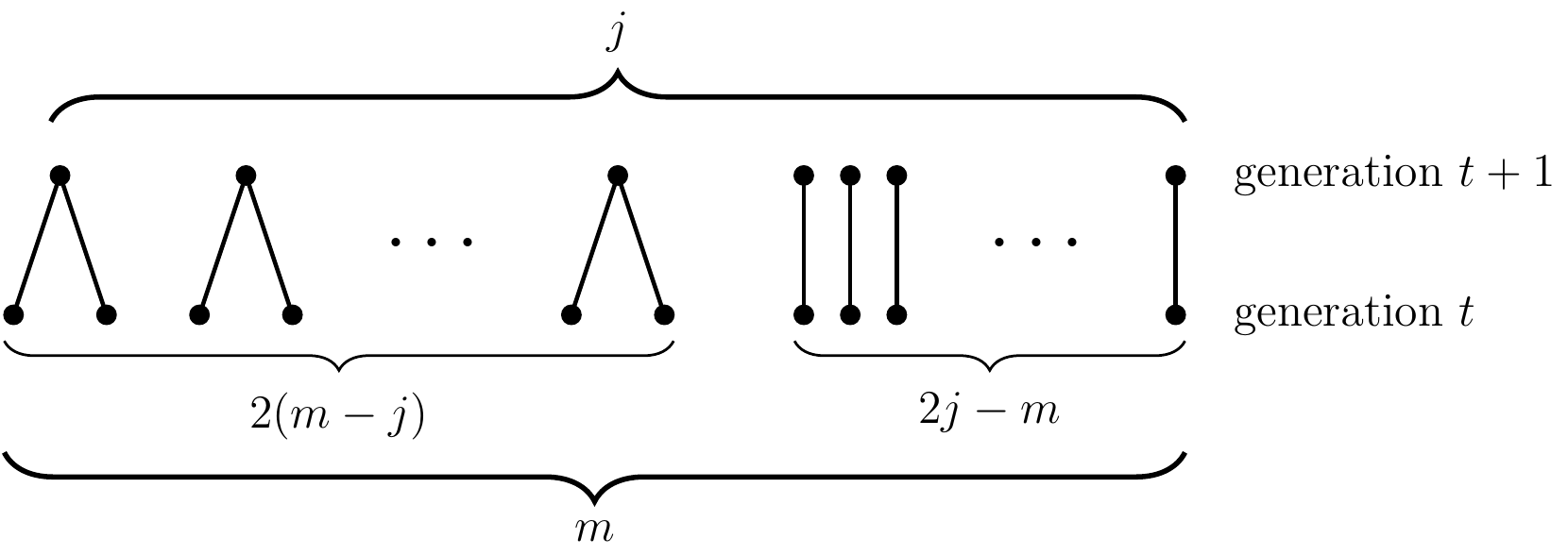}
	\caption{{\bf An example generation of an $(m-j)$-pairwise-simultaneous merger during which $m$ lineages in generation $t$ find $j$ parents in generation $t+1$ (backwards in time).} Each of $m-j$ pairs of lineages in generation $t$ finds a different common parent in generation $t+1$, while the remaining $2j-m$ lineages of generation $t$ each have a different parent in generation $t+1$.}
	\label{fig:pairwise_simultaneous_merger}
\end{figure*}

\begin{figure*}[ht]
	\begin{center}
	\subfigure[]{\label{fig:fraction_lineages_lost_to_simultaneous_mergers_non_2drops_model1}
		\includegraphics[width=0.48\textwidth, trim=8mm 5mm 12mm 5mm, clip]{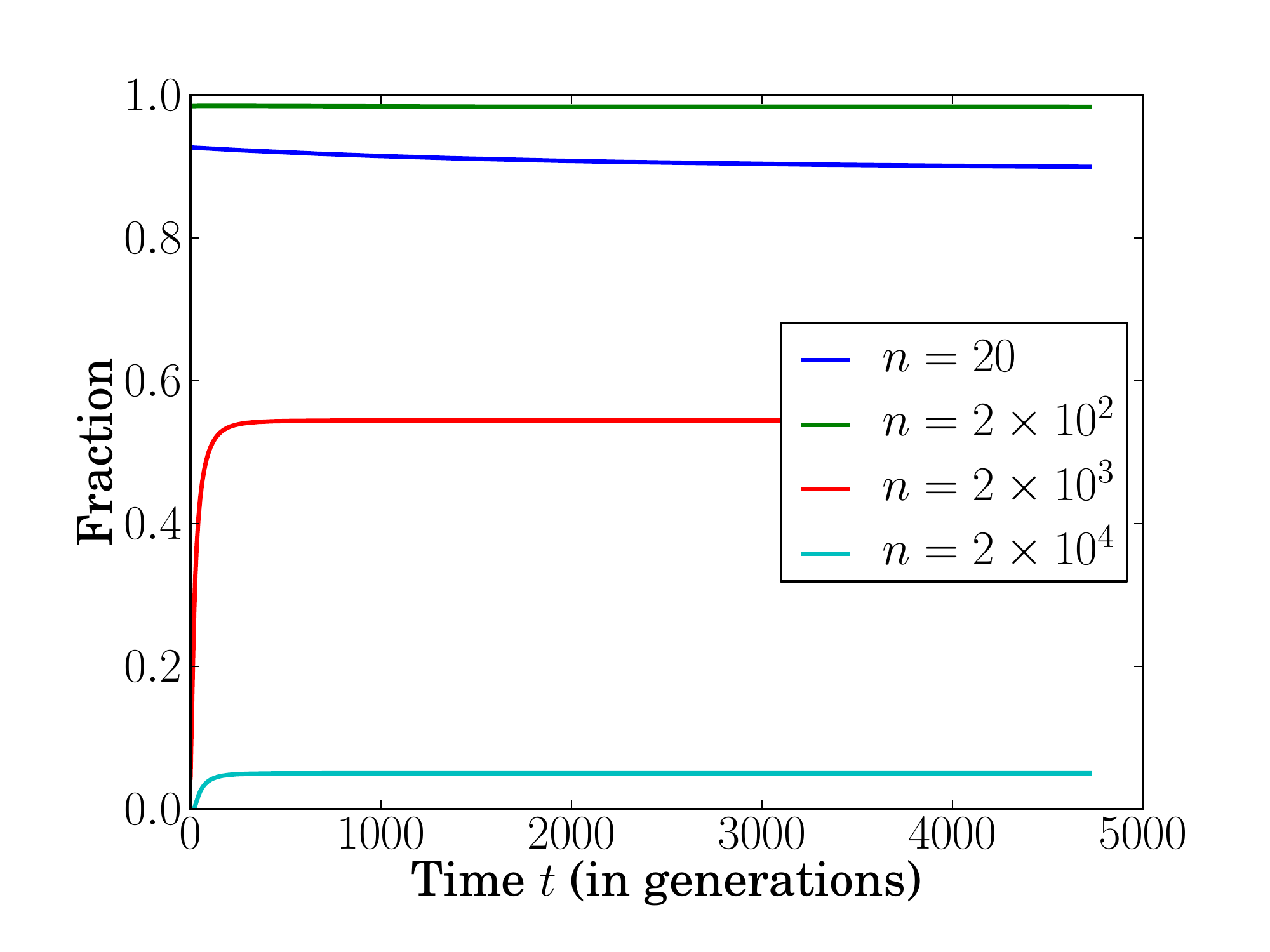}
	}
	\subfigure[]{\label{fig:fraction_lineages_lost_to_simultaneous_mergers_non_2drops_model2}
		\includegraphics[width=0.48\textwidth, trim=8mm 5mm 12mm 5mm, clip]{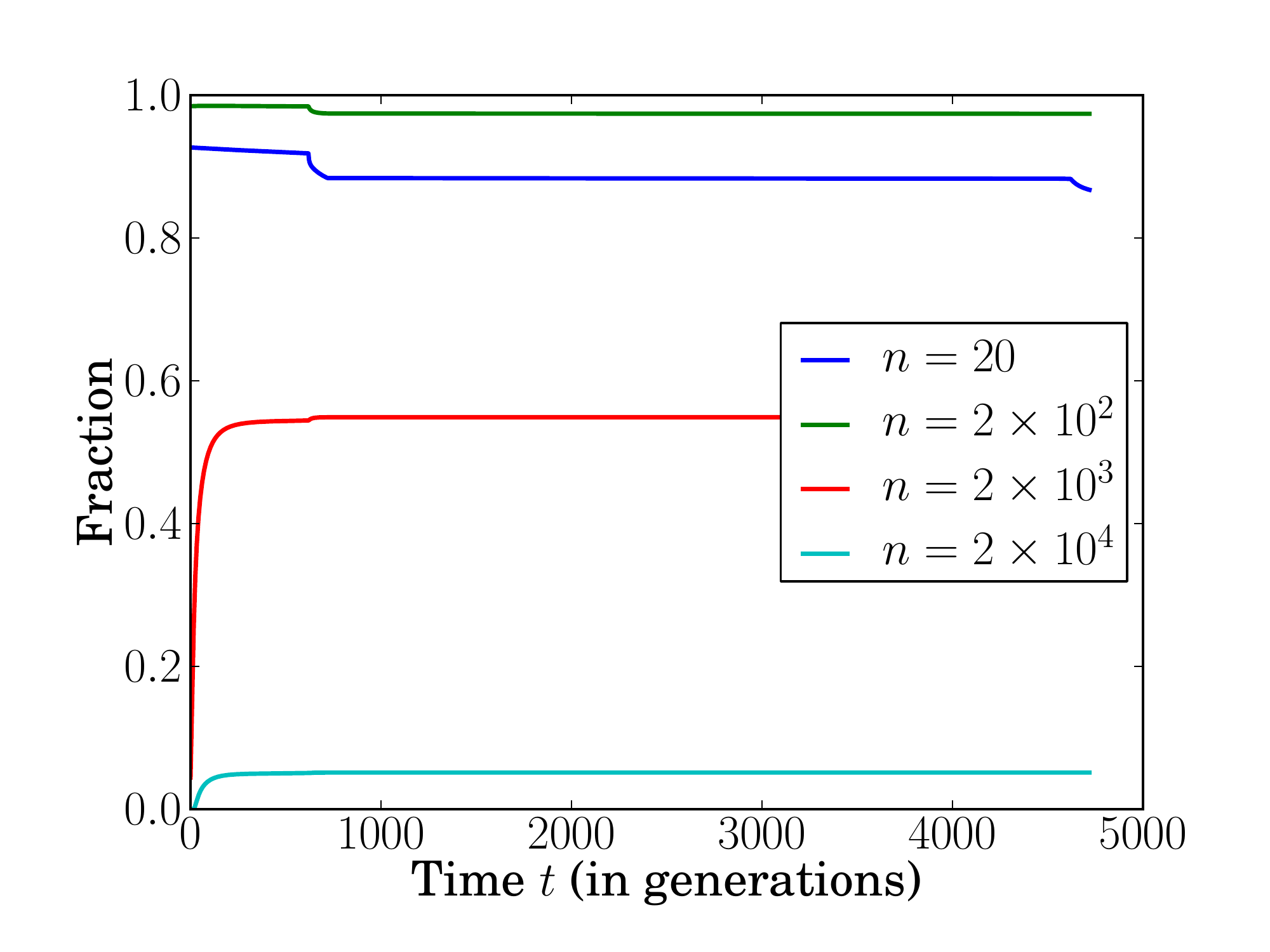}
	}
	\subfigure[]{\label{fig:fraction_lineages_lost_to_simultaneous_mergers_non_2drops_model3}
		\includegraphics[width=0.48\textwidth, trim=8mm 5mm 12mm 5mm, clip]{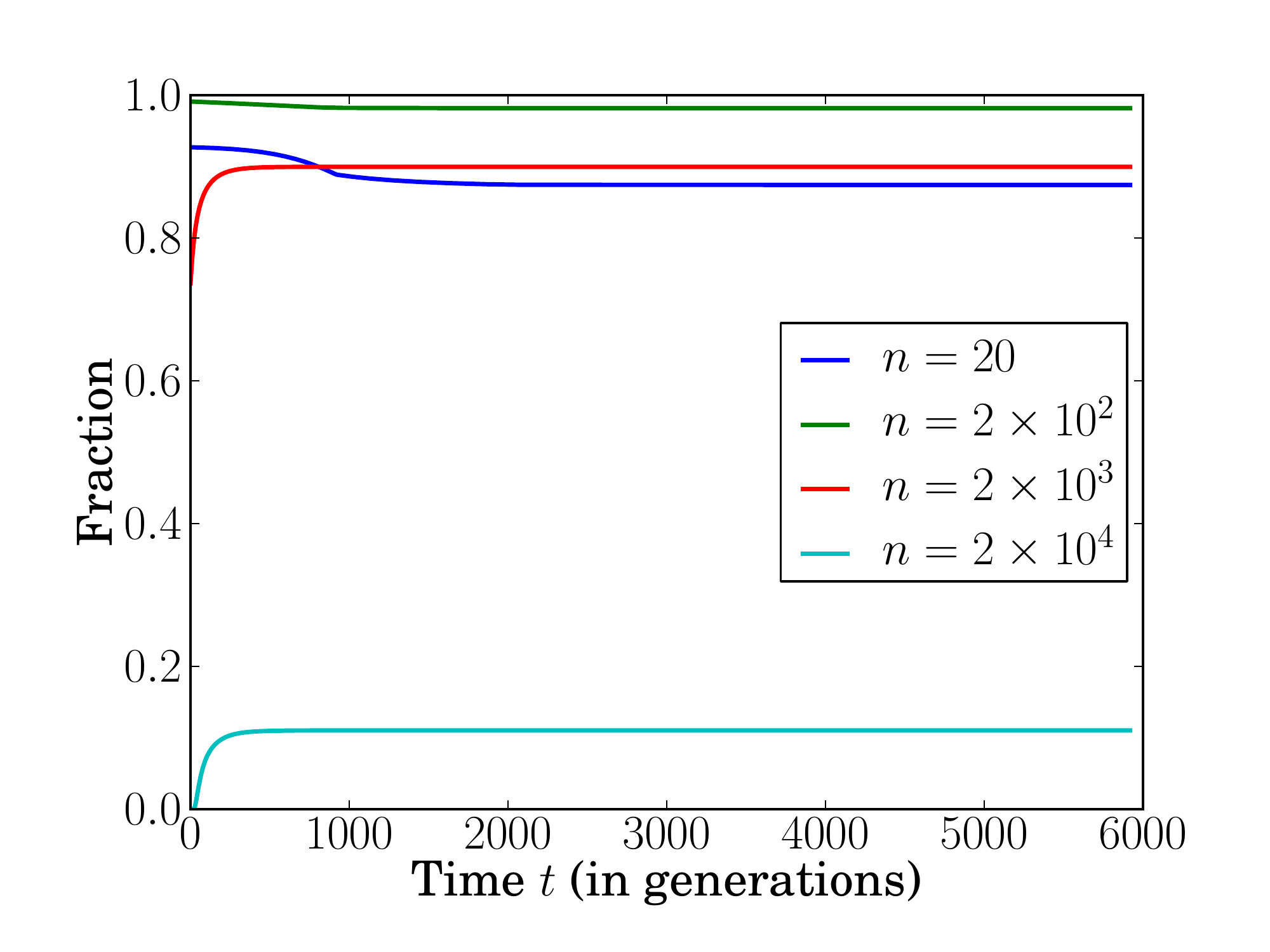}
	}
	\subfigure[]{\label{fig:fraction_lineages_lost_to_simultaneous_mergers_non_2drops_model4}
		\includegraphics[width=0.48\textwidth, trim=8mm 5mm 12mm 5mm, clip]{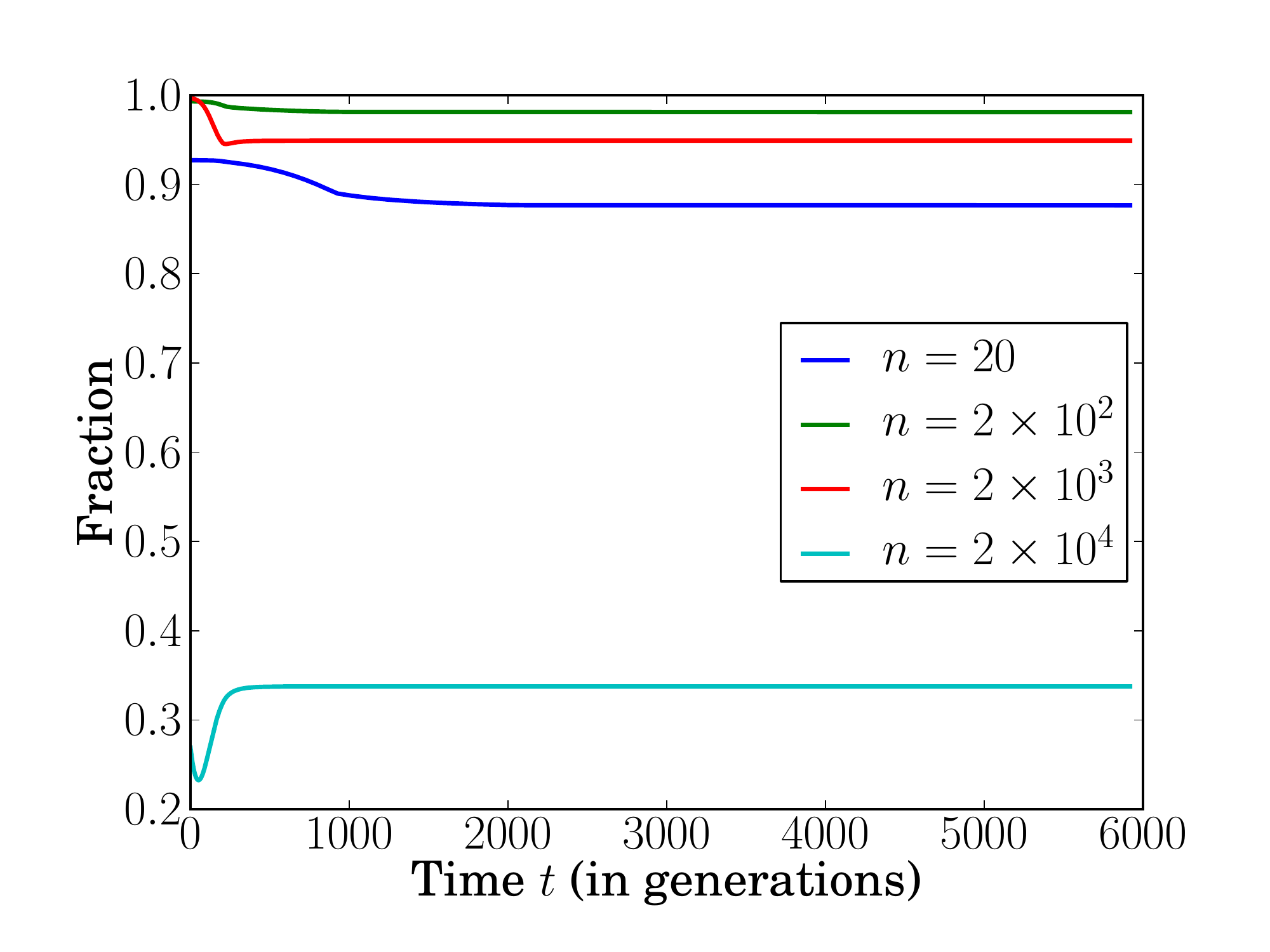}
	}
	\end{center}
	\caption{{\bf Ratio of the sum of the expected number of lineages lost due to $k$-pairwise-simultaneous mergers up to time $t$, for $k \geq 2$, to the respective results shown in \fref{fig:fraction_lineages_lost_to_non_2drops}.} This quantity measures the proportion of lineages lost in generations where more than one lineage is lost due to pairwise simultaneous merger events (and no multiple merger events).	The plots correspond to
	 \subref{fig:fraction_lineages_lost_to_simultaneous_mergers_non_2drops_model1} \modelOne, \subref{fig:fraction_lineages_lost_to_simultaneous_mergers_non_2drops_model2} \modelTwo, \subref{fig:fraction_lineages_lost_to_simultaneous_mergers_non_2drops_model3} \modelGravel, and \subref{fig:fraction_lineages_lost_to_simultaneous_mergers_non_2drops_model4} \modelTennessen.
	A substantial fraction of the lineages that are lost in generations with multiple lost lineages are due to pairwise simultaneous mergers.
	}
	\label{fig:fraction_lineages_lost_to_simultaneous_mergers_non_2drops}
\end{figure*}

\begin{figure*}[ht]
	\begin{center}
	\subfigure[]{\label{fig:fraction_3mergers_function_of_time_model1}
		\includegraphics[width=0.48\textwidth, trim=8mm 5mm 12mm 13mm, clip]{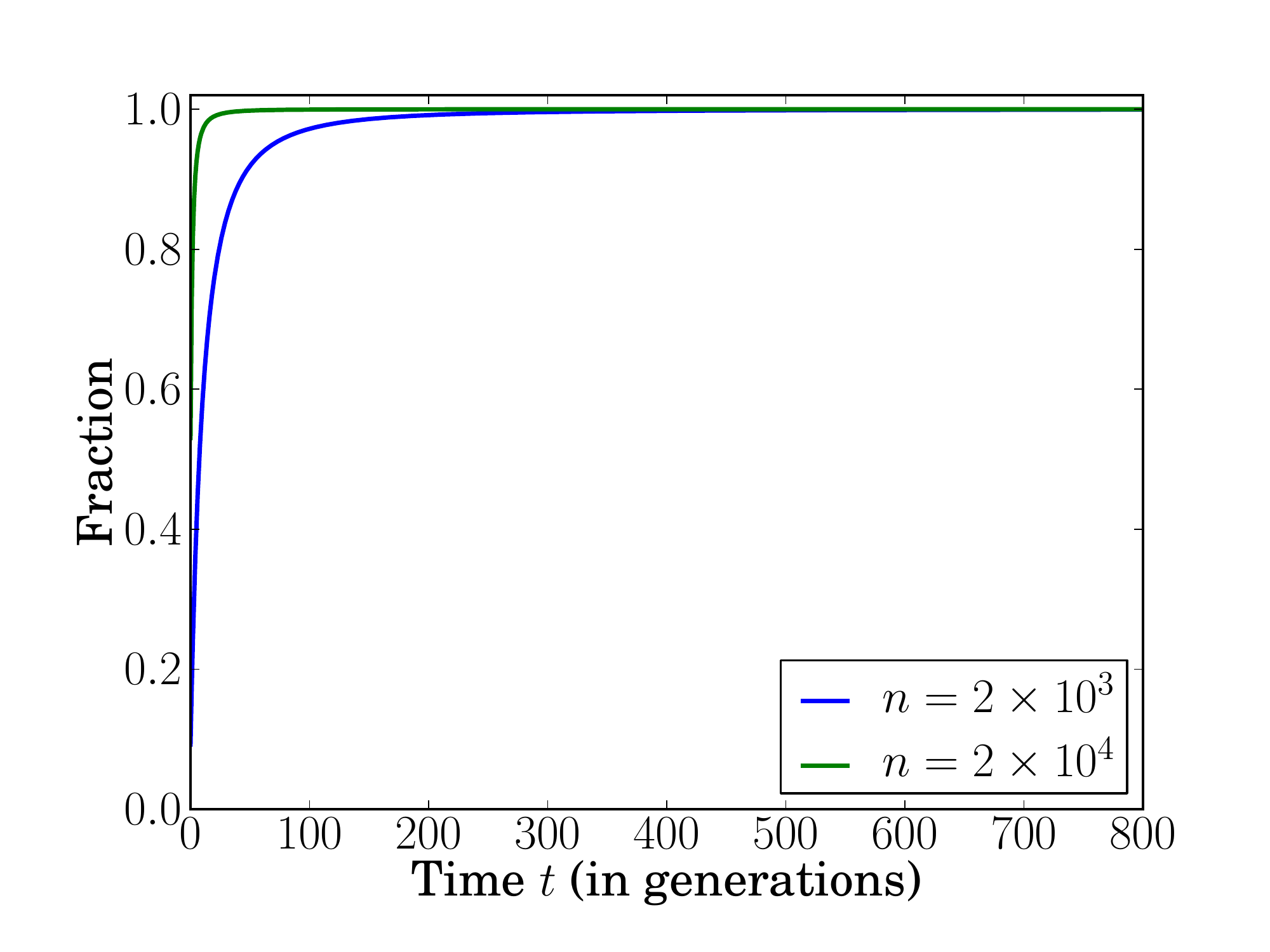}
	}
	\subfigure[]{\label{fig:fraction_3mergers_function_of_time_model2}
		\includegraphics[width=0.48\textwidth, trim=8mm 5mm 12mm 13mm, clip]{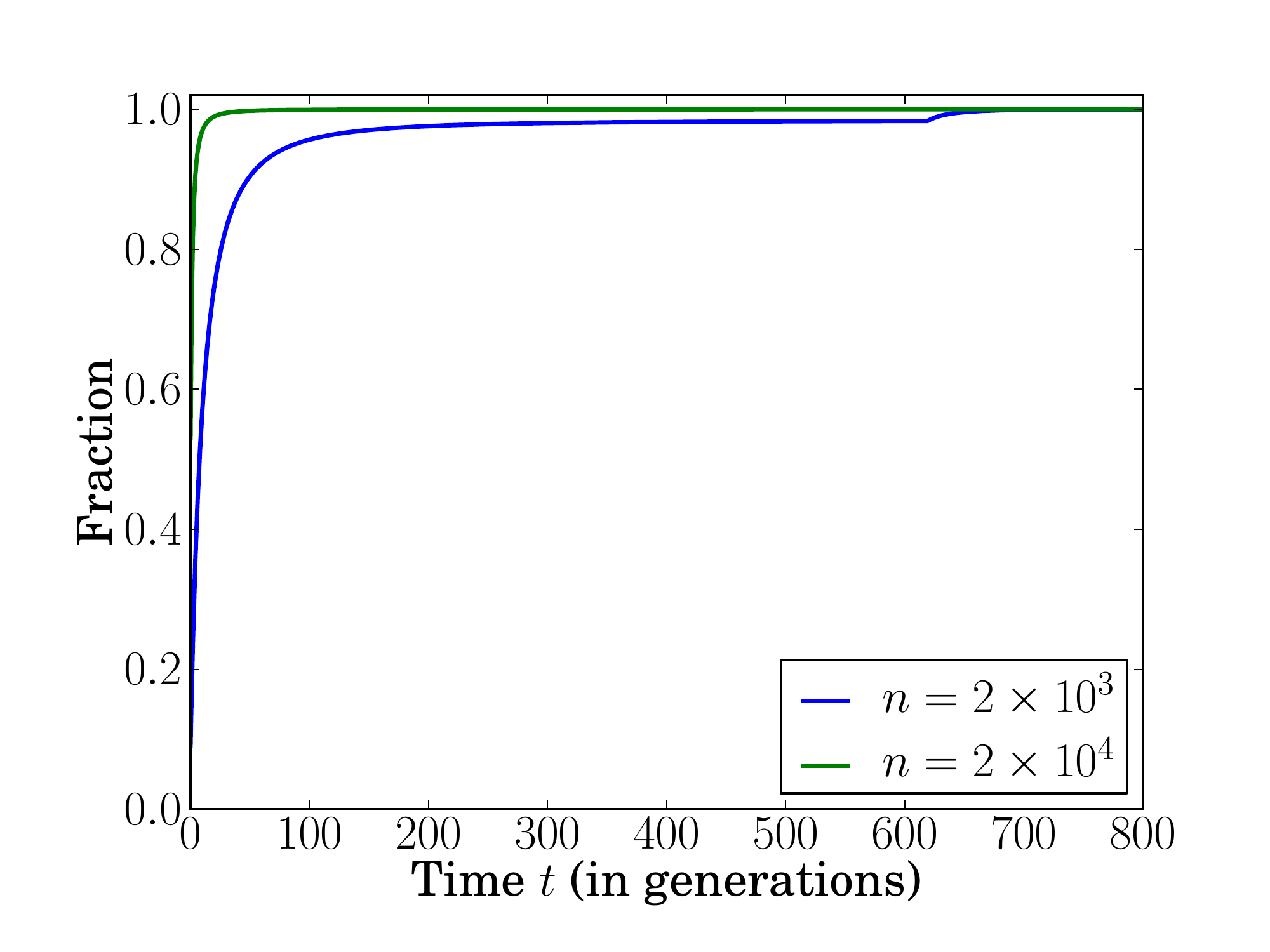}
	}
	\subfigure[]{\label{fig:fraction_3mergers_function_of_time_model3}
		\includegraphics[width=0.48\textwidth, trim=8mm 5mm 12mm 13mm, clip]{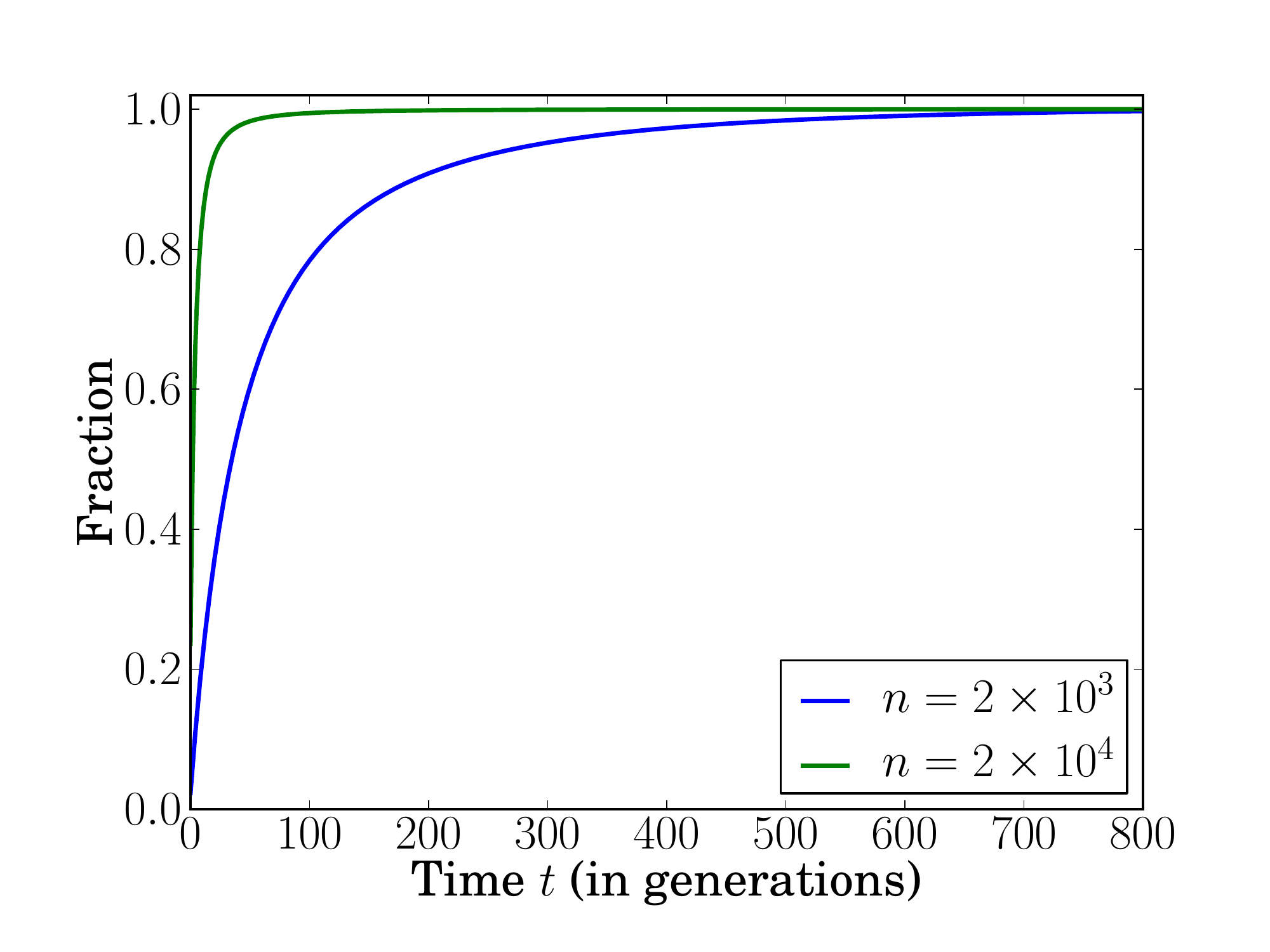}
	}
	\subfigure[]{\label{fig:fraction_3mergers_function_of_time_model4}
		\includegraphics[width=0.48\textwidth, trim=8mm 5mm 12mm 13mm, clip]{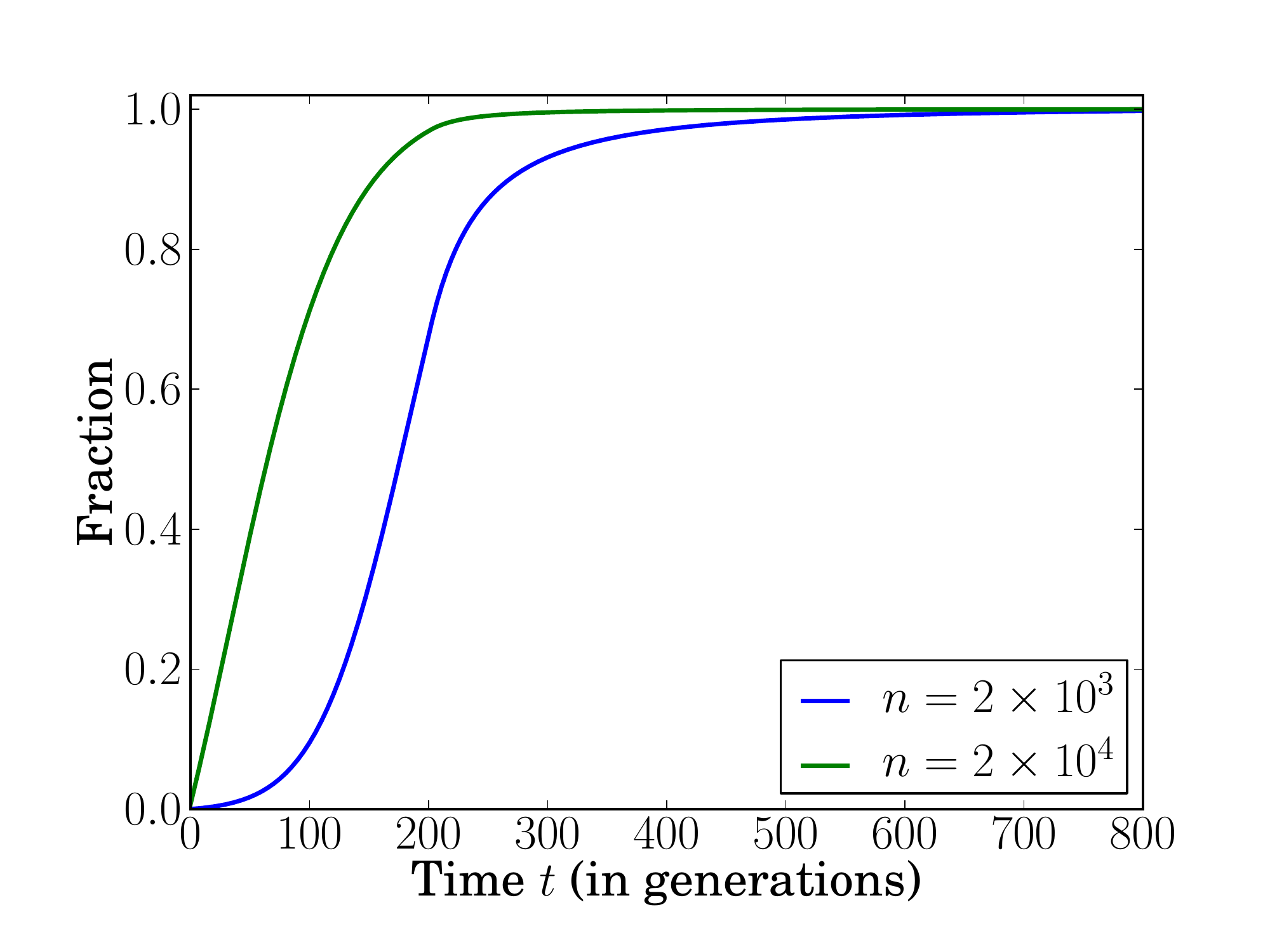}
	}
	\end{center}
	\caption{{\bf Ratio of the expected number of 3-mergers until time $t$ to the total expected number of 3-mergers.} The plots correspond to \subref{fig:fraction_3mergers_function_of_time_model1} \modelOne, \subref{fig:fraction_3mergers_function_of_time_model2} \modelTwo, \subref{fig:fraction_3mergers_function_of_time_model3} \modelGravel, and \subref{fig:fraction_3mergers_function_of_time_model4} \modelTennessen.
	In \modelOne~and \modelTwo, due to the large sample size relative to the population size at time $0$, a substantial portion of the 3-mergers take place very early when the number of surviving lineages drops quickly.  Even in \modelTennessen~where there is a rapid exponential population growth in the most recent 204 generations, more than 25\% of the expected 3-mergers for $n = 2 \times 10^4$ occur in the most recent 32 generations when the effective population size is at least $5.5 \times 10^5$ haploids.
	}
	\label{fig:fraction_3mergers_function_of_time}
\end{figure*}

\begin{figure*}[ht]          
	\begin{center}
 	\subfigure[]{\label{fig:wf_fraction_lineages_surviving_model1} 
	    \includegraphics[width=0.48\textwidth, trim=0mm 0mm 0mm 0mm,clip]{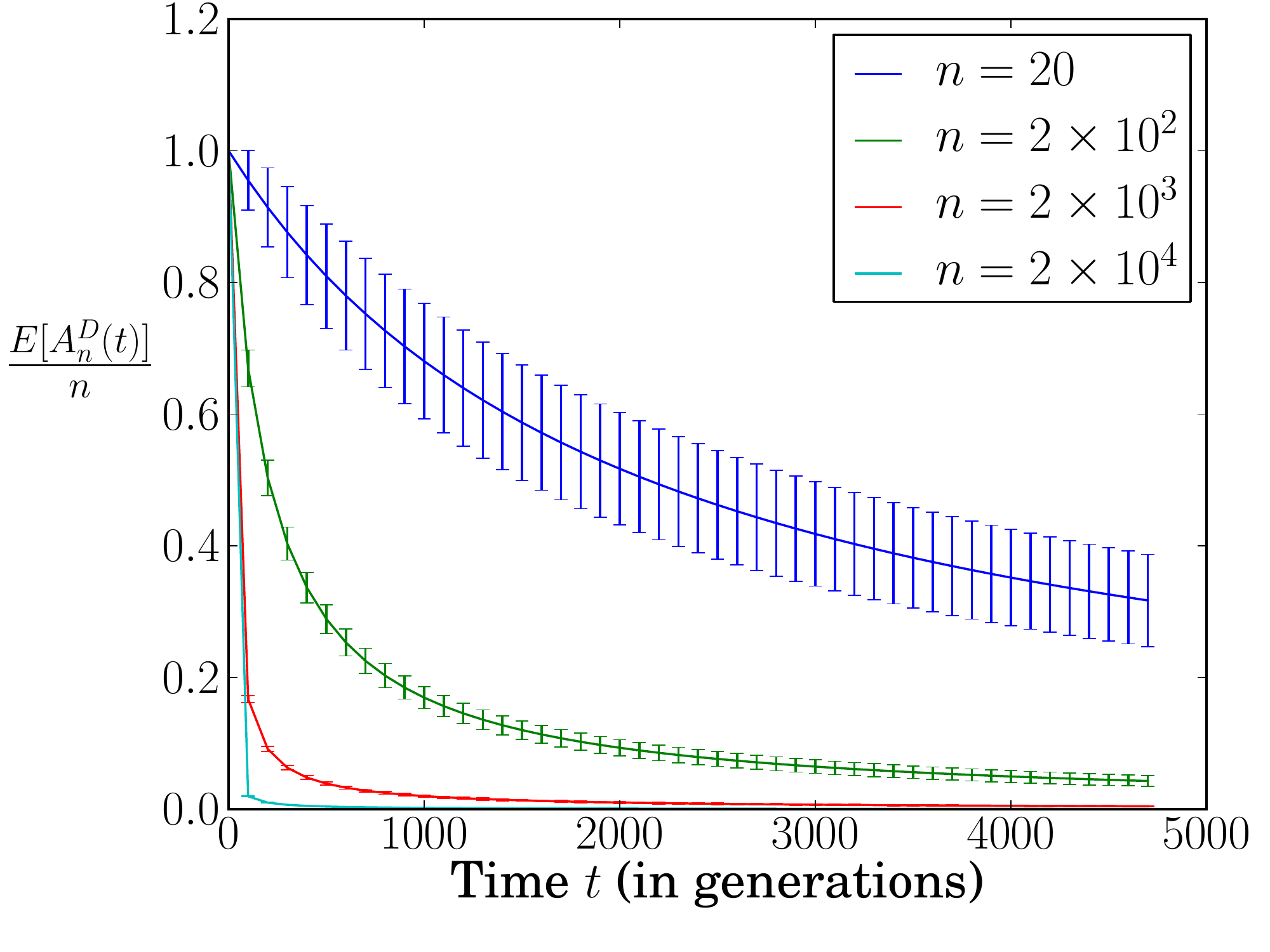}	  
	}
	\subfigure[]{\label{fig:wf_fraction_lineages_surviving_model2}
		\includegraphics[width=0.48\textwidth, trim=0mm 0mm 0mm 0mm,clip]{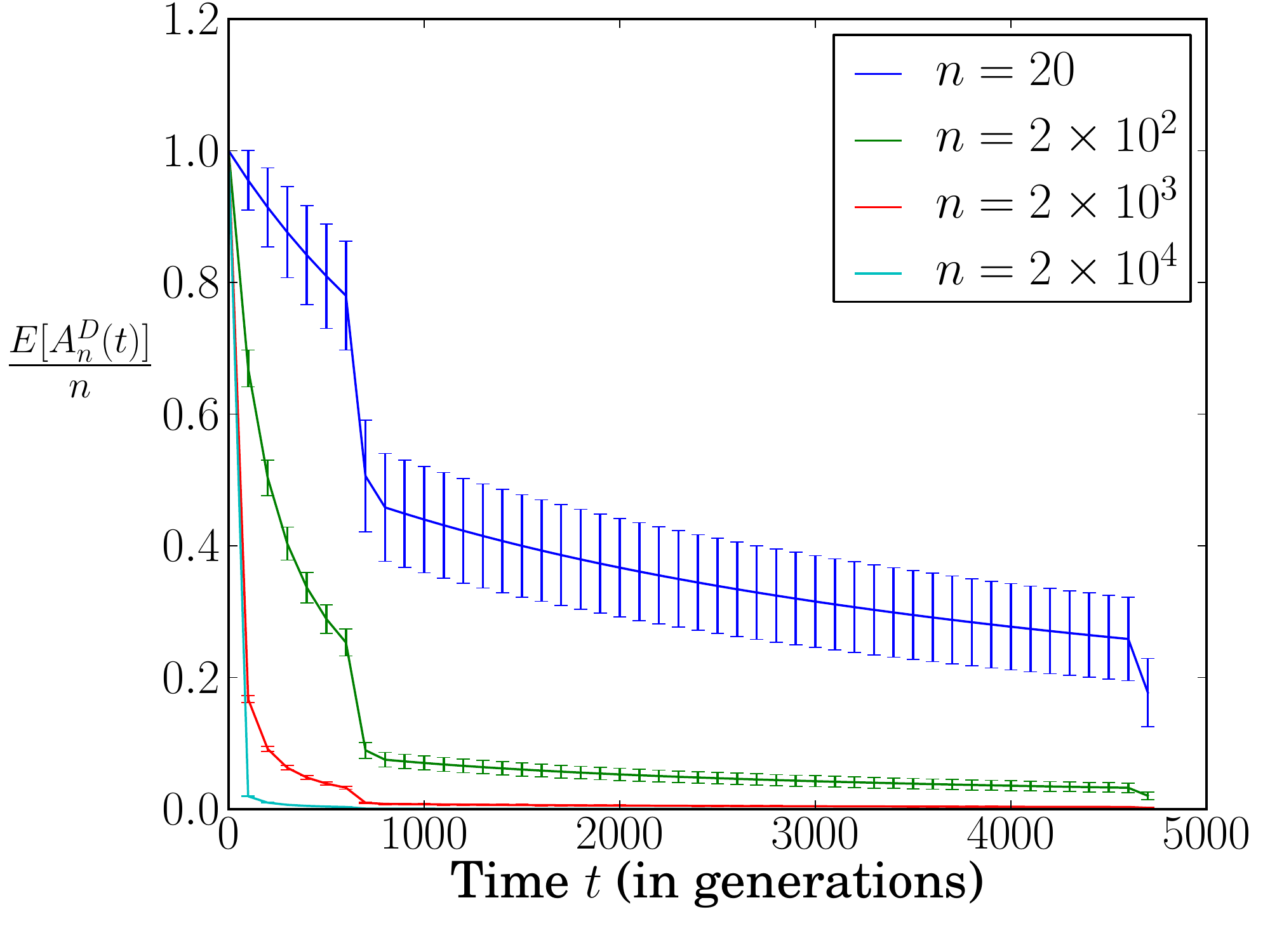}
	}
	\subfigure[]{\label{fig:wf_fraction_lineages_surviving_model3}
		\includegraphics[width=0.48\textwidth, trim=0mm 0mm 0mm 0mm,clip]{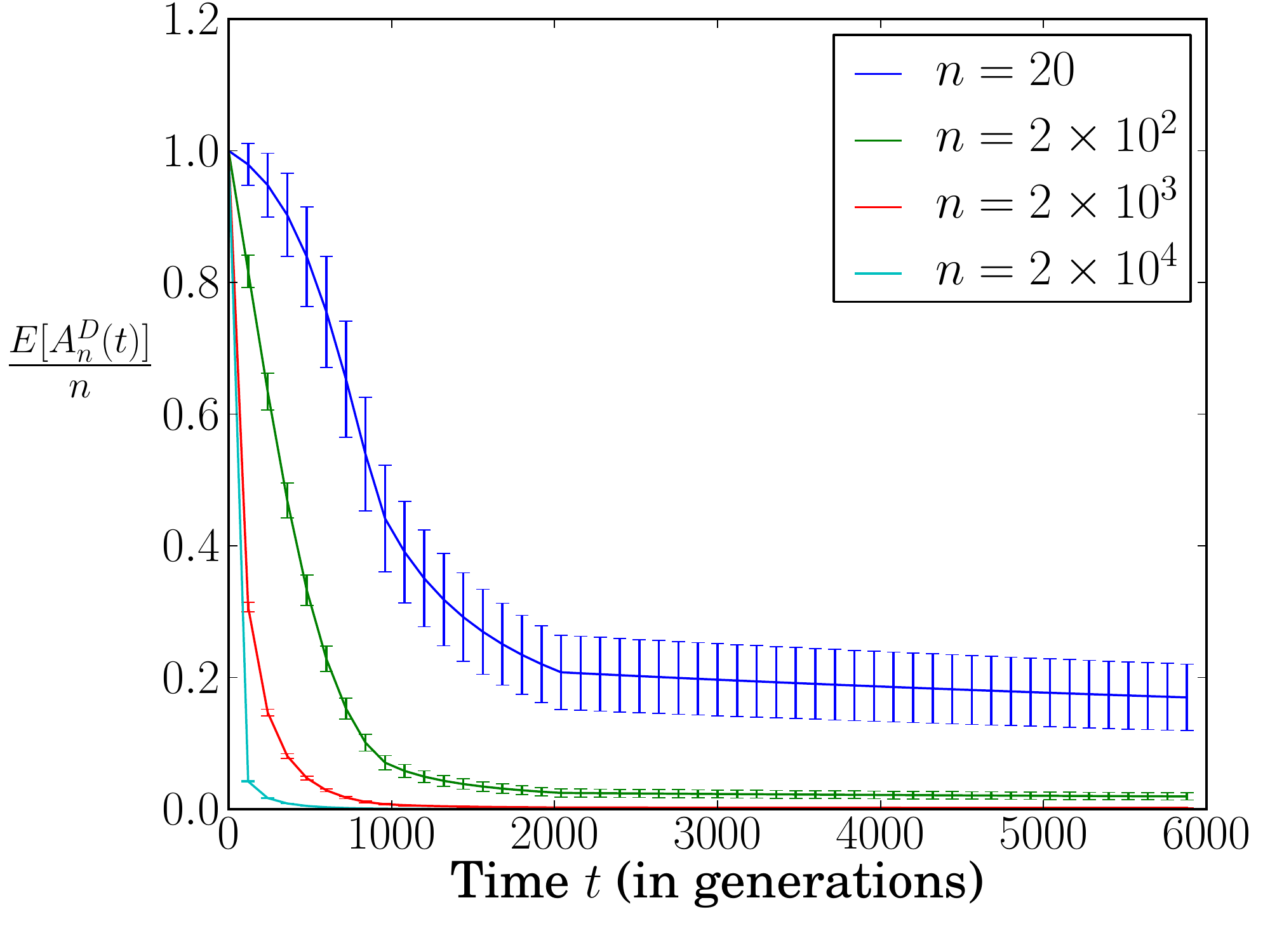}
	}
	\subfigure[]{\label{fig:wf_fraction_lineages_surviving_model4}
		\includegraphics[width=0.48\textwidth, trim=0mm 0mm 0mm 0mm,clip]{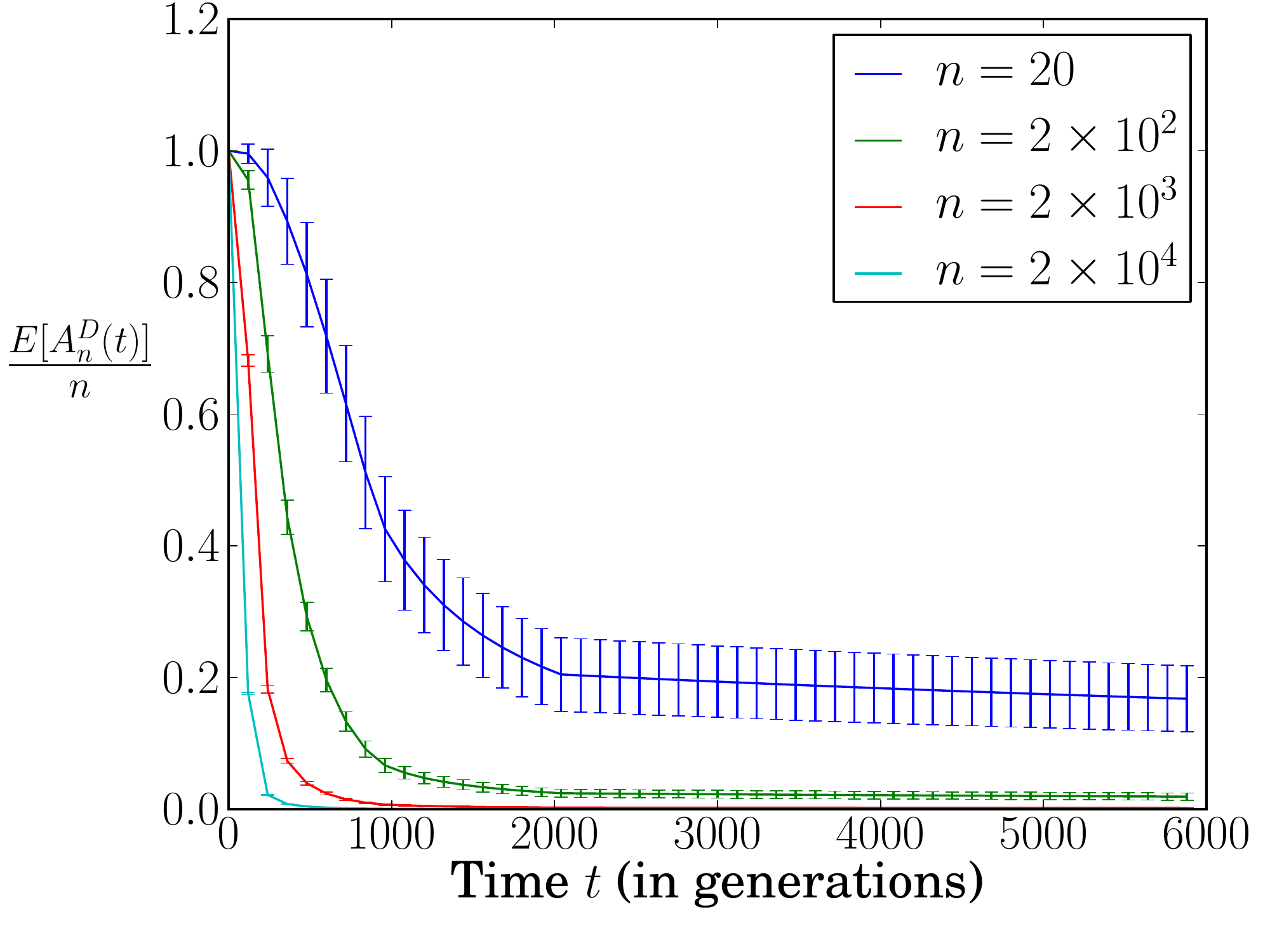}
	}
		
	\end{center}
    \caption{{\bf The expectation (solid line) and standard deviation (vertical whiskers) of the NLFT under the DTWF model.}
\subref{fig:wf_fraction_lineages_surviving_model1} \modelOne.
\subref{fig:wf_fraction_lineages_surviving_model2} \modelTwo. 
\subref{fig:wf_fraction_lineages_surviving_model3} \modelGravel.
\subref{fig:wf_fraction_lineages_surviving_model4} \modelTennessen.}
    \label{fig:wf_fraction_lineages_surviving}
\end{figure*}

\begin{figure*}[ht]       
	\begin{center}    
	\subfigure[]{\label{fig:fraction_lineages_surviving_WF_vs_coalescent_model1} 
		\includegraphics[width=0.48\textwidth]{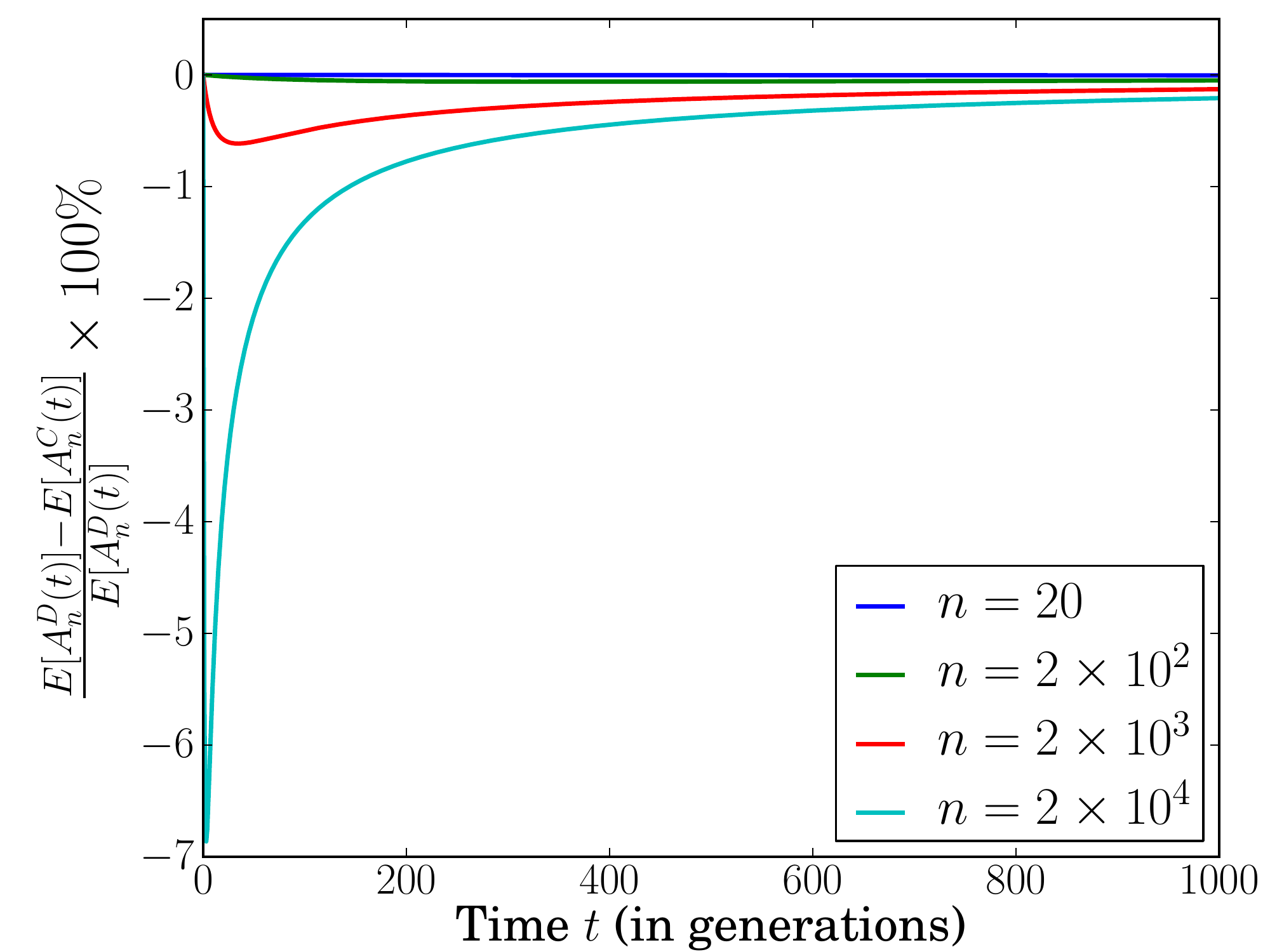}	
	}
	\subfigure[]{\label{fig:fraction_lineages_surviving_WF_vs_coalescent_model2}
		\includegraphics[width=0.48\textwidth]{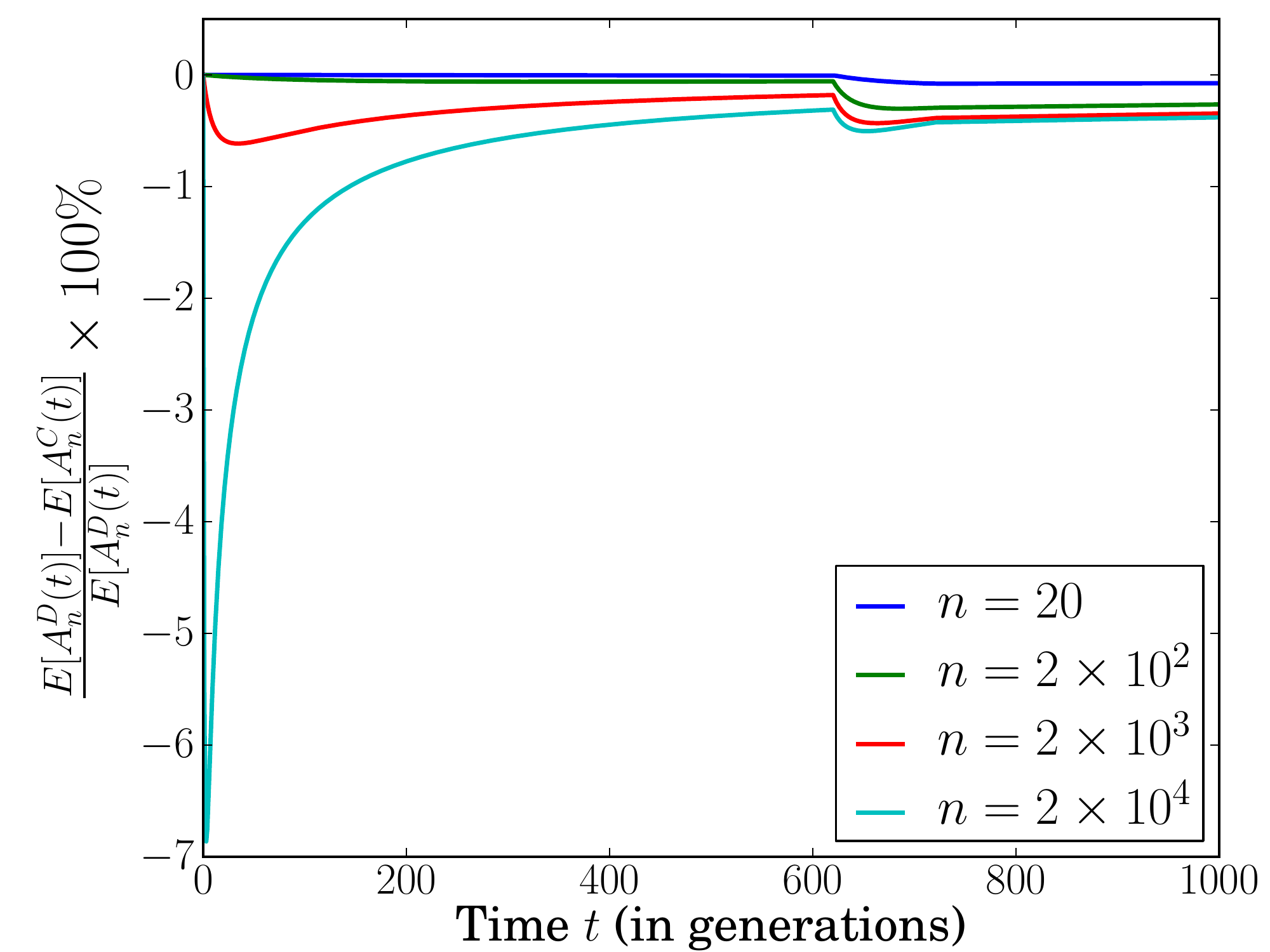}
	}
	\subfigure[]{\label{fig:fraction_lineages_surviving_WF_vs_coalescent_model3}
		\includegraphics[width=0.48\textwidth]{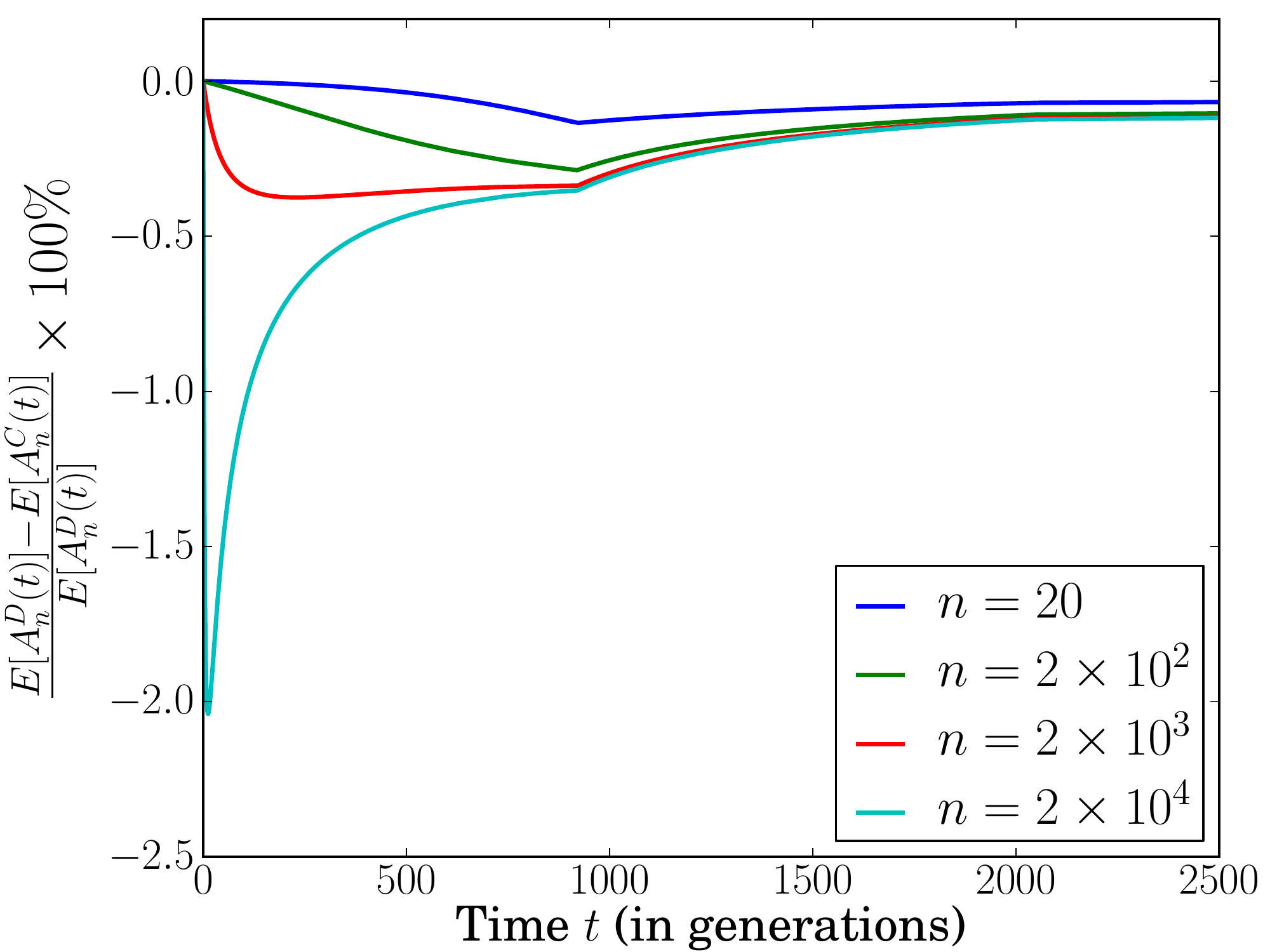}
	}
	\subfigure[]{\label{fig:fraction_lineages_surviving_WF_vs_coalescent_model4}
		\includegraphics[width=0.48\textwidth]{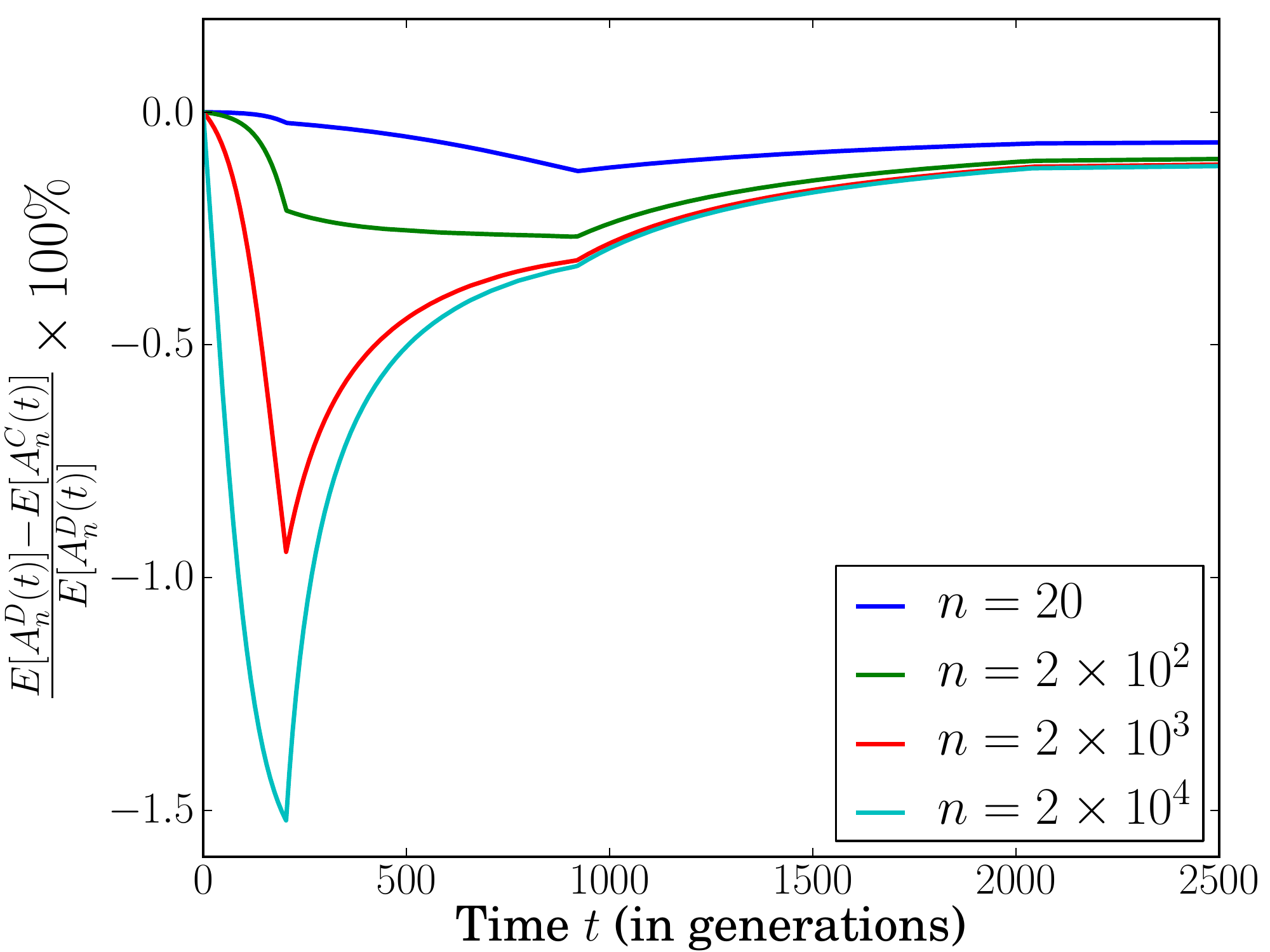}
	}
\end{center}
    \caption{{\bf The percentage difference in the expected NLFT between the coalescent and DTWF models, for a haploid sample of size $n \in \{20, 2 \times 10^2, 2 \times 10^3, 2 \times 10^4\}$.} 
The plots correspond to \subref{fig:fraction_lineages_surviving_WF_vs_coalescent_model1} \modelOne,
\subref{fig:fraction_lineages_surviving_WF_vs_coalescent_model2} \modelTwo,
\subref{fig:fraction_lineages_surviving_WF_vs_coalescent_model3} \modelGravel, and
\subref{fig:fraction_lineages_surviving_WF_vs_coalescent_model4} \modelTennessen.
For all demographic models, lineages are lost at a faster rate in the DTWF model than in the coalescent, consistent with the fact that there are a substantial number of 3-mergers in the DTWF model for large sample sizes. This deviation is more pronounced for larger sample sizes and for \textsc{Models 1--3} where the sample size is comparable to the current population size.
}
    \label{fig:fraction_lineages_surviving_WF_vs_coalescent}
\end{figure*}

\begin{figure*}[ht]       
	\begin{center}    
	\subfigure[]{\label{fig:stddev_lineages_surviving_WF_vs_coalescent_model1} 
		\includegraphics[width=0.48\textwidth]{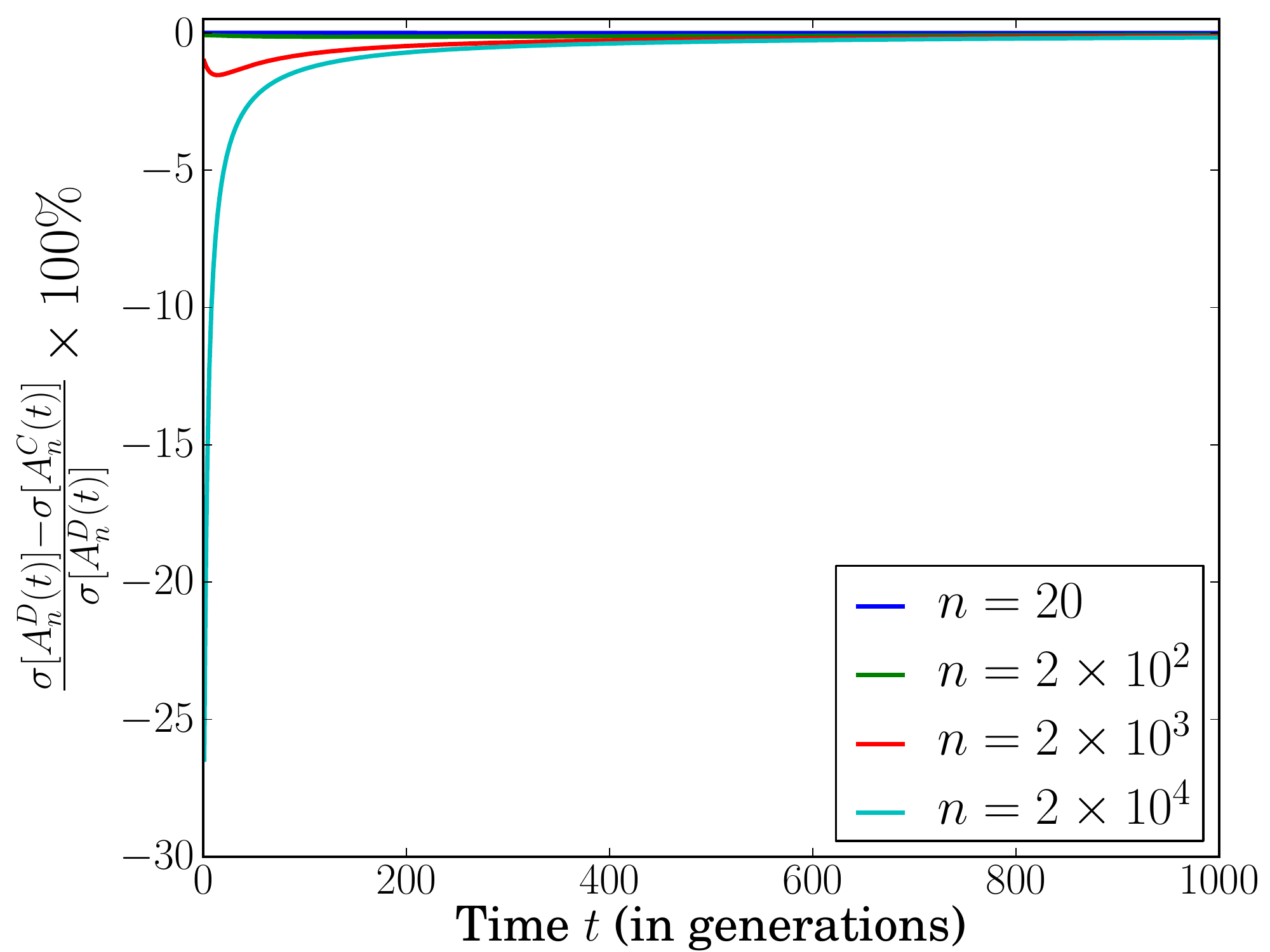}	
	}
	\subfigure[]{\label{fig:stddev_lineages_surviving_WF_vs_coalescent_model2}
		\includegraphics[width=0.48\textwidth]{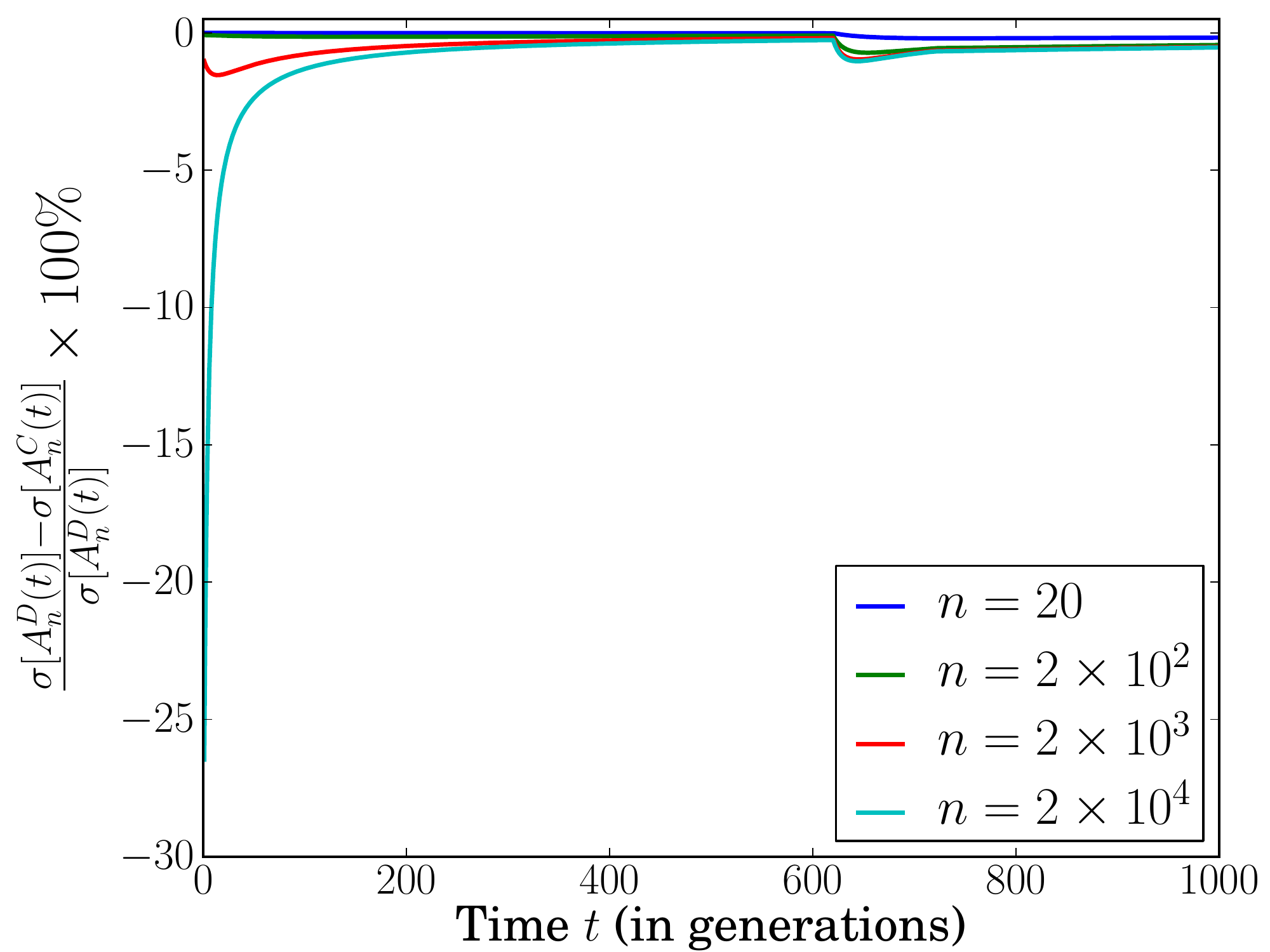}
	}
	\subfigure[]{\label{fig:stddev_lineages_surviving_WF_vs_coalescent_model3}
		\includegraphics[width=0.48\textwidth]{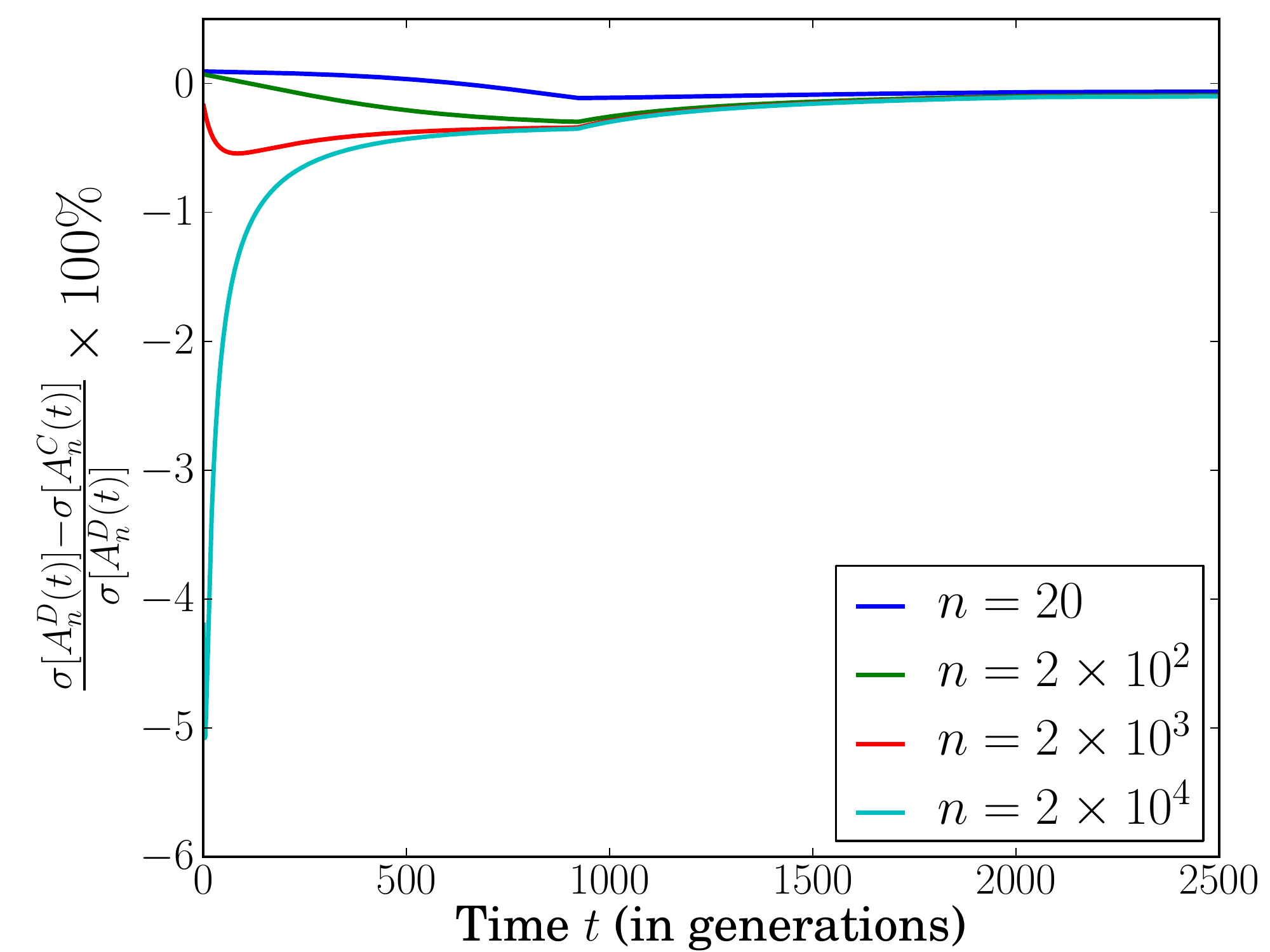}
	}
	\subfigure[]{\label{fig:stddev_lineages_surviving_WF_vs_coalescent_model4}
		\includegraphics[width=0.48\textwidth]{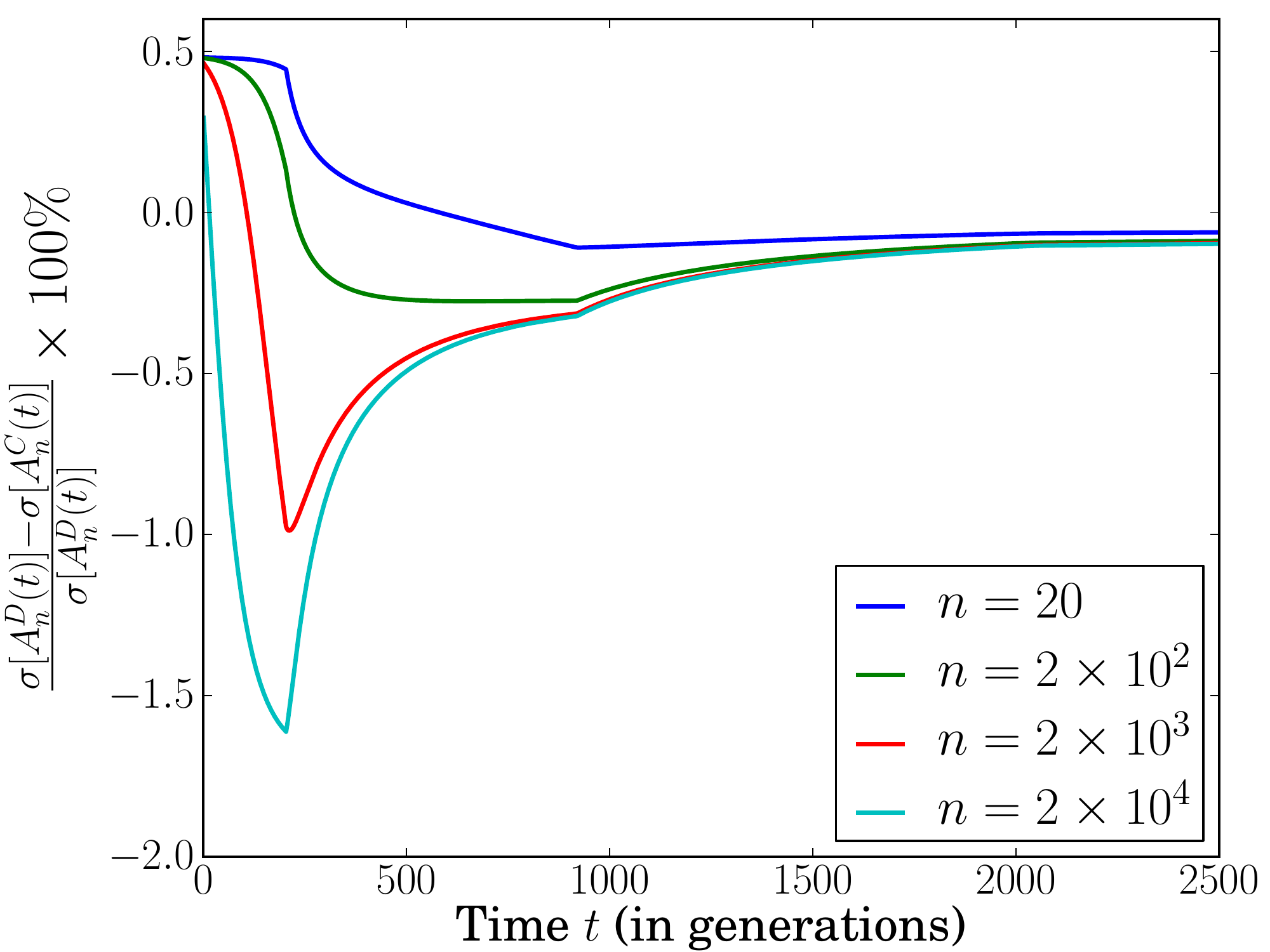}
	}
\end{center}
    \caption{{\bf The percentage difference in the standard deviation of the NLFT between the coalescent and DTWF models, for a haploid sample of size $n \in \{20, 2 \times 10^2, 2 \times 10^3, 2 \times 10^4\}$.} 
The plots correspond to \subref{fig:stddev_lineages_surviving_WF_vs_coalescent_model1} \modelOne,
\subref{fig:stddev_lineages_surviving_WF_vs_coalescent_model2} \modelTwo,
\subref{fig:stddev_lineages_surviving_WF_vs_coalescent_model3} \modelGravel, and
\subref{fig:stddev_lineages_surviving_WF_vs_coalescent_model4} \modelTennessen.
	}
    \label{fig:stddev_lineages_surviving_WF_vs_coalescent}
\end{figure*}

\begin{figure*}[ht]    
	\begin{center}
 	\subfigure[]{\label{fig:cdf_freq_spectrum_WF_model1} 
	    \includegraphics[width=0.48\textwidth]{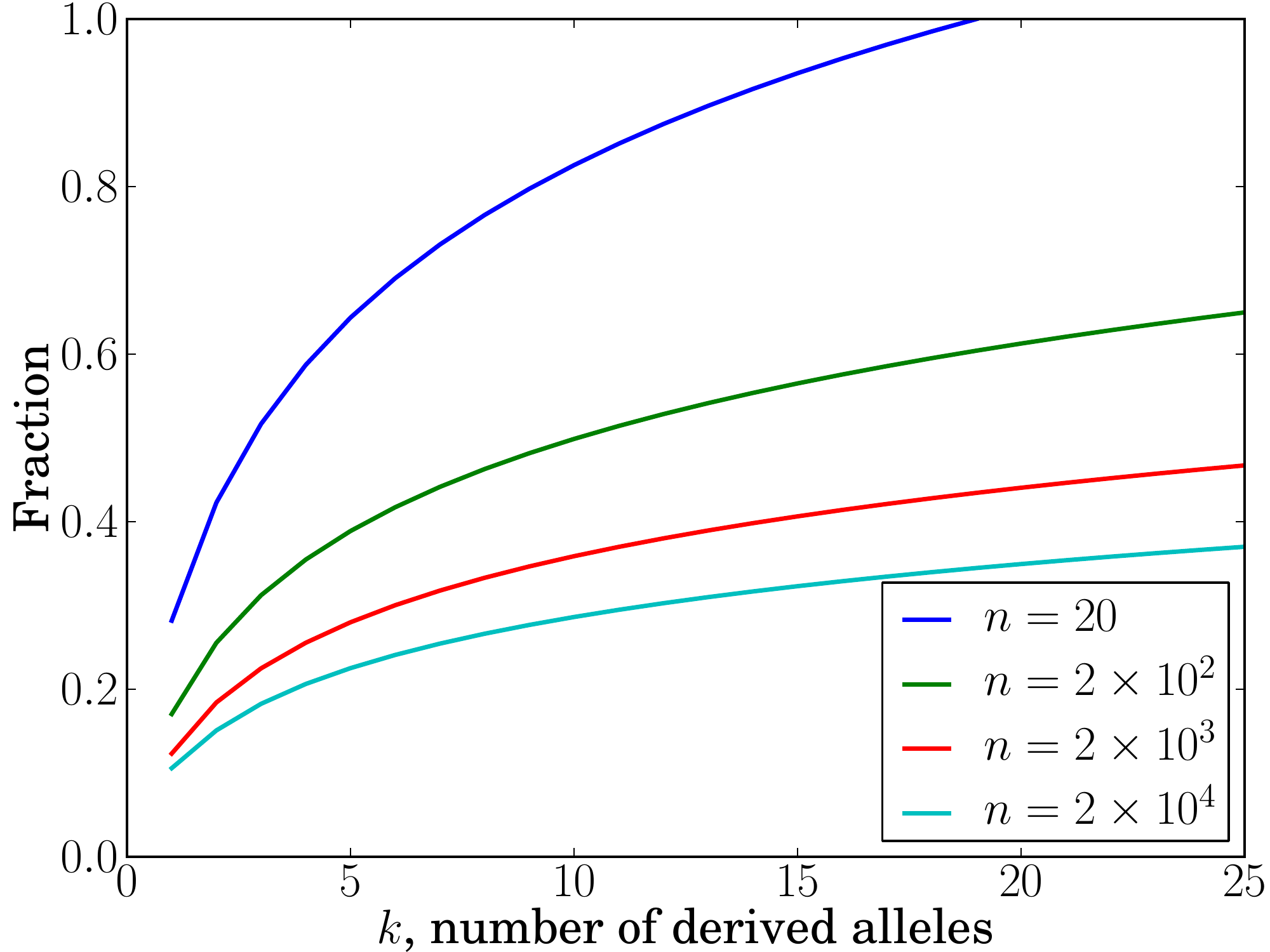}	    
	}
	\subfigure[]{\label{fig:cdf_freq_spectrum_WF_model2} 
		\includegraphics[width=0.48\textwidth]{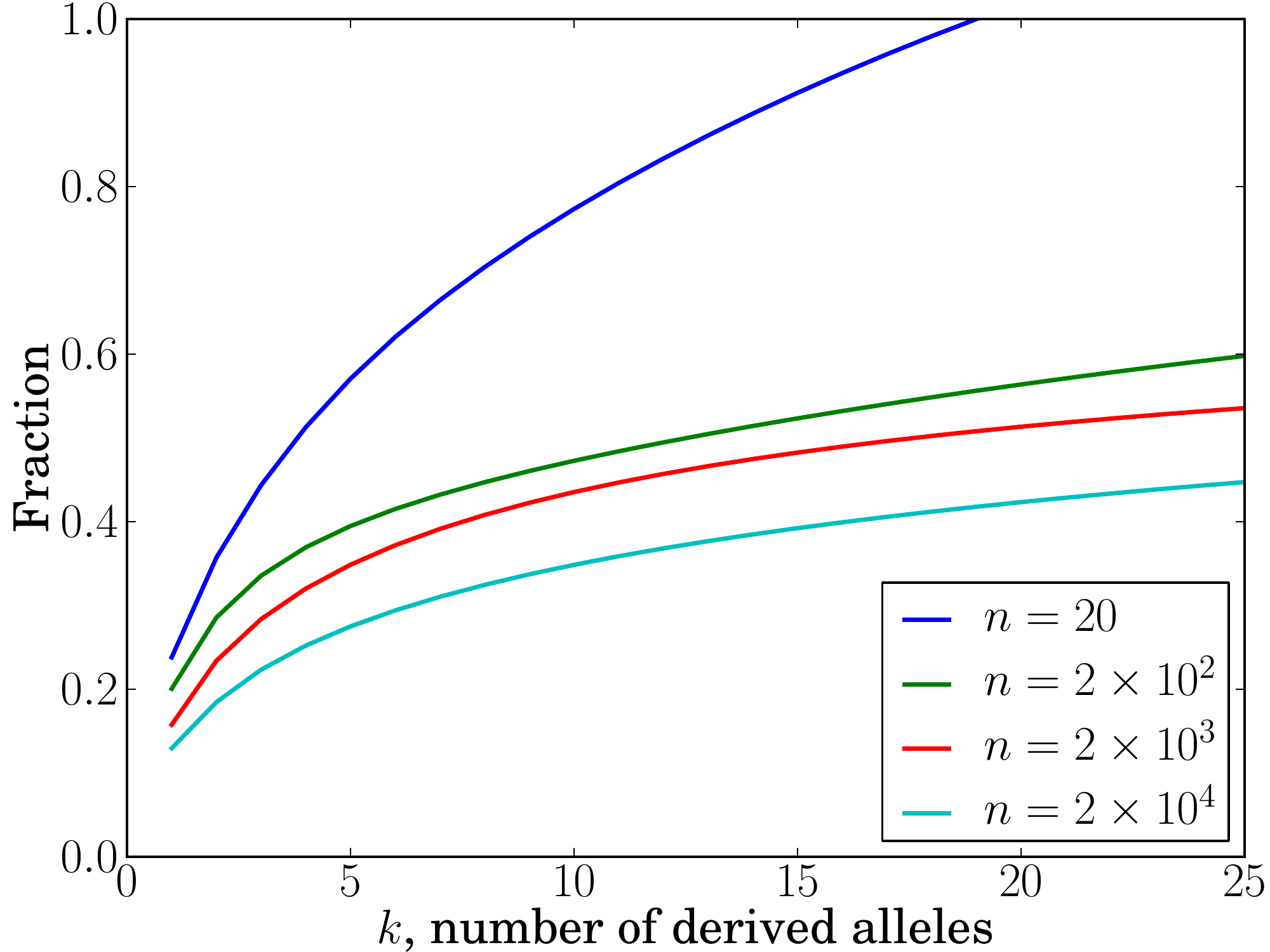} 
	}
	\subfigure[]{\label{fig:cdf_freq_spectrum_WF_model3}
		\includegraphics[width=0.48\textwidth]{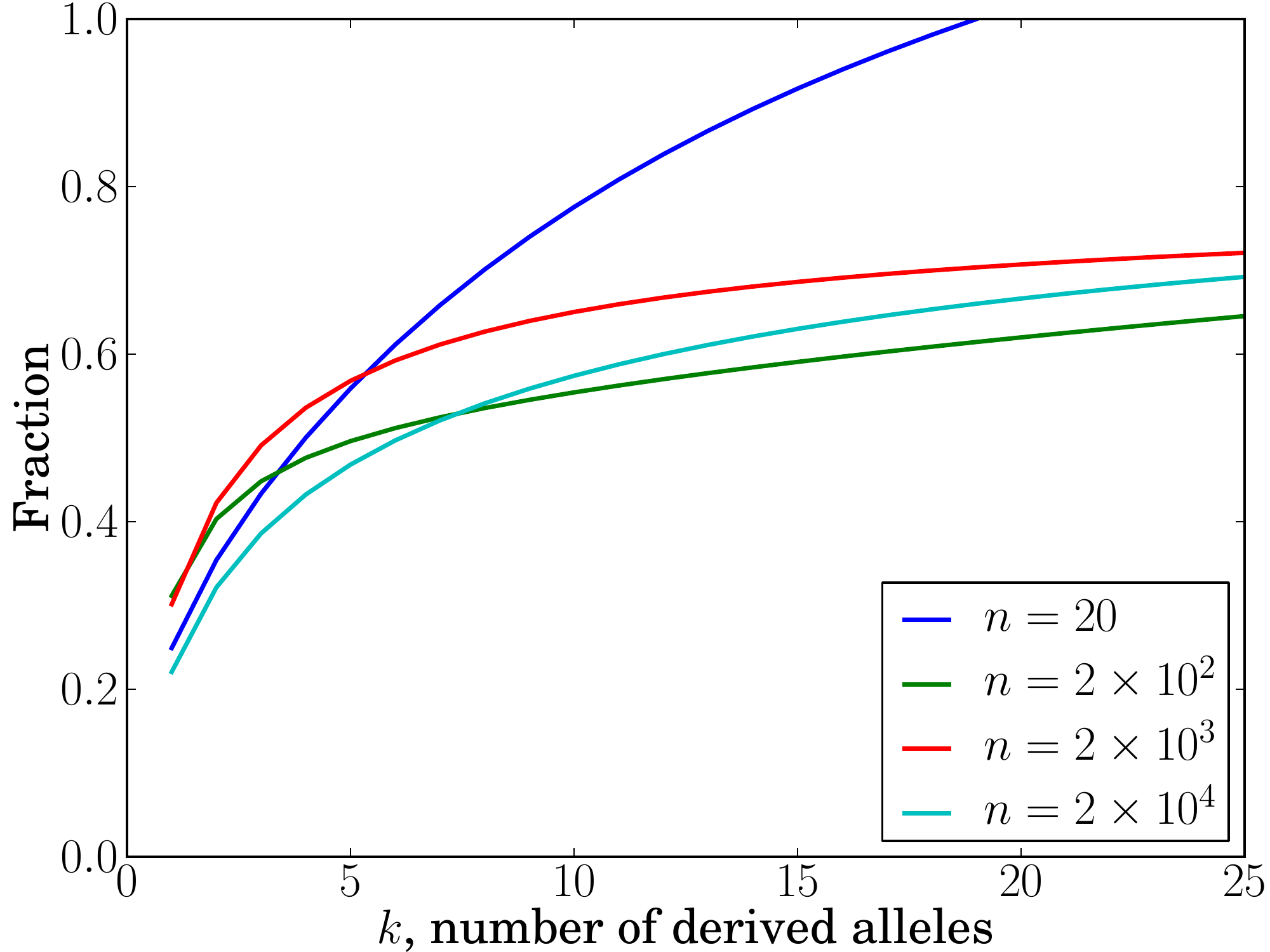}
	}
	\subfigure[]{\label{fig:cdf_freq_spectrum_WF_model4}
		\includegraphics[width=0.48\textwidth]{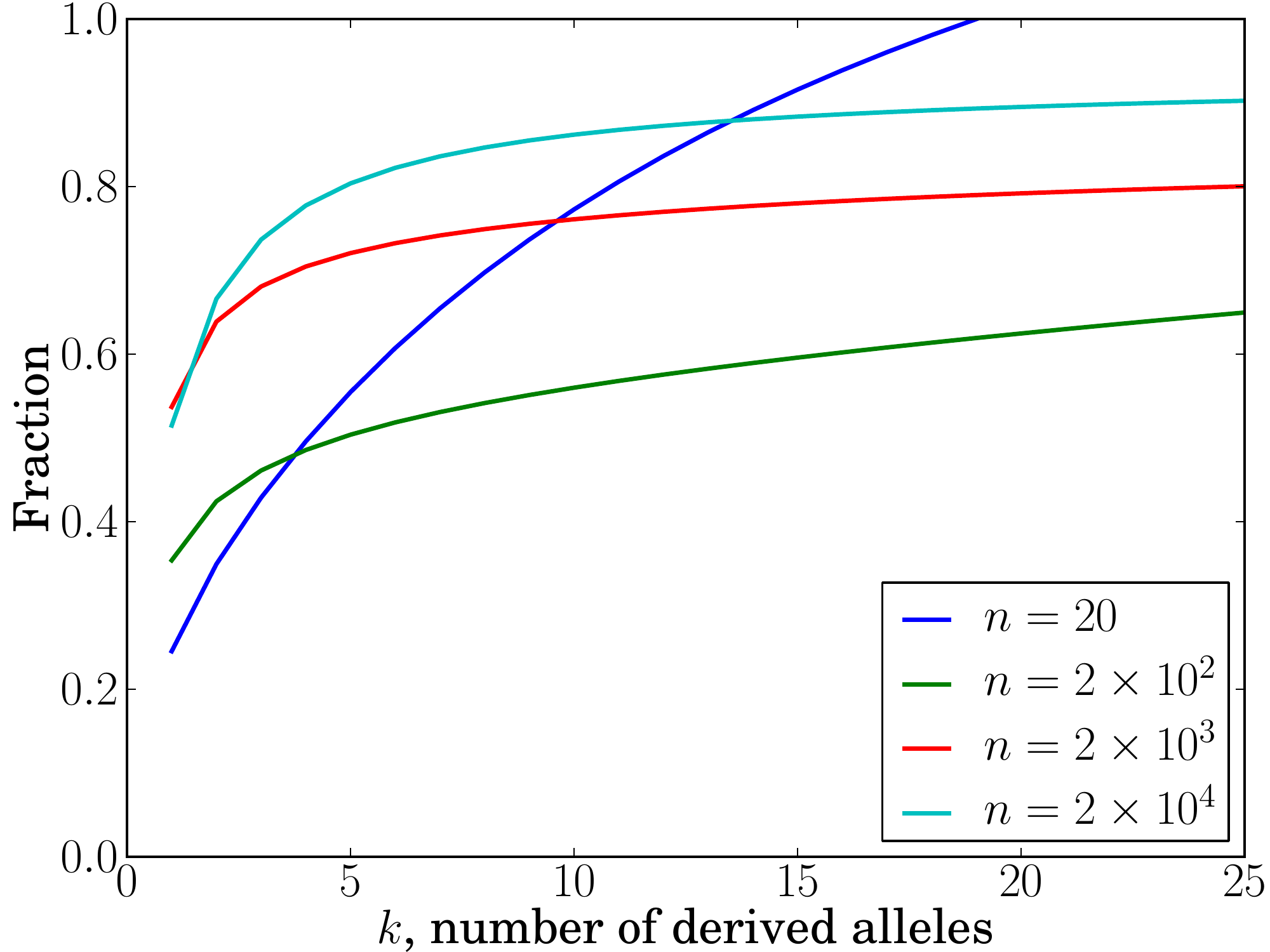}
	}
	\end{center}
	\caption{{\bf The cumulative distribution function of the frequency spectrum, as a function of the number of derived alleles, $k$, in the DTWF model with sample sizes $n \in \{20, 2 \times 10^2, 2 \times 10^3, 2 \times 10^4 \}$.} For each value of $k$ on the x-axis, the y-axis is the proportion of segregating sites with at most $k$ copies of the derived allele in the DTWF model, $\frac{\sum_{j=1}^{k} \tau^D_{n,j}}{\sum_{i=1}^{n-1} \tau^D_{n,i}}$.
The plots correspond to \subref{fig:cdf_freq_spectrum_WF_model1} \modelOne,
    \subref{fig:cdf_freq_spectrum_WF_model2} \modelTwo,
    \subref{fig:cdf_freq_spectrum_WF_model3} \modelGravel, and
	\subref{fig:cdf_freq_spectrum_WF_model4} \modelTennessen.
	In \modelTennessen~where the population grows rapidly in the recent past, about $51\%$ of the segregating sites are singletons in a sample of size $n = 2 \times 10^4$. In \modelGravel, the fraction of singleton sites is lower than in \modelTennessen~because the population growth rate is lower.
	}
	\label{fig:cdf_freq_spectrum_WF}
\end{figure*}

\begin{figure*}[ht]
	\begin{center}
		\includegraphics[width=0.75\textwidth, trim=0 0 0 0]{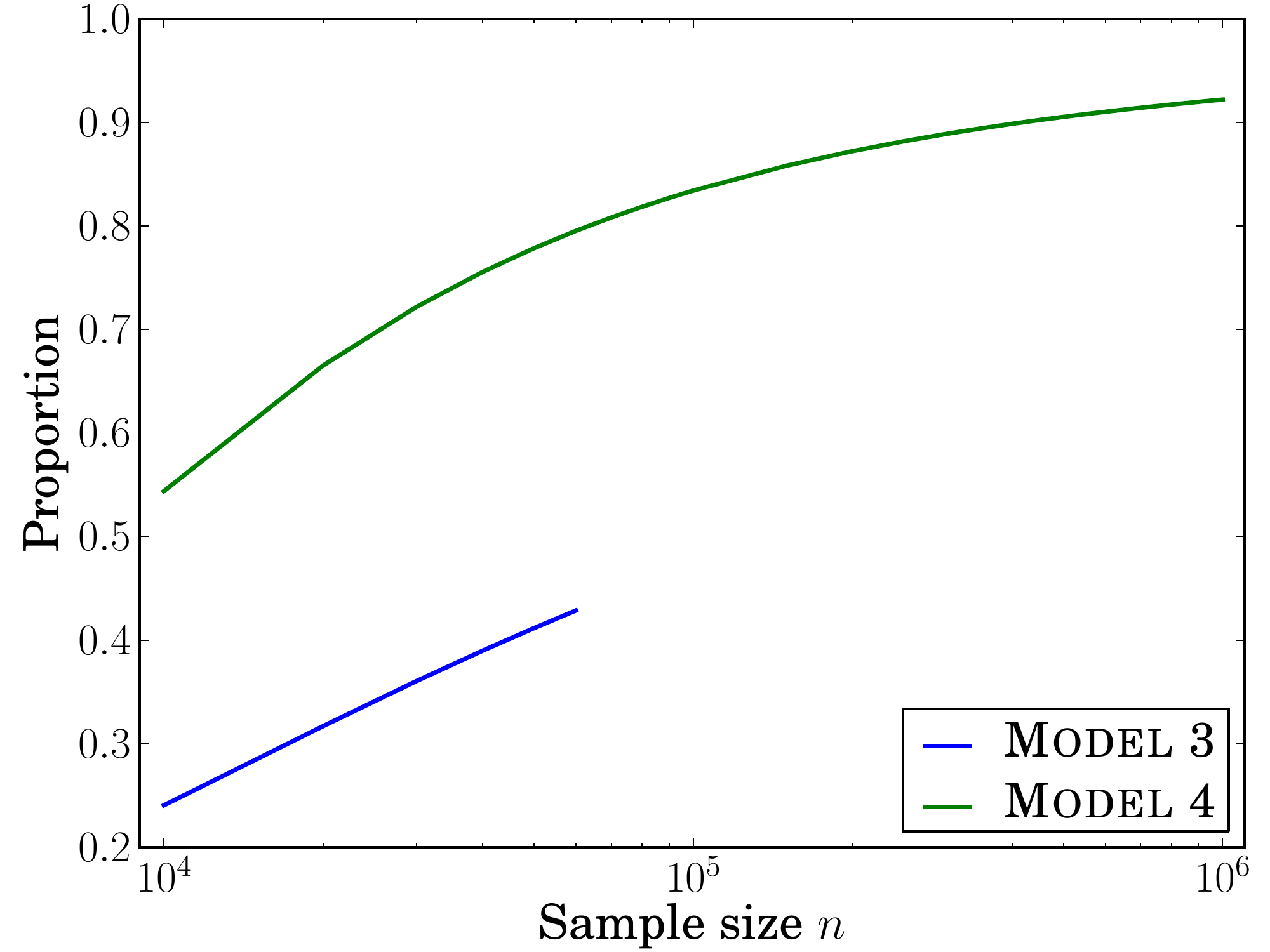}
	\end{center}
	\caption{{\bf Expected proportion of segregating sites with derived allele frequency $\leq 0.01\%$, as a function of the sample size $n$ in the coalescent for } \textsc{Models 3} {\bf and} \textsc{4}.
	The frequency of such rare variants is increasing in the sample size and in the case of \modelTennessen, rare variants practically account for all the variants for very large sample sizes.}
	\label{fig:coal_freq_spectrum_top_entries}
\end{figure*}     

\begin{figure*}[ht]
	\begin{center}
		\includegraphics[width=0.75\textwidth, trim=0 0 0 0]{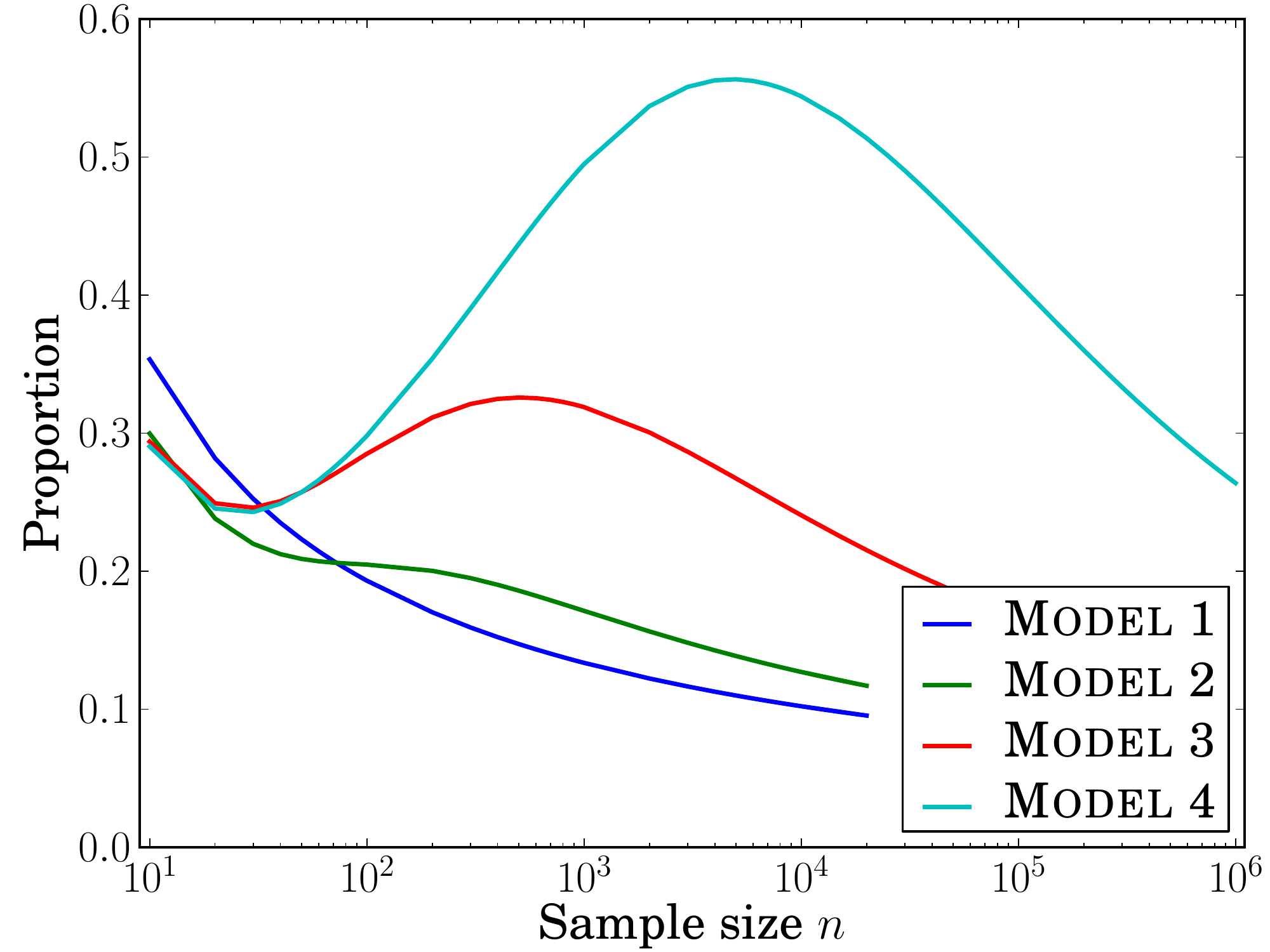}
	\end{center}
	\caption{{\bf Expected proportion of segregating sites that have only one copy of the derived allele (i.e., singletons), as a function of the sample size $n$ in the coalescent for} \textsc{Models 1--4}.
	The difference in the number of singletons in \modelTennessen~ (which incorporate recent population expansion) and the number of singletons in \textsc{Models 1} and \textsc{2} (without a recent expansion) rapidly increases for sample sizes beyond a few hundred individuals, indicating that large sample sizes are needed to infer demographic models of recent population growth from frequency spectrum data.
	}
	\label{fig:coal_freq_spectrum_singletons}
\end{figure*}     

\end{document}